\documentclass[fleqn,usenatbib]{mnras}

\usepackage[T1]{fontenc}

\DeclareRobustCommand{\VAN}[3]{#2}
\let\VANthebibliography\thebibliography
\def\thebibliography{\DeclareRobustCommand{\VAN}[3]{##3}\VANthebibliography}

\usepackage{graphicx}	
\usepackage{amsmath}	
\usepackage{amssymb}	
\usepackage{xcolor}

\usepackage{newtxtext,newtxmath}

\newcommand{\revision}[1]{#1}

\newcommand{\Msun}{\,\mathrm{M}_\odot}
\newcommand{\epsff}{\epsilon_{\mathrm{ff}}}
\newcommand{\epsffinf}{\bar{\epsilon}_{\mathrm{ff}}}
\newcommand{\epsint}{\epsilon_{\mathrm{int}}}
\newcommand{\fboost}{f_{\mathrm{boost}}}
\newcommand{\fhno}{f_{\mathrm{HN, 0}}}
\newcommand{\fhn}{f_{\mathrm{HN}}}

\newcommand{\hmol}{H$_2$}

\newcommand{\tl}{\texttt{Thelma \& Louise}}
\newcommand{\rj}{\texttt{Romeo \& Juliet}}
\newcommand{\huiic}{\texttt{Isolated MW}}
\newcommand{\um}{\textsc{UniverseMachine}}

\defcitealias{li_etal_18_paper2}{L18}

\graphicspath{{./}{figures/}}

\title[Stellar Feedback]{Testing Feedback from Star Clusters in Simulations of the Milky Way Formation}

\author[G. Brown \& O. Y. Gnedin]{
Gillen Brown\href{https://orcid.org/0000-0002-9114-5197}{\includegraphics[scale=0.04]{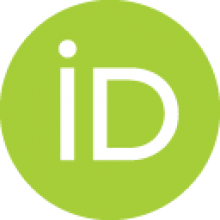}}\thanks{E-mail: gillenb@umich.edu} and
Oleg Y. Gnedin\href{https://orcid.org/0000-0001-9852-9954}{\includegraphics[scale=0.04]{orcid.png}}
\\
Department of Astronomy, University of Michigan, Ann Arbor, MI 48109, USA
}

\date{Accepted XXX. Received YYY; in original form ZZZ}

\pubyear{2021}

\begin{document}
\label{firstpage}
\pagerange{\pageref{firstpage}--\pageref{lastpage}}
\maketitle

\begin{abstract}
We present a suite of galaxy formation simulations that directly model star cluster formation and disruption. Starting from a model previously developed by our group, here we introduce several improvements to the prescriptions for cluster formation and feedback, then test these updates using a large suite of cosmological simulations of Milky Way mass galaxies. We perform a differential analysis with the goal of understanding how each of the updates affects star cluster populations. Two key parameters are the momentum boost of supernova feedback $\fboost$ and star formation efficiency per freefall time $\epsff$. We find that $\fboost$ has a strong influence on the galactic star formation rate, with higher values leading to less star formation. The efficiency $\epsff$ does not have a significant impact on the global star formation rate, but dramatically changes cluster properties, with increasing $\epsff$ leading to a higher maximum cluster mass, shorter age spread of stars within clusters, and higher integrated star formation efficiencies. We also explore the redshift evolution of the observable cluster mass function, finding that most massive clusters have formed at high redshift $z>4$. Extrapolation of cluster disruption to $z=0$ produces good agreement with both the Galactic globular cluster mass function and age-metallicity relation. Our results emphasize the importance of using small-scale properties of galaxies to calibrate subgrid models of star cluster formation and feedback.
\end{abstract}


\section{Introduction}

Most stars form in clustered environments \citep{lada_lada_03}, and young massive clusters (YMCs) are found in the Milky Way and other star-forming galaxies. The properties of young clusters are sensitive tracers of the star formation process. Young clusters show a well-defined mass function typically described as a \citet{schechter_76} function with a low-mass slope of $-2$ \citep{bastian_08,portegies_zwart_etal_10}. The cutoff mass scales with the star formation rate of the host galaxy, as does the maximum cluster mass \citep{larsen02}.

Globular clusters (GCs) are also ubiquitous within galaxies, as they are found in all nearby galaxies with stellar masses above $10^9\Msun$ \citep{brodie_strader_06}. GCs are typically old, with ages above 10~Gyr \citep{puzia_etal_05, strader_etal_05}, and have sizes of a few parsecs that are consistent with YMCs found in the local universe \citep{brown_gnedin_21b}. This naturally leads to the hypothesis that GCs are the surviving subset of a larger population of YMCs that formed at high redshift. However, the mass function of GCs is well characterized by a lognormal distribution with a peak mass of around $2\times10^5\Msun$ \citep{harris_91,jordan_etal_07}, in contrast to the \citet{schechter_76} function commonly used to describe YMCs. This transformation of the mass function over cosmic time requires a preferential destruction of low-mass clusters \citep{fall_zhang_01,vesperini_etal_03,prieto_gnedin_08,elmegreen_10,kruijssen_15}. 

The dynamical evolution of clusters results from a combination of stellar evolution, two-body relaxation, tidal truncation, and tidal shocks \citep{spitzer_1958,elmegreen_10,gnedin_ostriker97,gieles_renaud_16}. While stellar evolution and internal two-body relaxation can be well understood by studying isolated clusters, dynamical evolution depends on the tidal field and requires more detailed modeling. Throughout their lifetime, clusters experience tidal interactions with their natal giant molecular clouds (GMCs), the galactic structure, and other GMCs, leading to complex evolution that is not well-approximated by idealized models \citep{meng_gnedin_22}. 

Simulations of galaxy formation are well-suited for a detailed investigation of cluster formation and disruption \citep{renaud_etal_17,pfeffer_etal_18}. By situating clusters within their galactic context, their formation and evolution can be realistically tracked. However, few cosmological simulations have the resolution required to directly resolve cluster formation \revision{or disruption}, so they must rely on subgrid models (although see 
\citealt{kim_etal_18,lahen_etal_19,benincasa_etal_20,ma_etal_20,hislop_etal_22}). As cluster formation is terminated by feedback from the newly-formed stars, simulations must self-consistently determine this feedback to obtain reliable properties of star clusters. Prescriptions for stellar feedback, particularly supernova feedback, have undergone many revisions over the years as they are calibrated against observations \citep{katz_92,springel_hernquist_03,stinson_etal_06,agertz_etal_11,agertz_etal_13,hopkins_etal_14,keller_etal_14,hopkins_etal_18}. However, these feedback models are often only tested against galaxy-scale properties, such as the global star formation rate or Kennicutt-Schmidt relation \citep{schmidt_59,kennicutt_98}. To properly model star cluster formation, these feedback prescriptions must be calibrated on smaller scales.

In \citet{li_etal_17_paper1} and \citet{li_etal_18_paper2} (hereafter L18), our group introduced a suite of cosmological simulations that directly models star cluster formation and disruption. The high spatial resolution of these simulations (3-6~pc) allows us to resolve GMCs where star cluster formation occurs. Star particles are seeded within GMCs and accrete material from their surroundings until feedback from the newly-formed stars stops further accretion. The final masses of these star particles are set self-consistently and represent the masses of individual star clusters. These simulations were able to reproduce many aspects of the observed young cluster populations, including the shape of the initial cluster mass function, the total mass of stars contained in bound clusters, the relationship between the maximum cluster mass and the star formation rate surface density, and the formation timescales of star clusters. Some of the central clusters in satellite galaxies have properties consistent with nuclear star clusters in the local universe \citep{brown_gnedin_18}. Star formation sites in the modeled galaxies at high redshift are temporarily contained in giant clumps, which dissolve after $\sim$100~Myr \citep{meng_gnedin_20}. In addition, these simulations resolve dense irregular structures within the thick galactic disk \citep{meng_gnedin_21}, allowing for an accurate calculation of the tidal field and therefore the tidal disruption of clusters \citep{li_gnedin_19_paper3,meng_gnedin_22}.

While these simulations advanced our modeling of star cluster formation in cosmological simulations, they still had several limitations. First, they reached only redshift $z=1.5$. This precludes a direct comparison to the GCs of the Milky Way (MW), as the disruption up to $z=0$ must be estimated.
Second, these simulations include only one MW-mass galaxy and its satellites, decreasing the statistical power of the results and potentially making the results dependent on the specific initial condition (IC) used.

In this paper, we present the next generation of simulations based on the prescriptions of \citetalias{li_etal_18_paper2}. These simulations use two Local Group-like ICs, with the goal of reaching $z=0$ with four MW-mass galaxies. In Section~\ref{sec:code_updates} we describe improvements to the formation and feedback schemes, then describe the new suite of simulations. This suite includes nine runs using Local Group-like ICs and 20 using the Milky Way-like IC from \citetalias{li_etal_18_paper2}. These runs vary a wide range of feedback and cluster formation parameters, allowing us to explore how different prescriptions affect the resulting cluster properties in Section~\ref{sec:results}. We perform a differential analysis, systematically exploring each of the parameters we vary. In Section~\ref{sec:jwst} we present an application of these simulations by presenting the redshift evolution of the observable cluster mass function. We discuss remaining uncertainties and compare our results with observations in Section~\ref{sec:discussion}, then summarize our results in Section~\ref{sec:conclusions}. 

\section{Simulation code and setup}
\label{sec:code_updates}

\revision{In this section we describe the ART code and the properties of the simulations. Throughout this section we introduce several parameters of the code, which we list for convenience in Table~\ref{tab:parameters}.}

\begin{table}
    \caption{\revision{Key parameters of the star formation and feedback prescriptions with the values used in this paper.}}
    \begin{tabular}{ll}
        \hline
        Parameter & Value(s) \\
        \hline
        Molecular fraction threshold for cluster creation & 50\% \\
        Maximum virial parameter for cluster creation & 10 \\
        Density threshold for cluster creation and growth & 1000~cm$^{-3}$ \\
        Star formation efficiency per freefall time ($\epsff$) & 1\%, 10\%, 100\% \\
        Radius of GMC & 5~pc \\
        Clumping factor ($C_\rho$) & 3, 10, 30 \\
        Stellar IMF range & 0.08--50 $\Msun$ \\
        Stellar mass range for SNII & 8--50~$\Msun$ \\
        Stellar mass range for HN & 20--50~$\Msun$ \\
        Initial hypernova fraction ($\fhno$) & 0\%, 5\%, 20\%, 50\% \\
        SNII momentum boost ($\fboost$) & 1, 2, 3, 5 \\
        Stellar mass range for AGB & 0.08--8~$\Msun$ \\
        Number of SNIa per unit stellar mass & $1.6\times10^{-3} \Msun^{-1}$ \\
        \hline
    \end{tabular}
    \label{tab:parameters}
\end{table}

\subsection{The ART code}

For our simulations we use the Adaptive Refinement Tree (ART) code \citep{kravtsov_etal_97,kravtsov_99,rudd_etal_08,li_etal_17_paper1,li_etal_18_paper2}. The ART code includes many physical processes that are important for modeling the formation of galaxies. Radiative transfer is calculated using an improved version of the Optically Thin Variable Eddington Tensor method \citep{gnedin_abel01}, which has been revised to minimize numerical diffusion \citep{gnedin_14}. Radiation from both stars and the extragalactic background \citep{haardt_madau_01} are included. A non-equilibrium chemistry network of molecular hydrogen is used to identify star-forming regions within GMCs.  It was calibrated using observations in nearby galaxies \citep{gnedin_kravtsov_11} and updated to include line overlap in computing self-shielding of molecular hydrogen \citep{gnedin_draine_14}.  This chemical network also calculates the ionization states of hydrogen and helium. This model uses the local abundance of all these species to calculate the heating and cooling functions self-consistently, without any assumptions of photoionization equilibrium or collisional equilibrium. The ART code also includes a subgrid-scale (SGS) model for numerically unresolved turbulence developed by \citet{semenov_etal_16}, which follows the results of the MHD simulations of \citet{padoan_etal_12}. 

A particularly novel aspect of the ART code is the direct modeling of time-resolved star cluster formation \citep{li_etal_17_paper1,li_etal_18_paper2,li_gnedin_19_paper3}. Star cluster particles are seeded in dense gas, and accrete gas from a surrounding region until feedback from the new cluster terminates gas accretion. This region, which we refer to as the ``GMC,'' has a radius of 5~pc and is fixed in physical size at all cosmic epochs. With the maximum spatial resolution of our simulations being set in the range of 3-6~pc, the GMC can extend past the central cell, allowing the cluster to accrete gas from neighbor cells. Specifically, the growth rate of a given cluster is 
\begin{equation}
    \dot{M} = \frac{\epsff}{t_\mathrm{ff}} \sum_\mathrm{cell} f_\mathrm{GMC}\, V_\mathrm{cell}\, f_{\mathrm{H}_2}\, \rho_\mathrm{gas}
    \label{eq:m_dot}
\end{equation}
where $\epsff$ is the local star formation efficiency per freefall time $t_\mathrm{ff}$, $f_\mathrm{GMC}$ is the fraction of cell volume $V_\mathrm{cell}$ included within the GMC sphere, $f_{\mathrm{H}_2}$ is the local mass fraction of molecular gas, and $\rho_\mathrm{gas}$ is the local total gas density. This mass growth is accumulated at each local timestep, which is typically in range of $10^2-10^3$ years. As long as the local gas density is above the threshold, clusters can continue accreting gas. This accretion stops either when it has accreted no material in the last 1~Myr or when it has reached an age of 15~Myr.  

To avoid the spurious creation of many small clusters, we impose a threshold such that clusters must have an expected mass (defined as the initial $\dot{M}$ times the maximum allowed formation time of 15~Myr) of at least 6000~$\Msun$. As clusters typically form over a few Myr, rather than the full 15~Myr, this results in the elimination of small clusters below about 1000~$\Msun$.

Due to the complex dynamical evolution that occurs throughout the process of cluster formation, not all stars in a given star-forming region will be bound to the fully-formed cluster. To model this, star cluster particles include a variable tracking the fraction of mass that is gravitationally bound. This is set at cluster formation (see Section \ref{sec:code_cluster_formation}) and is updated as clusters undergo dynamical disruption throughout their lifetime. 

\subsection{Updates to the cluster formation modeling}
\label{sec:code_cluster_formation}

We implement several updates to the ART code to improve the star cluster formation algorithm. In the implementation of \citetalias{li_etal_18_paper2}, a cluster particle is created if the gas density in a cell reaches $n_H > 1000\,\mathrm{cm}^{-3}$ and the local $H_2$ mass fraction is larger than 0.5, meaning the cell contains mostly dense molecular gas. Here we introduce an additional criterion based on the local virial parameter of the gas, intended to select gravitationally bound gas. Generally, the virial parameter is
\begin{equation}
    \alpha_{\rm vir} = \frac{5 \sigma^2 R}{3 G M}    
    \label{eq:virial_1}
\end{equation}
where $\sigma$ is the local gas velocity dispersion, $R$ is the radius of the sphere we consider, and $M$ is the mass within this sphere. We calculate this locally in any cell meeting the other star formation criteria, assuming a sphere with a diameter equal to the size of the cell ($l = 2R$), giving
\begin{equation}
    \alpha_{\rm vir} = \frac{5\sigma^2}{\pi G \rho_\mathrm{gas} l^2}.
    \label{eq:virial_2}
\end{equation}
We use both the turbulent velocity and sound speed when calculating the velocity dispersion ($\sigma^2 = v_{\rm turb}^2 + c_s^2$), but do not include cell-to-cell velocity differences. We require $\alpha_{\rm vir} < 10$ to seed star clusters. This threshold is near the typical value for observed GMCs in the Milky Way \citep{mivilledeschenes_etal_17}. Star formation is allowed on the four finest refinement levels. 

We also use a new prescription for the initial bound fraction of star clusters, as determined by \citet{li_etal_19}. These authors performed simulations of 80 isolated molecular clouds with a range of mass, size, velocity configuration, and feedback strength. After feedback terminates star formation, they calculate the integrated star formation efficiency $\epsint$, which is the fraction of the initial gas mass that formed stars, as well as the fraction of stars that are bound to the final cluster $f_{\rm bound}$. They then determine the relation between these two parameters:
\begin{equation}
    f_{\rm bound} = \left[ \rm{erf} \left(\sqrt{\frac{3 \epsint}{\alpha_\star}}\right) - \sqrt{\frac{12 \epsint}{\pi \alpha_\star}}\exp\left(-\frac{3 \epsint}{\alpha_\star}\right)  \right] f_{\rm sat} 
    \label{eq:bound_fraction}
\end{equation}
where $\alpha_\star=0.48$ and $f_{\rm sat}=0.94$ are free parameters the authors fitted. Determining $\epsint$ in our simulations is not trivial. The initial gas mass when the cluster was seeded is not an accurate representation of the available gas mass, as GMCs accrete material over time. To account for this, we define $\epsint$ as the ratio of the final stellar mass to the maximum value of the stellar mass plus gas mass at any time during cluster formation:
\begin{equation}
    \epsint = \frac{M_{\star, {\rm final}}}{\max \left( M_{\star}(t) + M_{g}(t) \right)}
    \label{eq:eps_int}
\end{equation}
We then use this directly in Equation~\ref{eq:bound_fraction} to calculate the initial bound fraction for each star cluster.

\subsection{Cluster disruption modeling}
\label{sec:cluster_disruption}

Our model for cluster disruption is unchanged from that described in detail in \citet{li_gnedin_19_paper3}, but we summarize the key points here. At each global timestep of the simulation (the length of the global timestep is typically a few Myr, with a maximum of 50~Myr), we calculate the tidal tensor around all fully formed clusters using the second-order finite difference of the gravitational potential across a $3\times3\times3$ cell cube centered on the star particle. To determine cluster disruption in runtime, we calculate the three eigenvalues of the tidal tensor $\lambda_1 > \lambda_2 > \lambda_3$, which describe the strength of the tidal field in the direction of their corresponding eigenvectors. We use the maximum of the absolute value of the eigenvalues to determine the dynamical timescale within the Roche lobe of the cluster:
\begin{equation}
    \Omega_{\rm tid}^2(t) = \frac{\lambda_m}{3}
    \label{eq:omega_tid}
\end{equation}
where 
\begin{equation}
    \lambda_m \equiv \max_i | \lambda_i |
\end{equation}
We then use it to determine the cluster disruption timescale:
\begin{equation}
    t_{\rm tid} = 10 \ {\rm Gyr} \left(\frac{M(t)}{2 \times 10^5 \Msun}\right)^{2/3} \frac{100 \ {\rm Gyr}^{-1}}{\Omega_{\rm tid}(t)}
    \label{eq:disruption_timescale}
\end{equation}
Finally, we use this cluster disruption timescale to decrease the mass bound to each cluster. We track it with the variable $f_{\rm dyn}$, which describes the fraction of cluster mass \revision{bound to the cluster after accounting for dynamical disruption}. At the $n$-th global timestep of length $dt_n$, we update this fraction as follows:
\begin{equation}
    f_{\rm dyn}^{n+1} = \exp{\left(-dt_n / t_{\rm tid}\right)} \, f_{\rm dyn,}^{n}
\end{equation}

We also output the full tidal tensor for each star particle at each global timestep, allowing us to postprocess star cluster disruption and explore how different prescriptions for tidal disruption, including capturing tidal shocks, may change cluster properties.

\subsection{Updates to the stellar feedback modeling}
\label{sec:feedback}

\subsubsection{Abundances of individual elements}

We have implemented runtime tracking of most important individual elements (C, N, O, Mg, S, Ca, Fe) and ejecta of AGB stars. This gives 10 total fields tracking chemical enrichment (C, N, O, Mg, S, Ca, Fe, Z$_{\rm SN Ia}$, Z$_{\rm SN II}$, and Z$_{\rm AGB}$) in both gas and stars. These elements are some of the most abundant in the universe, have reliable yields, and enable comparisons with both gas-phase and stellar abundance measurements at a variety of redshifts. N, O, and S are commonly used to measure gas-phase metallicity \citep[e.g.][]{kewley_dopita_02,maiolino_mannucci_19}. Fe, Mg, and Ca are commonly measured in stellar spectra, with Fe representing total metallicity and Mg and Ca being representative $\alpha$ elements \citep{gallazii_etal_05,kirby_etal_13,hayden_etal_15}. 

\subsubsection{Discrete supernova events}

We have updated the supernova (SN) feedback prescriptions in the ART code to include discrete SN explosions at rate calculated from the stellar lifetimes, IMF, and total stellar mass of the particle. Conceptually, we use the stellar lifetimes to calculate the mass range of stars leaving the main sequence during a given timestep, then integrate the IMF over this range to determine the total number of stars leaving the main sequence. We explode an integer number of these as SN, leaving any fractional SN to accumulate to the next timestep. This leads to only an integer number of SN exploding in a given timestep while also appropriately conserving the total number of SN over the life of the stellar population.

We calculate the number of SN in a given timestep:
\begin{align}
    N_{\rm SN}(\tau) + &N_{\rm SN,leftover}(\tau+dt) = \nonumber \\
    &M_\star(\tau) \int_{\mathcal{M}(\tau)}^{\mathcal{M}(\tau+dt)} \Phi(\mathcal{M}) d\mathcal{M} + N_{\rm SN,leftover}(\tau)
    \label{eq:imf_integral}
\end{align}
where $\tau$ is the age of the stellar population (discussed in more detail in Section~\ref{sec:timing}), $dt$ is the length of the current timestep, $M_\star$ is the total mass of the cluster particle, $\mathcal{M}(\tau)$ is the mass of the star leaving the main sequence at age $\tau$, $\Phi(\mathcal{M})$ is the IMF normalized such that $M_\star = \int \mathcal{M}\Phi(\mathcal{M})d\mathcal{M}$, and $N_{\rm SN,leftover}$ is the fractional number of SN not exploded in the previous timestep. $N_{\rm SN}$ is always an integer value, and  $0 \leq N_{\rm SN,leftover} < 1$. We use a \citet{kroupa_01} IMF with a mass range of 0.08 to $50\Msun$, and use $8\Msun$ as the minimum mass to explode as a SN. We use the metallicity-dependent analytic stellar lifetimes from \citet{raiteri_etal_96}.

When SN explode, we inject energy and mass into the surroundings. The mass of different elements is taken directly from the stellar yield tables of \citet{kobayashi_etal_06}. We use the yield for a star of mass $\mathcal{M} = 0.5\left(\mathcal{M}(\tau) + \mathcal{M}(\tau+dt) \right)$, and use the metallicity of the star particle. We linearly interpolate the yield tables in both mass and metallicity to determine the yields at arbitrary stellar masses and metallicities.

\subsubsection{Introduction of hypernovae}

\begin{figure}
    \includegraphics[width=\columnwidth]{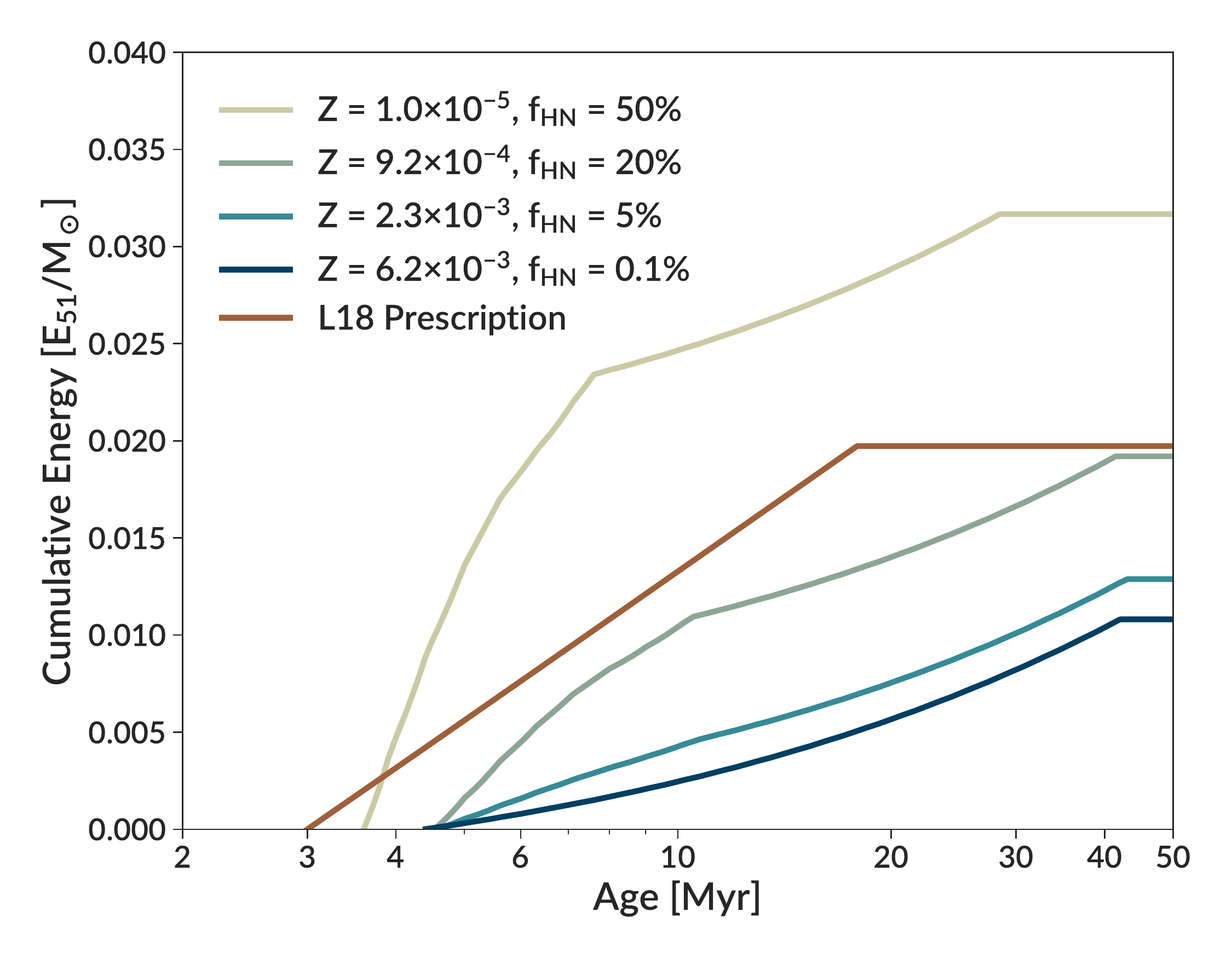}
    \vspace{-6mm}
    \caption{Cumulative energy injected by SN per unit stellar mass in units of $10^{51}$ erg $\Msun^{-1}$ as a function of time since beginning of star formation for different prescriptions. Four lines show the model used in this set of simulations, while the last shows that used by \citetalias{li_etal_18_paper2}. The new prescription is plotted at several metallicities, with HN fractions following Equation~\ref{eq:hn_fraction} with $\fhno=0.5$. The metallicity-dependent stellar lifetimes also change the time of the onset of SN. The line with $f_{\mathrm{HN}}=0.1$\% is visually indistinguishable from a line with $f_{\mathrm{HN}}=0$.}
    \label{fig:sn_energy}
\end{figure}

Hypernovae (HN) are SN explosions with significantly more energy than a typical SN, and may be associated with gamma ray bursts \citep[e.g.][]{iwamoto_etal_98}. The \citet{kobayashi_etal_06} yield tables include stellar yields and energies for HN so we include them in our feedback model. We model both the energy and yields from HN self consistently. SN with progenitor stellar masses above $20\Msun$ are eligible to explode as HN. Each explosion is randomly assigned to be either HN or SN, depending on a metallicity-dependent HN fraction. We use the functional form proposed by \citet{grimmett_etal_2020}:
\begin{equation}
    \fhn = \max \left( \fhno \exp \left(- \frac{Z}{0.001} \right), 0.001 \right)
    \label{eq:hn_fraction}
\end{equation}
These authors suggest that $\fhno=0.5$, but we leave it as a free parameter to test how varying it affects galaxy properties. SN explosions always inject $E_{51} \equiv 10^{51}$ ergs of energy, while for HN we use the mass-energy relation from \citet{kobayashi_etal_06}, where the energy ranges from 10 to 30 $E_{51}$, with high mass stars releasing the most energy. We linearly interpolate the energy released by HN for stellar masses between those given in \citet{kobayashi_etal_06}. Increasing $\fhn$ significantly changes the energy injected into the simulation. Figure~\ref{fig:sn_energy} shows the cumulative energy injected from SN as a function of cluster age. Different lines show different metallicity and therefore different $\fhn$. As HN are only active for stars with masses above $20\Msun$, the difference in $\fhn$ is apparent at early times, while at later times SN energy injection is the same. As our stellar lifetimes are metallicity-dependent, the age of the onset of SN and the age at which HN end changes as well. Of note, the \citet{raiteri_etal_96} lifetimes give an onset of SN in this new prescription that is always later than the constant 3~Myr onset adopted by \citetalias{li_etal_18_paper2}.

\subsubsection{Momentum boost}

To model SN feedback, we use the prescriptions from \citet{martizzi_etal_15}. They used simulations of inhomogeneous turbulent medium to parametrize the partition of the SN remnant energy into the thermal, kinetic, and turbulent components. The resulting energy and momentum input depend on the ambient gas density and spatial resolution of the simulation. However, their simulations of isolated SN explosions underestimate the effect for star clusters. Cluster-forming regions usually produce a large number of massive stars that undergo simultaneous SN explosion. \citet{gentry_etal_17} found that such clustering of SN can enhance momentum feedback by an order of magnitude relative to that delivered by an isolated SN. \citetalias{li_etal_18_paper2} tested a boost to the momentum feedback from SN remnants by a factor $\fboost=3-10$ and found that the value $\fboost=5$ can reproduce the galactic star formation history expected from the abundance matching technique. As $\fboost$ is a key parameter of our feedback model, we explore its ideal value in our new simulations below in Section~\ref{sec:feedback_strength}. The momentum created by stellar particles is distributed spherically to 26 nearest neighbor cells surrounding the parent cell of the particle, as in  \citet{li_etal_17_paper1}. 

\subsubsection{Supernovae type Ia}

We have updated the SNIa feedback prescription, implementing discrete SN and a new delay time distribution (DTD). We use the power-law DTD for field galaxies from \citet{maoz_graur_17}:
\begin{equation}
    \frac{dN_{\rm SNIa}}{dt} \propto \tau^{-1.13}
\end{equation}
normalized to produce $1.6 \times 10^{-3}$ SNIa per $\Msun$ of stellar mass. Similarly to how we integrate over the IMF to produce the number of SNII, we integrate over the DTD to produce the number of SNIa. We model these as discrete events as we do for SNII, and use the yields from \citet{nomoto_leung_18}. The feedback from SNIa is modeled simply as an injection of $2 E_{51}$ of thermal energy.

\subsubsection{AGB feedback}

Our final addition to the feedback prescription is chemical enrichment from AGB stars, defined to be the last stages of evolution of stars with masses below $8\Msun$. The prescription for AGB stars is analogous to that for SNII as described by Equation~\ref{eq:imf_integral}. However, we abandon the requirement for integer numbers and simply use the full integral in Equation~\ref{eq:imf_integral}. This is justified by the fact that this phase of stellar evolution is not instantaneous like a SN. We use the yields from \citet{ritter_etal_18}. We only inject mass from AGB feedback. We do not inject the energy or momentum, as their wind velocities are small and have little impact on the total feedback budget \citep{goldman_etal_17,hopkins_etal_18}.

\medskip\noindent
We also include two other sources of feedback, which are unchanged from the implementation of \citetalias{li_etal_18_paper2}: radiation pressure from massive stars using the analytical fit by \citet{gnedin_14}, and momentum from stellar winds as an analytical fit to the results of \citet{leitherer_etal_92}.

\subsubsection{Timing of cluster feedback}
\label{sec:timing}

Since our star cluster particles accrete material over time, defining a single age to use in the above feedback prescriptions is not trivial. Without storing the full cluster growth histories, which are prohibitively large, we must make some assumptions. One choice would be to simply use the time $t$ since the star particle was seeded: $\tau_{\rm birth}(t) = t$. We refer to this as the ``birth approach'', since it treats all stars as forming at the same time as the first one in the cluster. This prescription is problematic if the cluster has significant star formation after the onset of SN at about 4~Myr. For example, consider some stars formed 6~Myr after the birth of that cluster particle. The birth approach assigns all stars in the cluster an age of 6~Myr, including these newly formed stars with a true age of zero. As these newly formed stars never had an age in the 0--6~Myr range, the feedback they should  contribute during that age range is skipped (particularly SN feedback from 4--6~Myr). This prescription also gets the timing of feedback wrong, as the assumption that all the mass of the cluster formed at the initial time is incorrect. 

An alternative is to adjust the age based on the mass-averaged time of cluster formation: $\tau_{\rm ave}(t) = t - t_{\rm ave}(t)$. This average time for cluster formation is calculated in runtime as
\begin{equation}
    t_{\rm ave}(t) \equiv \frac{\int_0^{t} t\, \dot{M}(t) dt}{\int_0^{t} \dot{M}(t) dt}
\end{equation}
where $\dot{M}$ is the cluster star formation rate at time $t$ \citepalias{li_etal_18_paper2}. This approach, which we refer to as the ``average approach'', does a much better job of reproducing the total amount of feedback. However, this approach pushes back the onset of SN feedback, allowing some clusters (particularly massive ones) to have unphysically long formation timescales before their growth is terminated by feedback.

\begin{figure}
    \includegraphics[width=\columnwidth]{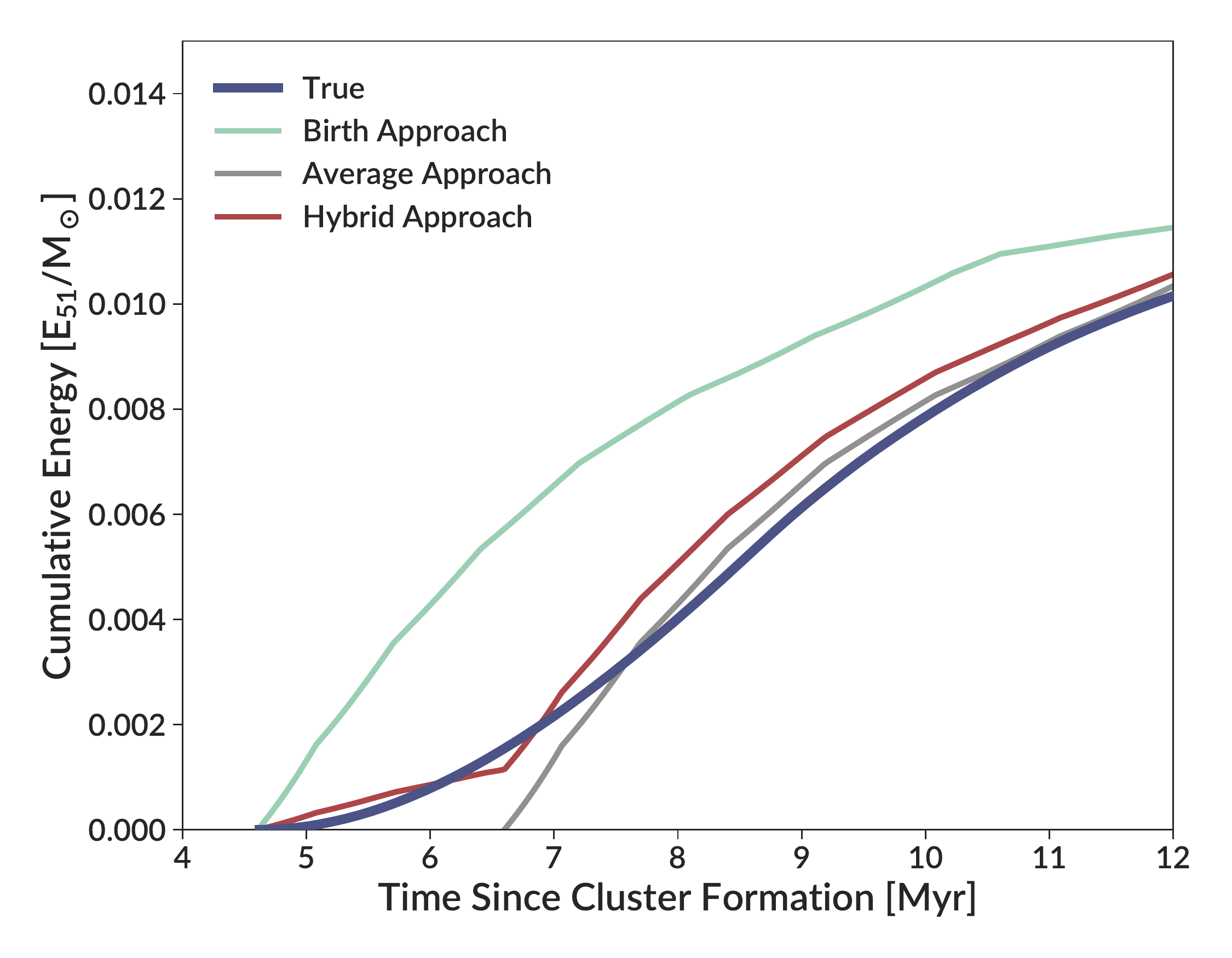}
    \vspace{-6mm}
    \caption{The cumulative energy injected by SN per unit stellar mass as a function of time since beginning of cluster formation for three approaches for the timing of SN. The input star formation history is a constant star formation rate for 4~Myr. The ``True'' line shows the actual energy injection produced by this stellar population, while the other lines show the energy injection for different ways of treating this star formation history as a simple stellar population, as described in the text. The hybrid approach is a weighted combination of the other two.}
    \label{fig:sn_timing}
\end{figure}

To solve this problem, we introduce a hybrid approach, where we allocate a fraction of cluster feedback to use the birth approach and the rest to use the average approach. Denoting the amount of feedback generally as $\mathcal{F}$, we set
\begin{equation}
    \mathcal{F}_{\rm tot}(t) = f_{\rm birth}(t) \  \mathcal{F}\left(\tau_{\rm birth}(t)\right) + [1 - f_{\rm birth}(t)] \mathcal{F}\left(\tau_{\rm ave}(t)\right)
\end{equation}
such that $f_{\rm birth}$ is the fraction of the cluster mass assigned to the birth approach. This hybrid approach gives the best of both worlds, as it gives the correct delay before the first SN explodes while also accurately reproducing the total amount of feedback. Using idealized test cases, we find that clusters with a larger age spread require a larger $f_{\rm birth}$. Conceptually, this is because clusters with a large age spread have a larger fraction of their feedback that comes from stars formed away from the mean cluster age. We use the following parametrization:
\begin{equation}
    f_{\rm birth}(t) = \frac{\tau_{\rm spread}(t)}{20~{\rm Myr}}
    \label{eq:fbirth}
\end{equation}
where the 20~Myr scale parameter was determined from idealized test cases, and $\tau_{\rm spread}$ is the cluster age spread calculated in runtime as
\begin{equation}
    \tau_{\rm spread}(t) \equiv \frac{M(t)}{\langle \dot{M} \rangle} = \frac{M^2(t)}{\int_0^{t} \dot{M}^2(t) dt}
    \label{eq:age_spread}
\end{equation}
\revision{where $\langle \dot{M} \rangle$ is the mass-weighted star formation rate:
\begin{equation}
    \langle \dot{M} \rangle 
    = \frac{\int_0^{t} \dot{M}(t) dM}{\int_0^{t} dM} = \frac{\int_0^{t} \dot{M}^2(t) dt}{M(t)}
\end{equation}
where $dM = \dot{M}(t) dt$.} As cluster age spreads are typically a few Myr, this gives no more than 20\% of the feedback coming early, with the majority using the average age. Figure~\ref{fig:sn_timing} shows an example of this prescription for the feedback from a toy cluster consisting of a 4~Myr period of constant star formation rate. \revision{To calculate the true energy injection rate that this toy cluster would be expected to give, we represent it with many simple stellar populations spaced evenly between 0 and 4~Myr. The total energy injection is then the sum of the energy injected by each simple stellar population. We also compute $t_{\rm ave}(t)$ and $\tau_{\rm spread}(t)$ to compute the feedback that would result when using the birth, average, and hybrid approaches. In Figure~\ref{fig:sn_timing},} the delayed onset of SN when using the average approach is clear, as is the increased energy output when assuming all stars formed at the birth of the cluster. As this hybrid approach is a weighted sum of the two other approaches, there is a break in the hybrid approach between 6 and 7~Myr due to the onset of SN in the average approach. 

\subsection{Hydrodynamics}
\label{sec:hydro}

When updating the code from the version used in \citetalias{li_etal_18_paper2} to \revision{a newer version of the ART code (ART 2.0),}
we updated the modeling of the energy equation that governs how thermal energy is calculated in the presence of subgrid turbulence. This update more accurately tracks thermal energy in shocks. It has little effect in the disc of the galaxy, as the thermal energy generated by shocks is subdominant to other process that govern energy balance such as heating, cooling, and stellar feedback. However, we find that the circumgalactic medium is affected by this update. In our new runs, there is significantly more hot gas in the halo. This in turn leads to less cold gas accreting onto the galaxy, leading to less star formation. The decrease in the amount of cold gas requires changes to the parameters governing star formation and feedback as we describe below. We describe the update to the hydrodynamics in more detail in Appendix~\ref{appendix:hydro}. In our suite of simulations, we used both this updated energy-based approach and the new entropy-conserving scheme of \citet{semenov_etal_21}. These authors found that the entropy-conserving scheme is able to more accurately evolve nonthermal energy components. They ran simulations of an $L_\star$ galaxy and found differences between the energy-based and entropy-based schemes. However, these differences are much smaller than those we find between the energy-based schemes of \citetalias{li_etal_18_paper2} and \revision{this paper}. 

\subsection{Initial Conditions}

In this work we use three different ICs. One is the IC used by \citetalias{li_etal_18_paper2}, a periodic comoving box of size 4~Mpc that contains a single central galaxy with a total mass of $10^{12}\,\Msun$ at $z=0$, which we refer to as \huiic. We also use two zoom-in ICs from the ELVIS project \citep{garrison_kimmel_etal_14}: \tl\ and \rj. Both of these ICs contain a Local Group analog with two Milky Way-mass galaxies, which we describe in more detail below. The \huiic\  box is much less computationally expensive to run than the zoom-in runs, so we use it to explore a broader range of parameter space.

\tl\ is a desirable IC as it has qualitative agreement with the accretion histories of the MW and M31. The less massive (MW-like) halo has a quieter accretion history \citep{hammer_etal07}, with no significant mergers after $z \approx 5$, while the more massive (M31-like) halo has more mergers at later times as expected from observations \citep{dsouza_bell_18}. \rj\ has two galaxies with much quieter merger histories. Including two different sets of ICs allows us to explore how our results vary with galaxy merger histories. 

\begin{figure}
    \includegraphics[width=\columnwidth]{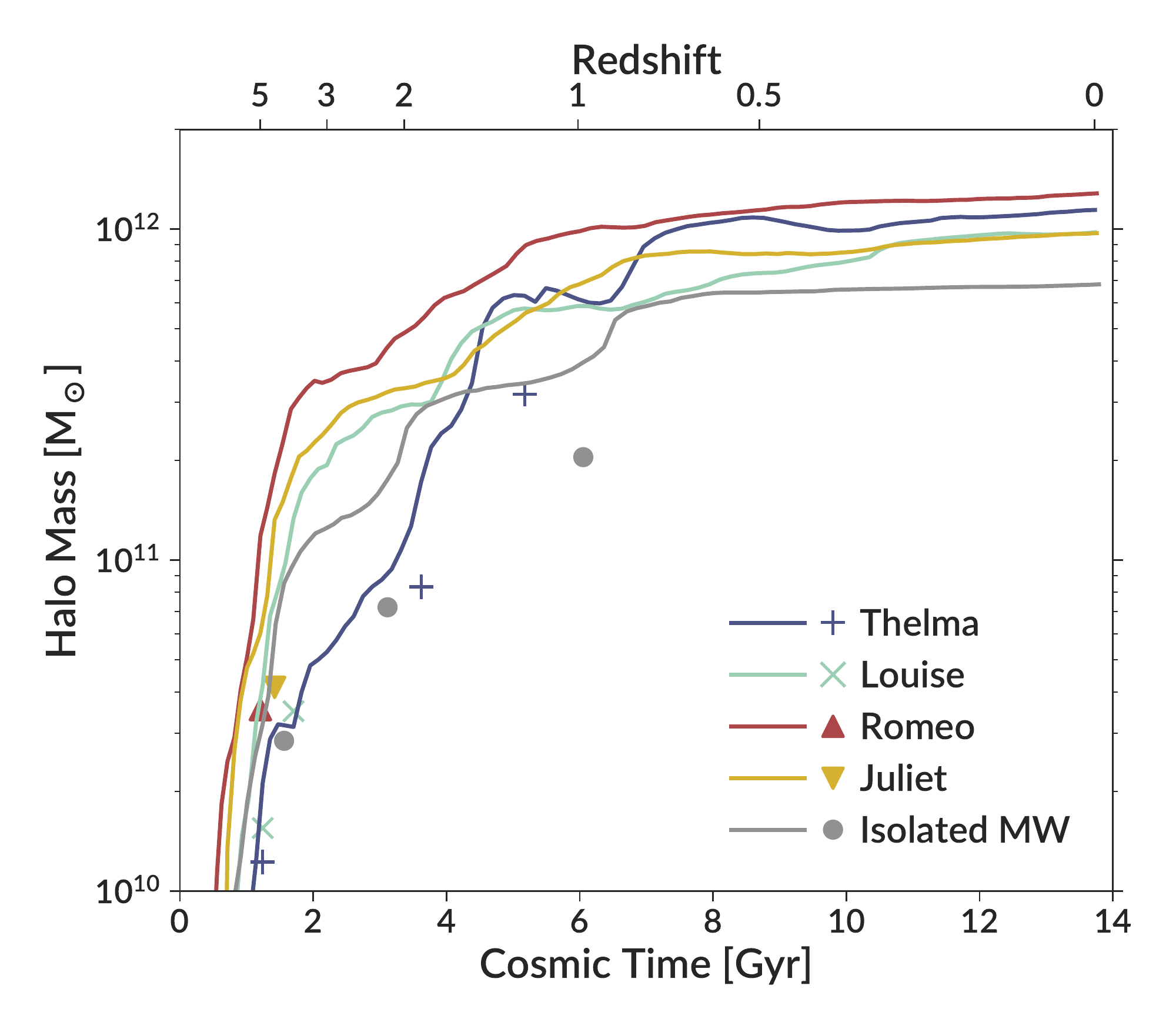}
    \vspace{-6mm}
    \caption{Mass growth of the central halos from collisionless runs with three initial conditions. Markers show major mergers with a mass ratio less than 4:1, and are placed at the maximum virial mass of the satellite and the time at which it reached this maximum mass before merging with the central galaxy. Note that \texttt{Thelma} and \huiic\ have major mergers at $z<2$, while the other three galaxies have quiet merger histories.}
    \label{fig:halo_growth}
\end{figure}

To improve computational performance with the ART code, we modify these zoom-in ICs following the prescription of \citet{brown_gnedin_21a}. Our initial conditions have a small zoom region in a large box (50-100~Mpc). This large box size with a small zoom region is difficult for the ART code to parallelize well, so our method decreases the box size and increases the resolution of the root grid. 
\revision{The initial conditions are created using the \texttt{MUSIC} software \citep{hahn_abel_11}, where a white noise field is convolved with the matter power spectrum to produce realistic matter overdensities. We regenerate the original white noise field at higher resolution, then cut out a smaller volume of interest. This smaller white noise cube is then convolved with the matter power spectrum to produce the density within a smaller volume. As the white noise is what seeds the resulting structures, this method reduces the box size while preserving large scale structure and enforcing periodic boundary conditions.}
To avoid disturbing the zoom region, the particles from this region are transplanted into the new box with a velocity offset to match the systemic velocity of this region in the new box. \revision{We refer readers to Figures~1 and 2 of \citet{brown_gnedin_21a} for a visual representation of the method.} We find that these modifications improve performance while minimally changing central galaxy properties. Table~\ref{tab:ics} details some key properties of these ICs, and Figure~\ref{fig:halo_growth} shows the halo mass growth of these galaxies in collisionless runs.

We run an initial suite of nine simulations with these zoom-in ICs, varying $\epsff$, $\fhno$, and $\fboost$. Table~\ref{tab:simulation_suite_lg} details the parameters of these runs. We also ran a large suite of 20 simulations on the \huiic\ initial condition varying many aspects of stellar feedback, which are detailed in Table~\ref{tab:simulation_suite_oldic}.

\begin{table*}
    \caption{Description of key properties of the initial conditions used here. For zoom-in ICs, the mass resolution quantities refer to the zoom region.}
    \begin{tabular}{lccccccc}
        \hline
        Initial Condition & Box Size  & Dark Matter Particle  & Typical Baryonic & $\Omega_m$ & $\Omega_\Lambda$ & $\Omega_b$ & $h$ \\
        & (comoving $h^{-1}$ Mpc) & Mass ($\Msun$) & Cell Mass ($\Msun$) \\
        \hline
        \tl & 25.0 & $1.57\times 10^5$ & $2 \times 10^4$ & 0.266 & 0.734 & 0.0449 & 0.71 \\
        \rj & 23.12 & $1.53\times 10^5$ & $2 \times 10^4$ & 0.31 & 0.69 & 0.048 & 0.68 \\ 
        \huiic & 4.0 & $1.0\times10^6$ & $4 \times10^4$  & 0.304 & 0.696 & 0.0479 & 0.681 \\ \hline
    \end{tabular}
    \label{tab:ics}
\end{table*}

\begin{table}
    \caption{The runs using the Local Group initial conditions included in this simulation suite. $z_{\rm last}$ is the redshift of the last output of each run.  All runs use average approach for SN timing, \revision{$C_\rho=10$}, and the \revision{energy-based hydrodynamics scheme}. \revision{The clumping factor $C_\rho$ will be discussed further in Section~\ref{sec:molecular}.}}
    \begin{tabular}{lccccc}
        \hline
        Initial Condition & $\epsff$ & $\fboost$ & $\fhno$  & $z_{\rm last}$ \\
        \hline
        \tl & 1\%   & 5 & 20\% & 3.32 \\
        \tl & 10\%  & 5 & 20\% & 2.36 \\
        \tl & 100\% & 1 & 0\%  & 3.17 \\
        \tl & 100\% & 3 & 0\%  & 2.80 \\
        \tl & 100\% & 5 & 0\%  & 1.83 \\
        \tl & 100\% & 5 & 5\%  & 1.86 \\
        \tl & 100\% & 5 & 20\% & 2.66 \\
        \rj & 10\%  & 5 & 20\% & 2.78 \\ 
        \rj & 100\% & 5 & 20\% & 1.87 \\ 
        \hline
    \end{tabular}
    \label{tab:simulation_suite_lg}
\end{table}

\begin{table*}
    \caption{The runs using the \huiic\ initial conditions included in this simulation suite. In the ``Hydro Scheme'' column, ``S21'' refers to the entropy-based scheme of \citet{semenov_etal_21}, ``Energy'' refers to the \revision{updated} energy-based scheme,
    and ``L18'' is the hydro scheme used in \citetalias{li_etal_18_paper2}. The schemes mentioned in the ``SN Timing'' column are described in Section~\ref{sec:timing}. Simulations are grouped by the attribute that is varied\revision{, although some simulations are used in multiple subsections.} All runs progressed to $z=1.5$ except for the two runs with $\epsff<100$\% and $\fboost=2$, which reached $z\approx 2$.}
    \begin{tabular}{ccccccl}
        \hline
        $\epsff$ & $\fboost$ & $\fhno$ & \revision{$C_\rho$} & SN Timing & Hydro Scheme & \revision{Other Comments} \\
        \hline
        100\% & 1 & 0 & \revision{10} & Average &  S21 & \revision{Used in all subsections below}  \\
        \hline
        
        \multicolumn{7}{c}{\revision{\textit{Section \ref{sec:effect_of_timing}}}} \\
        100\% & 1 & 0 & \revision{10} & Hybrid &  S21 &  \\ 
        100\% & 1 & 0 & \revision{10} & Birth &  S21 &  \\ 
        \hline
        
        \multicolumn{7}{c}{\revision{\textit{Section \ref{sec:discreteness}}}} \\
        100\% & 1 & 0 & \revision{10} & Average &  S21 & Continuous energy injection from SN \\
        \hline
        
        \multicolumn{7}{c}{\revision{\textit{Section \ref{sec:feedback_strength}}}} \\
        100\% & 1 & 50\% & \revision{10} & Average &  S21 \\ 
        100\% & 2 & 0 & \revision{10} & Average &  S21 \\ 
        100\% & 3 & 0 & \revision{10} & Average &  S21 \\
        100\% & 5 & 0 & \revision{10} & Average &  S21 \\ 
        \hline
        
        \multicolumn{7}{c}{\revision{\textit{Section \ref{sec:molecular}}}} \\
        100\% & 1 & 0 & \revision{3} & Average &  S21  \\ 
        100\% & 1 & 0 & \revision{3} & Average &  S21 & Changed shielding to \citet{gnedin_kravtsov_11} \\ 
        100\% & 1 & 0 & \revision{30} & Average &  S21 &  \\ 
        \hline

        \multicolumn{7}{c}{\revision{\textit{Section \ref{sec:results_epsff}}}} \\
        1\%  & 1 & 0 & \revision{10} & Average &  S21  \\ 
        10\% & 1 & 0 & \revision{10} & Average &  S21 \\ 
        1\%  & 2 & 0 & \revision{10} & Average &  S21 \\ 
        10\% & 2 & 0 & \revision{10} & Average &  S21 \\
        \hline
        
        \multicolumn{7}{c}{\revision{\textit{Section \ref{sec:virial}}}} \\
        100\% & 1 & 0 & \revision{10} & Average &  S21 & No virial parameter criterion for star formation \\
        \hline
        
        \multicolumn{7}{c}{\revision{\textit{Appendix \ref{appendix:hydro}}}} \\
        100\% & 5 & 0 & \revision{10} & Average & Energy & Analogous to Local Group runs \\ 
        100\% & 5 & 0 & \revision{10} & Average &  Energy & Feedback scheme of \citetalias{li_etal_18_paper2}. \\ 
        100\% & 5 & 0 & \revision{10} & Average &  L18 \\ 
        \hline

    \end{tabular}
    \label{tab:simulation_suite_oldic}
\end{table*}

\subsection{Run Setup}

We keep the spatial resolution of the finest grid level between \mbox{3--6} physical pc at all times. To accomplish this, we add refinement levels as the simulation progresses. The specific levels and when they are added depend on the initial condition. In the \huiic\ box, we start with 9 levels of refinement on the 128$^3$ root grid, then add levels at $z=9$, 4, and 1.5. For \tl, we allow 11 levels of refinement on the 256$^3$ root grid, then add additional levels at $z\approx10.2, 4.6, 1.8$, and 0.41. \rj\ also starts with 11 levels, but its slightly different box size requires adding levels at $z\approx9.8, 4.4, 1.7$, and 0.35. 

We use three criteria to determine when to refine the grid. In this section we will illustrate the refinement criteria using specific values from the \tl\ IC, but the principles are the same for all ICs. First, we use Lagrangian refinement for both gas and dark matter. Cells are refined when their gas mass exceeds approximately $1.6 \times 10^5 \Msun$ or dark matter exceeds $3.9 \times 10^6 \Msun$. The gas refinement is active on all levels, while the dark matter criterion is not active on the four finest levels. We also increase the dark matter mass refinement threshold above that from the simple baryon fraction scaling. These changes are for two reasons. First, the discrete dark matter particles (of mass $1.5\times10^5\Msun$) do not allow their mass to be distributed evenly, so their distribution cannot be trusted on small scales. Second, we find that there are times when the dark matter criterion will prevent a cell with very small gas mass from derefining. If stellar momentum feedback is imparted on this cell, it will acquire very high velocities due to its small mass, leading to small timesteps and a slower runtime of the simulation. Restricting the levels on which the dark matter Lagrangian criterion is active and increasing the mass threshold for dark matter-triggered refinement mitigates this situation. The final refinement criterion uses a local Jeans length. Cells are refined if their size exceeds twice the Jeans length. This criterion is applied only on the four deepest levels. We find that with these refinement criteria, cell gas masses remain around 2$\times10^4\Msun$. Table~\ref{tab:ics} includes the typical baryonic cell masses for all ICs.

While we do not record the level on which a star is formed in runtime, we postprocess the outputs to see the levels on which stars can form. In the runs using the Local Group ICs, we find that 15\% of the cells that satisfy the star formation criteria are on the highest refinement level with sizes of 3--6~pc, 60\% have sizes in the 6--12~pc range, 25\% are within 12--24~pc, and a very small fraction are on the fourth level with sizes of 24--48~pc. The lower mass resolution of the \huiic\ runs results in the corresponding fractions of 10\%, 35\%, 50\%, and 5\%, respectively.

The ART code uses adaptive time stepping, such that the finest levels have much shorter timesteps than the coarse root grid. For the \tl\ runs with $\epsff=100\%$, the global timestep of the root grid is restricted to be less than 10~Myr. We write outputs at each global timestep. For all other runs, the output spacing is allowed to be at most 50~Myr. The timestep for the finest level is similar for all runs, typically between 100--1000~years.

\section{Effects of cluster formation and feedback modeling}
\label{sec:results}

In this section, we analyze the large suite of simulations laid out in Tables~\ref{tab:simulation_suite_lg}~and~\ref{tab:simulation_suite_oldic} to test the implementation of code updates and explore how parameter variation affects our results. We will primarily focus on the galaxy star formation rate, cluster mass function, and the timescales of cluster formation. In this section we exclusively use the particle mass at the end of its star formation episode, which does not account for the initial bound fraction, stellar evolution, or dynamical disruption of a star cluster represented by that particle. We explore those quantities and the observable cluster mass function in Section~\ref{sec:jwst}. We also note that when examining star cluster populations, we include all clusters from the central galaxies in the simulations (the one MW-mass galaxy in \huiic, and the two galaxies in the Local Group-like environments of \tl\ and \rj). When plotting the star formation rate of these galaxies we plot the two central galaxies in the Local Group-like IC separately, but when plotting cluster properties of a given run we group these two galaxies together.

\subsection{Timing of supernova feedback}
\label{sec:effect_of_timing}

\begin{figure*}
    \includegraphics[width=\textwidth]{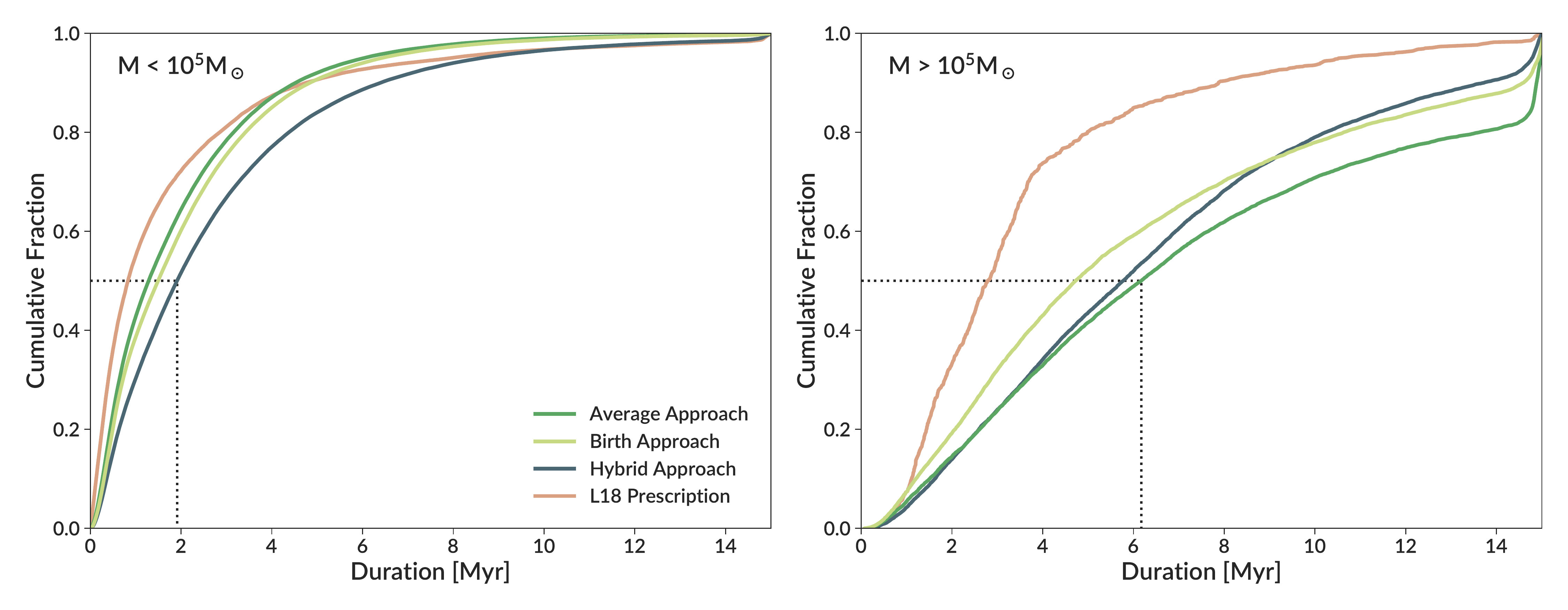}
    \vspace{-6mm}
    \caption{The cumulative distribution of the duration of cluster formation for different approaches to determining the timing of SN feedback, as described in Section~\ref{sec:timing}. The left panel shows clusters less massive than $10^5\Msun$, while the right panel shows clusters more massive than $10^5 \Msun$. The dotted line shows the longest median duration of cluster formation. Cluster growth is algorithmically truncated at 15 Myr. Note that here we use a new run with the L18 feedback model, not the L18 simulations themselves. The L18 prescription uses $\fboost=5$, while all other runs use $\fboost=1$. All runs use the \huiic\ IC, $\epsff=100$\%, $\fhno=0$, and show all clusters formed before $z=1.5$.}
    \label{fig:age_duration_sn_timing}
\end{figure*}

In Section~\ref{sec:timing} we describe how the finite length of cluster formation makes it difficult to create an accurate prescription for the timing of stellar feedback. We ran simulations with the birth approach, the average approach, and the hybrid approach. We also compared these to the feedback model of \citetalias{li_etal_18_paper2}, which has SNe that start earlier (see Figure~\ref{fig:sn_energy}). We found no significant differences in any galaxy-scale properties between these prescriptions. However, we did find that the cluster formation lifetimes were different between these prescriptions. In particular, the average approach gave significantly longer timescales for massive clusters. Figure~\ref{fig:age_duration_sn_timing} shows the cumulative distribution of the length of star formation within clusters formed using different timing choices, for the local efficiency $\epsff=100\%$. Note that the quantity we plot here is the duration of star formation, defined as the age difference between the birth of the cluster and its last accretion event. This is not $t_{\rm ave}$ or $\tau_{\rm spread}$ as defined in Section~\ref{sec:timing}. We use this quantity as it clearly demarcates when feedback ends cluster formation.

With all the approaches to SN feedback, the majority of low-mass clusters have finished their accretion before the onset of SNe at 3--4~Myr, leading to little difference in the durations between our approaches. Such short durations indicate that the other sources of feedback are able to terminate cluster formation before the start of SN feedback \citep{kruijssen_etal_19,grudic_etal_22}. SN feedback remains more relevant for massive clusters. 

We do see a difference in the high mass clusters. The feedback prescriptions of \citetalias{li_etal_18_paper2} produce the shortest durations of star formation. Among the three new models of determining the timing of SN feedback, the average approach produces clusters with the longest duration, the birth approach gives clusters with the shortest duration, and the hybrid approach is in the middle. As the birth approach has the most early feedback and the average approach has the least early feedback, these results indicate that delaying the start of SN feedback tends to increase the time over which massive clusters can accrete material. This matches what we see in the \citetalias{li_etal_18_paper2} model, which allows SN feedback begin earlier and stop cluster growth earlier. 

\begin{figure}
    \includegraphics[width=\columnwidth]{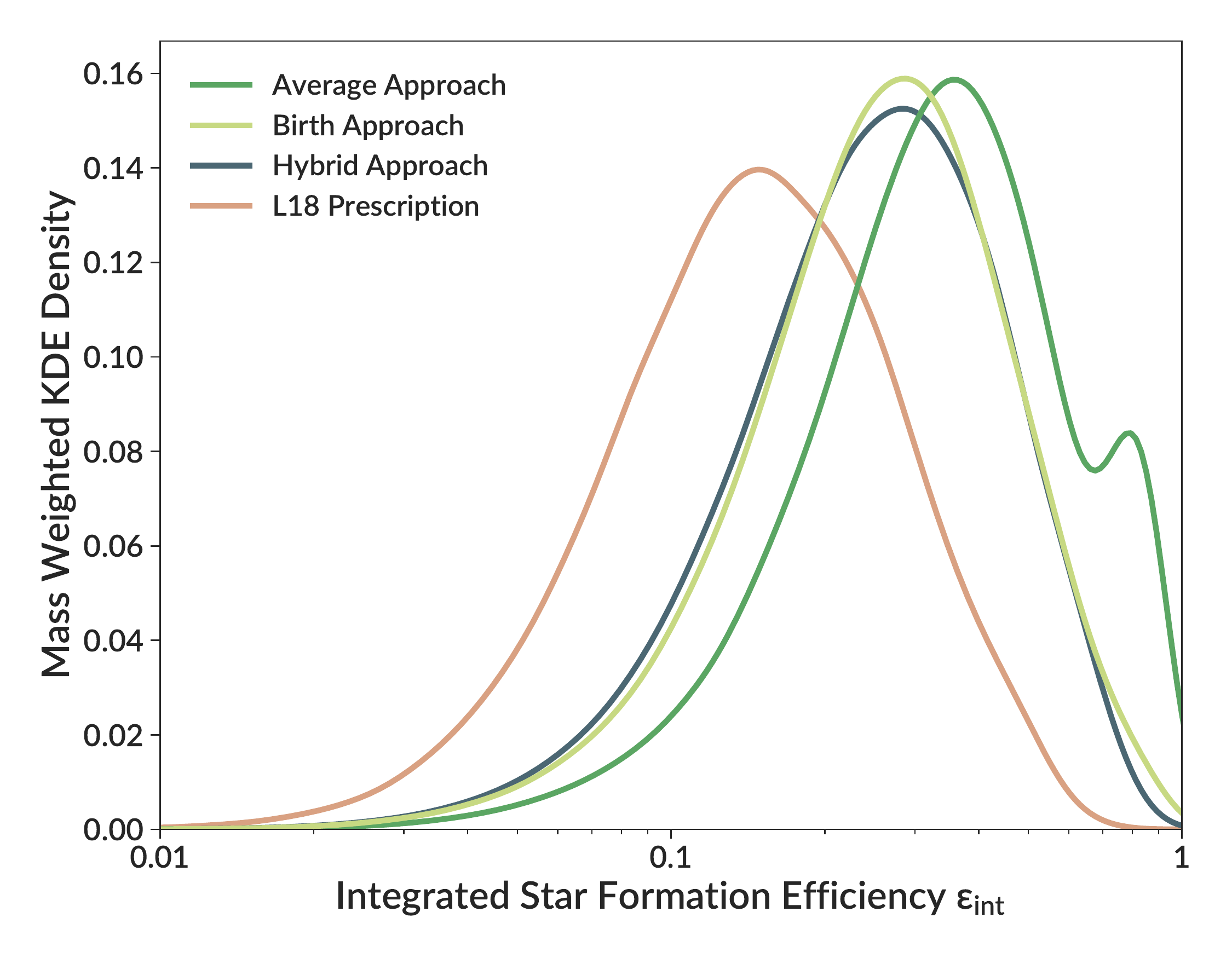}
    \vspace{-7mm}
    \caption{Kernel density estimation for the distribution of integrated star formation efficiency for clusters in the runs with variations in the timing of SN feedback, as described in Section~\ref{sec:timing}. We use a Gaussian kernel with a width of 0.05 dex. Each curve is normalized to the same area for comparison purposes. Note that here we use a new run with the L18 feedback model, not the L18 simulations themselves. The L18 prescription uses $\fboost=5$, while all other runs use $\fboost=1$. All runs use the \huiic\ IC, $\epsff=100$\%, $\fhno=0$, and show all clusters formed before $z=1.5$.}
    \label{fig:eps_int_sn_timing}
\end{figure}

These trends are also reflected in the integrated star formation efficiency $\epsint$, defined in Equation~\ref{eq:eps_int}. Figure~\ref{fig:eps_int_sn_timing} shows the distribution of $\epsint$ for the runs with variations in the timing of SN feedback. The \citetalias{li_etal_18_paper2} feedback model has the earliest SN feedback and the lowest mean value of $\epsint$ (15\%), while the average approach has the latest SN feedback and the highest mean value of $\epsint$ (35\%). Interestingly, the hybrid approach and birth approach are very similar, with mean values at $\epsff\approx25$\%. This may be because early SN feedback (present in both variations to some extent) is important for dispersing gas before it can be accreted by the cluster. \revision{While the simulation with the \citetalias{li_etal_18_paper2} model uses $\fboost=5$ instead of $\fboost=1$, other runs varying $\fboost$ show no little difference in either the duration of star formation or $\epsint$, indicating that the SN timing is responsible.}

Despite these differences in the duration of star formation and $\epsint$, we see no significant differences in the star particle mass functions. 

\subsubsection{Discreteness of supernova}
\label{sec:discreteness}

In addition to multiple runs with different prescriptions for SN feedback, we also ran one simulation with continuous energy injection from SN. This run uses the \huiic\ IC, $\epsff=100$\%, $\fboost=1$, and $\fhno=0$. The number of SN still follows the IMF integral as in Equation~\ref{eq:imf_integral}, but with the modification that we do not require there to be an integer number of SN in each timestep. We find that this change makes little difference to galaxy properties. The star formation rate was not affected, and neither were star cluster properties, including their mass function and age spread.

We note that the similarity between these two runs is despite real differences in how the energy is injected over time. The SN rate changes with time, but is within the range of $(2-6)\times 10^{-10}\, N_{\rm SN} \Msun^{-1}\, {\rm yr}^{-1}$. Our typical timesteps on the highest refinement levels are below $10^3$ yr, so even massive clusters with $M=10^6\Msun$ do not have a SN every timestep. Clusters of mass $M=10^3\Msun$ have only 10 SN over the $\sim40$~Myr timescale for SN feedback, resulting in significant gaps between SNe. The onset of SN can also be delayed in low-mass clusters, as the decrease in the normalization of the IMF means we need to integrate to lower stellar masses to reach one star (Equation~\ref{eq:imf_integral}). These results indicate that the total injected energy and the timing of the onset of SN cause larger differences than does discretizing SN events.

\medskip\noindent
To summarize, we find that different prescriptions to change the onset of SN (without changing the total energy injection) do not affect any galaxy-scale properties, but do affect the properties of star clusters. When SN feedback is delayed, massive clusters have longer formation timescales, and all clusters have higher $\epsint$. When comparing disretized SN to continuous energy injection, we find no significant differences.

\subsection{Strength of supernova feedback}
\label{sec:feedback_strength}

Our simulations have two main parameters to control the strength of SN feedback: $\fboost$ and $\fhno$. In this section, we explore how those parameters affect our results. 

\begin{figure}
    \includegraphics[width=\columnwidth]{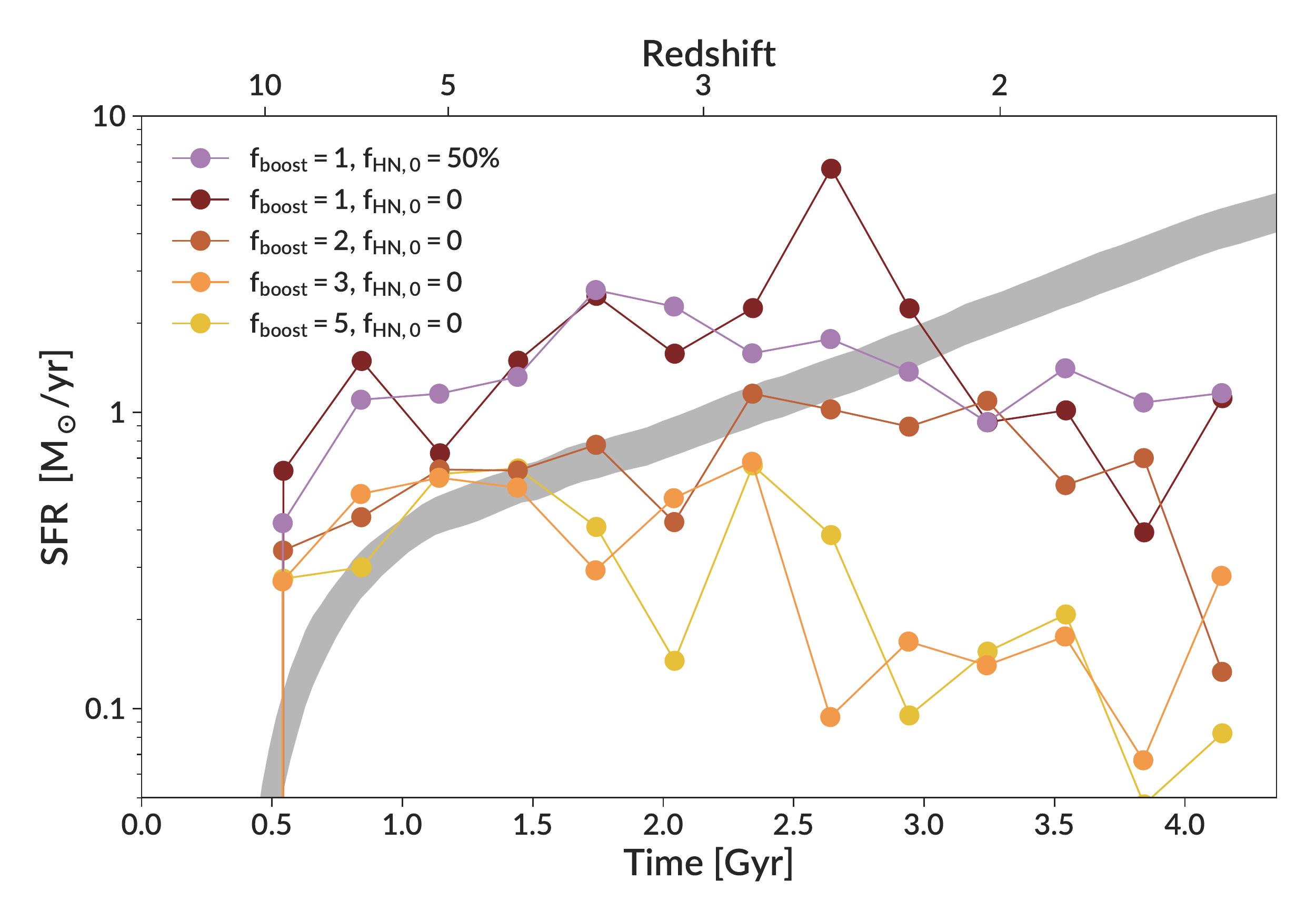}
    \vspace{-7mm}
    \caption{A comparison of the star formation history for the central galaxy in the \huiic\ IC when varying $\fboost$ and $\fhno$. The shaded region shows the expected star formation history as given by \um. All runs use $\epsff=100$\%.}
    \label{fig:sfh_feedback_oldic}
\end{figure}

\begin{figure}
    \includegraphics[width=\columnwidth]{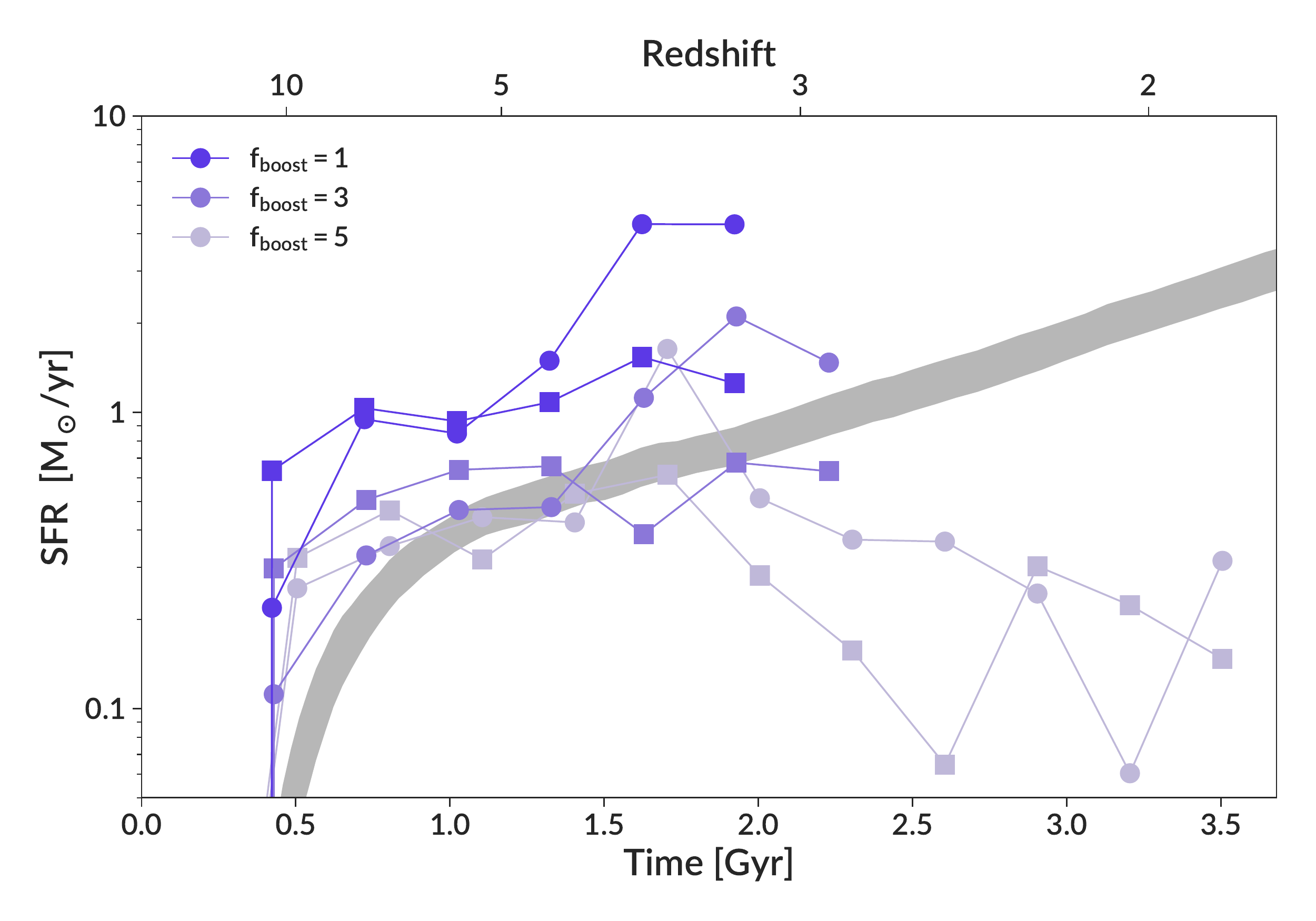}
    \vspace{-7mm}
    \caption{Same as Figure~\ref{fig:sfh_feedback_oldic}, but for the \tl\ IC and only showing variations in $\fboost$. There are two main galaxies in each run. Circles represent the MW analog, with squares representing M31. All runs use $\epsff=100$\% and $\fhno=0$.}
    \label{fig:sfh_feedback_lg}
\end{figure}

In Figures~\ref{fig:sfh_feedback_oldic} and \ref{fig:sfh_feedback_lg} we show the impact of these two parameters on the star formation history of the main galaxies. In Figure~\ref{fig:sfh_feedback_oldic} we show the star formation history of the single central galaxy of the \huiic\ IC, while in Figure~\ref{fig:sfh_feedback_lg} we show two lines for each run representing the two main galaxies in a Local Group-like environment. We also show the expected star formation history as given by \um\ \citep{behroozi_etal_19}. However, we note that the MW assembly history may be atypical for halos of its mass, as both the ancient merger of Gaia-Enceladus Sausage and the current infall of the LMC influence its evolution \citep{evans_etal_20}.

First, we find that $\fboost$ has a strong impact on the global star formation rate. Higher values of $\fboost$ result in generally lower star formation rates. In the \huiic\ runs shown in Figure~\ref{fig:sfh_feedback_oldic}, the run with $\fboost=5$ matches the \um\ prediction well until roughly $z\approx4$, at which point the star formation rates start to decline significantly. This is similar to what we see for the $\fboost=5$ run in \tl\ in Figure~\ref{fig:sfh_feedback_lg}. The star formation rate is reasonable until $z\approx4$, with a significant decline afterwards. A value of $\fboost=3$ matches \um\ more closely in both ICs, but in the \huiic\ IC the star formation rate drops off significantly after $z=3$. The $\fboost=3$ run using \tl\ has only progressed to $z=2.8$ at the time of writing, so it remains possible that its star formation rate will drop as it did in the \huiic\ run. However, we must be careful making direct comparisons between different ICs, as it is likely that they will have different star formation histories. In particular, \citet{santistevan_etal_20} found that Local Group-like galaxies form earlier than isolated galaxies. They conclude that the denser environment of Local Group-like pairs causes the initial collapse of halos to happen earlier \citep{gallart_etal_15}. This leads to more mass forming earlier, and this buildup of stellar mass may affect how feedback affects the galaxy at later times. 

A slightly lower value of $\fboost=2$ matches \um\ well up to $z\approx2$ before decreasing greatly. Finally, runs with $\fboost=1$ have the highest levels of star formation in both initial conditions. This high level has persisted in \tl\ until the last available output, but in \huiic\ the star formation rate dropped dramatically starting at $z=2$. Even this low value of $\fboost$ is not able to produce reasonable galactic star formation histories over the full time range spanned by these simulations. 

\citetalias{li_etal_18_paper2} calibrated $\fboost$ in their simulations, finding a preferred value of $\fboost=5$. The difference in our result is due to the changes in hydrodynamics. As described above in Section~\ref{sec:hydro}, that change led to a decrease in the amount of cold gas that reaches the galaxy. This requires changes to the feedback modeling to compensate. Without decreasing $\fboost$, the galaxies have lower total gas mass and less cold gas, which leads to less molecular gas. Since molecular gas is required by our star formation prescription, this decrease leads to less star formation.

While we find that $\fboost$ has a strong impact on the star formation rate, we find that $\fhno$ does not. In Figure~\ref{fig:sfh_feedback_oldic}, runs with $\fboost=1$ have similar star formation histories, regardless of the value of $\fhno$. While we do not show runs varying $\fhno$ in Figure~\ref{fig:sfh_feedback_lg} for clarity, runs with $\fhno=0, 5\%,$ and $20$\% all show similar star formation rates (all using $\fboost=5$). 

\begin{figure}
    \includegraphics[width=\columnwidth]{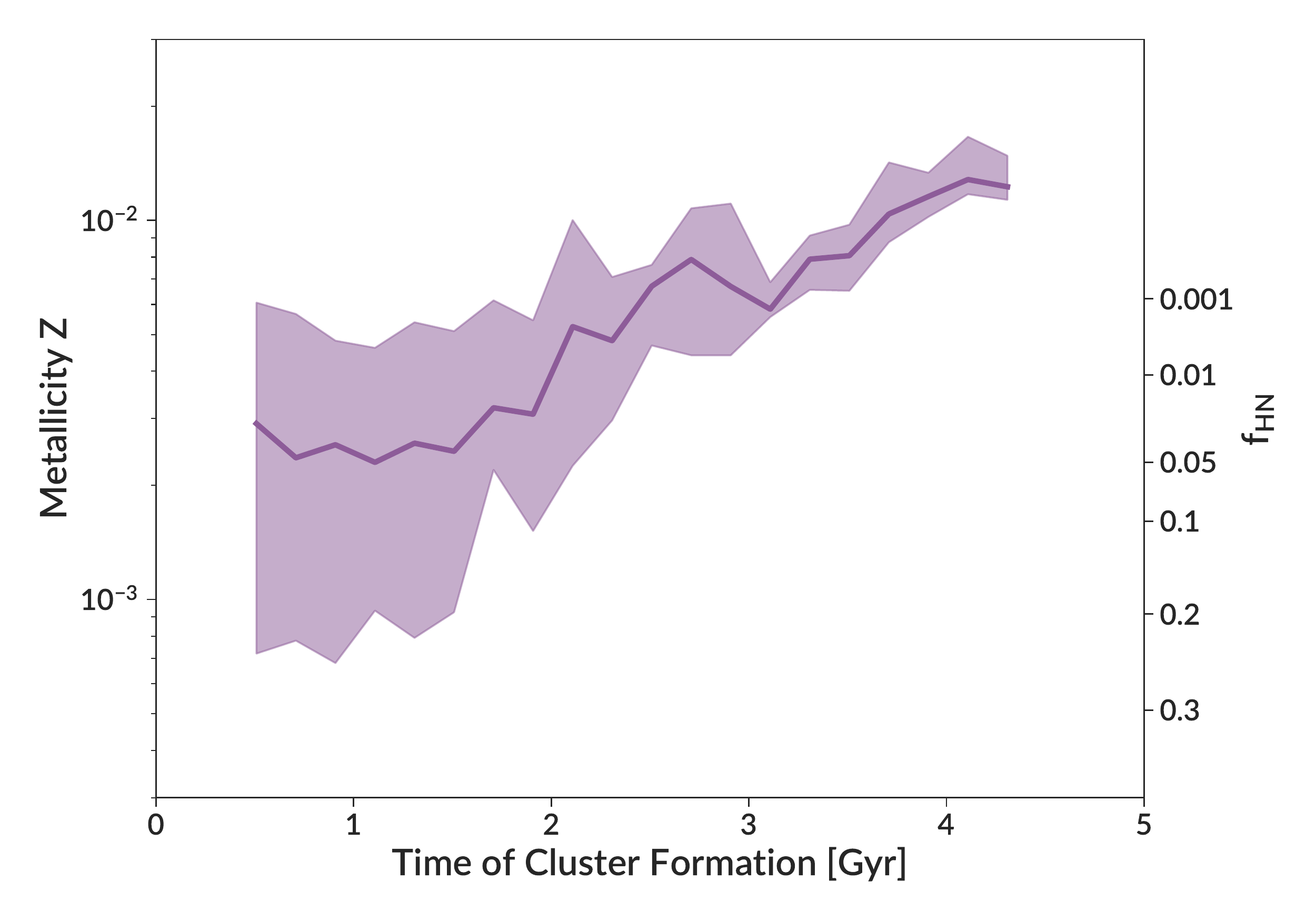}
    \vspace{-7mm}
    \caption{The stellar metallicity $Z$ (not scaled to solar metallicity) of clusters forming across cosmic time in the run using the \huiic\ IC, $\epsff=100$\%, $\fboost=1$, and $\fhno=50$\%. The shaded region shows the interquartile range at a given age, while the solid line shows the median. The right axis labels shown the hypernova fraction at a given metallicity.}
    \label{fig:age_metallicity_hn}
\end{figure}

This is likely due to the metallicity dependence of the hypernova fraction $\fhn$ (see Equation~\ref{eq:hn_fraction}). The value of $\fhn$ is highest at low metallicity, but decreases rather quickly with metallicity. Figure~\ref{fig:age_metallicity_hn} shows the metallicity of stars forming at different times and their $\fhn$. This plot uses the run on the \huiic\ IC with maximum $\fhno=50$\%, yet the quick enrichment means that the bulk of clusters have $\fhn < 10$\%. As shown in Figure~\ref{fig:sn_energy}, this small $\fhn$ produces energy injection rates not too dissimilar from $\fhn=0$. This small change is in contrast to the large changes in momentum feedback that come from varying $\fboost$ by a factor of 5, explaining why $\fboost$ has a strong impact on galactic properties while $\fhno$ does not.

\begin{figure}
    \includegraphics[width=\columnwidth]{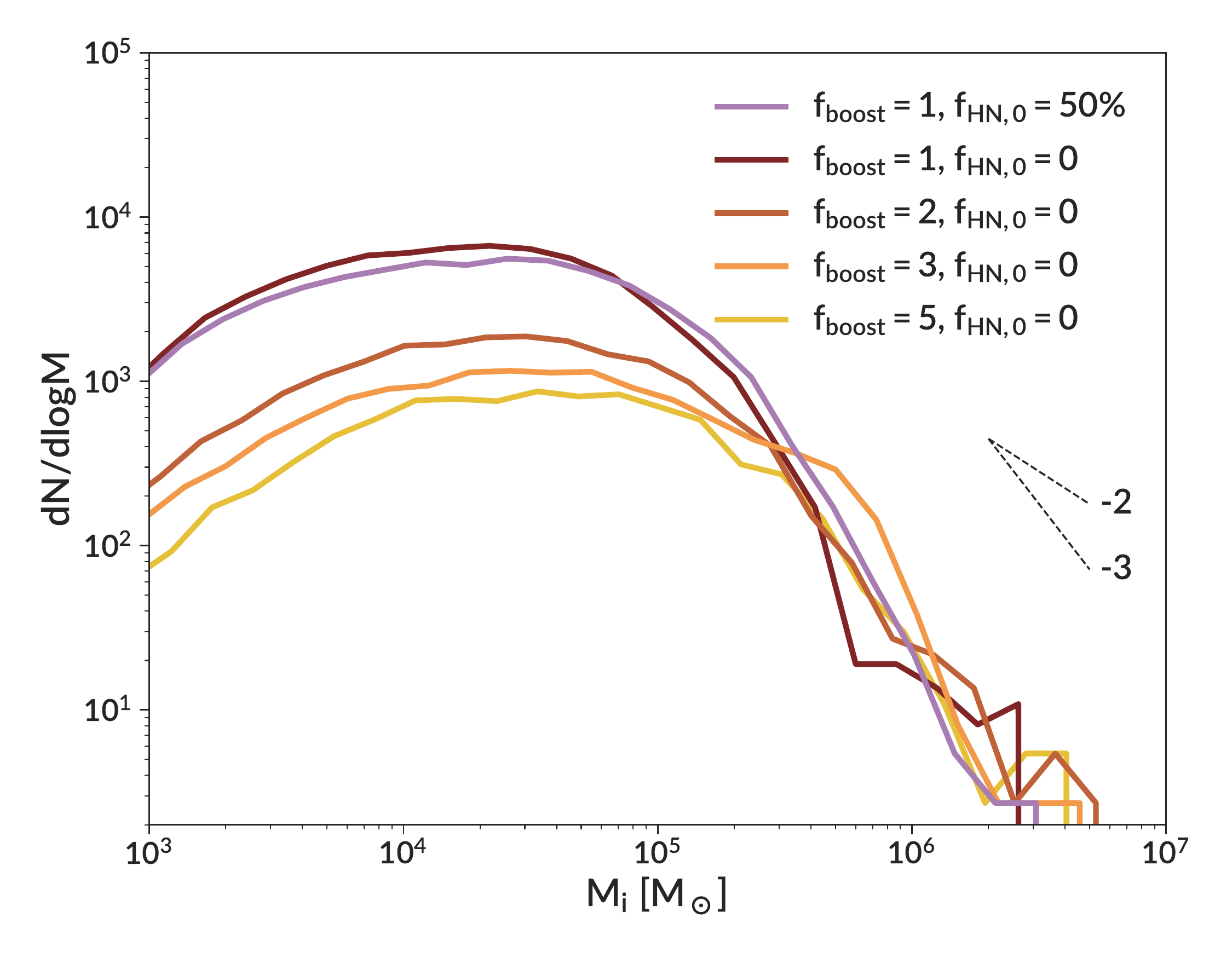}
    \vspace{-7mm}
    \caption{A comparison of the initial cluster mass function for runs with varied $\fboost$ and $\fhno$ at $z=4$. Black dashed lines indicate power-low slopes of $-2$ and $-3$. The lower limit of the plot corresponds to one cluster. All runs use the \huiic\ IC and $\epsff=100$\%.}
    \label{fig:cimf_fboost_oldic}
\end{figure}

While changes in $\fboost$ lead to dramatic changes in global galaxy properties, the changes to the cluster mass function are more subtle.  Figure~\ref{fig:cimf_fboost_oldic} shows the initial cluster mass function for the \huiic\ IC when varying $\fboost$ and $\fhno$. We show all clusters formed before $z=4$, as this higher redshift reduces the differences caused by variations in the star formation rate and includes a higher fraction of low-metallicity clusters where $\fhn$ could potentially make a difference. The normalization changes reflect the change in total stellar mass. Interestingly, the high-mass end is less affected by $\fboost$ than the low-mass end. A lower $\fboost$ serves to increase the number of low-mass particles without systematically increasing the number of massive clusters or the maximum cluster mass. While not shown in Figure~\ref{fig:cimf_fboost_oldic}, we see the same trends when examining the runs using the Local Group ICs. 

We find little difference in the cluster mass function when changing $\fhno$. Figure~\ref{fig:cimf_fboost_oldic} shows little difference between $\fhno=50$\% and $\fhno=0$ for $\fboost=1$ for masses below $10^6\Msun$. However, the run with $\fhno=0$ has several clusters with masses above $10^6\Msun$, while the run with $\fhno=50$\% does not. There are very few clusters in these mass ranges, so stochasticity may play a role in these results. We also examined the low-metallicity clusters separately, again finding no difference. This is true as well of the runs with the Local Group ICs.

To quantitatively evaluate the shape of the mass functions, we fit them with a power-law. As our mass functions do not show a power-law behavior down to low masses, we restrict our fit to masses above $10^5\Msun$ where it is approximately a power-law. Again we note that we are using the particle masses without including the bound fraction, so these results are not directly comparable to observations. Including the bound fraction generally makes the mass function shallower, as high mass clusters have a higher bound fraction (see Section~\ref{sec:jwst}). For $\fhno=0$, we find slopes of $-2.94$, $-2.48$, $-2.16$, and $-2.31$ for $\fboost=1, 2, 3$ and 5, respectively. For $\fhno=50$\% and $\fboost=1$, we find $-2.78$. Lower values of $\fboost$ tend to have steeper slopes due to the higher number of low mass clusters. The $\fboost=3$ run has the shallowest slope due to the large number of clusters at $\approx5\times10^5\Msun$ that deviate from a pure power law fit and draw the fit toward a shallower slope. This feature becomes less prominent at $z=1.5$ as more clusters form and fill out the mass function more evenly. We see similar trends in the Local Group runs, where the slope takes values of $-2.62$, $-2.40$, and $-2.22$ for $\fboost=1, 3$, and 5 respectively.

Lastly, we examined the visual appearance of the gas distribution in these galaxies. \citetalias{li_etal_18_paper2} found that reducing $\fboost$ to 3 led to a dramatic increase in star formation and the formation of an axisymmetric disc, while runs with $\fboost=5$ produced very irregular galaxies \citep{meng_etal_19}. Here, we find that all of our runs produce irregular galaxies, even with $\fboost=1$. 

In summary, we find that higher values of $\fboost$ can greatly decrease the galactic star formation rate by decreasing the number of low mass clusters that form, without changing the number of massive clusters. Increasing the initial hypernova fraction $\fhno$ has little effect on galaxy properties. The fraction quickly approaches zero as metallicity increases, leading to little difference in the injected energy and momentum.

\subsection{Molecular gas prescription}
\label{sec:molecular}

A key ingredient in modeling star formation in our simulations is the amount of molecular gas, as we require a given cell to have a mass fraction of molecular gas greater than 50\% to seed a cluster particle. 

As discussed in \citet{gnedin_draine_14} and Appendix A7 of \citet{gnedin_kravtsov_11}, the clumping factor $C_\rho$ is one of the tunable parameters of the molecular gas model. This factor accounts for the fact that gas is clustered on scales that are not resolved in a given simulation, so \hmol\ formation would be missed. Larger values of the clumping factor produce more molecular gas at a given surface density. Numerical simulations of turbulent molecular clouds find lognormal density distributions with widths that imply $C_\rho \approx 3 - 10$ \citep{mckee_ostriker_07}. \citet{gnedin_kravtsov_11} and \citet{gnedin_draine_14} calibrated the clumping factor in the ART code based on simulations, finding that values in the range of 10 to 30 work well. However, those simulations had lower resolution than our runs. This would imply that our runs should prefer a lower clumping factor, because they are resolving more substructure and leaving less on subgrid scales.

Motivated by the disagreement between our simulations and the observed global galactic star formation history (e.g. Figure~\ref{fig:sfh_feedback_oldic}), we explored a range of molecular gas prescriptions. We ran simulations with a range of clumping factors, using $C_\rho=3, 10,$ and $30$. \citetalias{li_etal_18_paper2} used 10, as do all other runs presented in this paper. We also used one run with $C_\rho=3$ where we changed the prescription for shielding from that of \citet{gnedin_draine_14} to that of \citet{gnedin_kravtsov_11}. The \citet{gnedin_draine_14} model includes the effects of line overlap in the Lyman-Werner bands, increasing self-shielding, which is particularly relevant in low metallicity environments with less dust shielding. However, both models for self-shielding were calibrated using runs with lower resolution than our runs ($>50$~pc compared to 3--6~pc) and with a different feedback model. These differences in simulation setup can affect the performance of the \hmol\ formation model, so we decided to explore both shielding prescriptions. All runs used $\fboost=1$, $\fhno=0$, and $\epsff=100$\%. 

As expected, the only significant differences caused by $C_\rho$ were in the amount of molecular gas. While the mass of molecular gas in each run varies greatly with time, we find a general trend that larger values of $C_\rho$ produce more molecular gas. We see little change in molecular gas masses when changing the shielding prescription. These differences in the amount of molecular gas when changing $C_\rho$ led to some differences in star formation histories. The total stellar mass at $z=1.5$ for the run with $C_\rho=3$ is $3 \times 10^9 \Msun$, while the mass for the run with $C_\rho=30$ is $6 \times 10^9 \Msun$. In particular, a higher clumping factor leads to more late-time star formation.

\subsection{Star formation efficiency}
\label{sec:results_epsff}

The local star formation efficiency per freefall time $\epsff$ is a key parameter of our model (see Equation~\ref{eq:m_dot}). As \citetalias{li_etal_18_paper2} showed, this parameter strongly influences many star cluster properties, particularly the mass function, while not strongly affecting the global galaxy properties. We continue that exploration here.

As $\epsff$ controls how fast star particles accrete material, we expect it to be reflected in the duration of cluster formation episodes. We find that to be the case. In particular, we find that runs with low values of $\epsff$ often fail to finish forming massive clusters before the algorithmic end to a star formation episode at 15~Myr. For example, in the run using the \huiic\ IC, $\epsff=1$\%, $\fboost=1$, and $\fhno=0$, only 20\% of clusters with masses above $10^5\Msun$ finished their formation before it was automatically stopped.

When this time cap is imposed, cluster formation ends even when gas is available to continue accreting onto the cluster. Therefore, we cannot interpret these particles as the end-products of cluster formation. Their masses are not self-consistently determined by their feedback. The masses we obtain are lower limits to the true masses that would have formed over longer timescales. However, as we will discuss more in Section~\ref{sec:discussion_epsff}, such long age spreads of stars within a single cluster are ruled out by observations. We define runs as having failed cluster formation if more than 50\% of clusters with masses above $10^5\Msun$ have durations longer than 14~Myr. This applies to all runs with $\epsff=1$\% and the run using the \huiic\ IC, $\epsff=10$\%, $\fboost=1$, and $\fhno=0$. \revision{While we still include these runs in plots, we indicate the cluster mass ranges where they are unreliable using dashed lines (namely Figures~\ref{fig:cimf_sfe_oldic}, \ref{fig:cimf_sfe_lg}, \ref{fig:bound_fraction_tl_sfe}, and \ref{fig:cimf_current_lg_sfe_z4}), or use completely dashed lines when mass is not an explicit variable (namely Figures~\ref{fig:eps_int_tl_sfe}, \ref{fig:eps_int_eps_ff_ratio}, and \ref{fig:h2_pdf})}. We defer a full investigation of this failed cluster formation to Section~\ref{sec:failed}. 

\begin{figure*}
    \includegraphics[width=\textwidth]{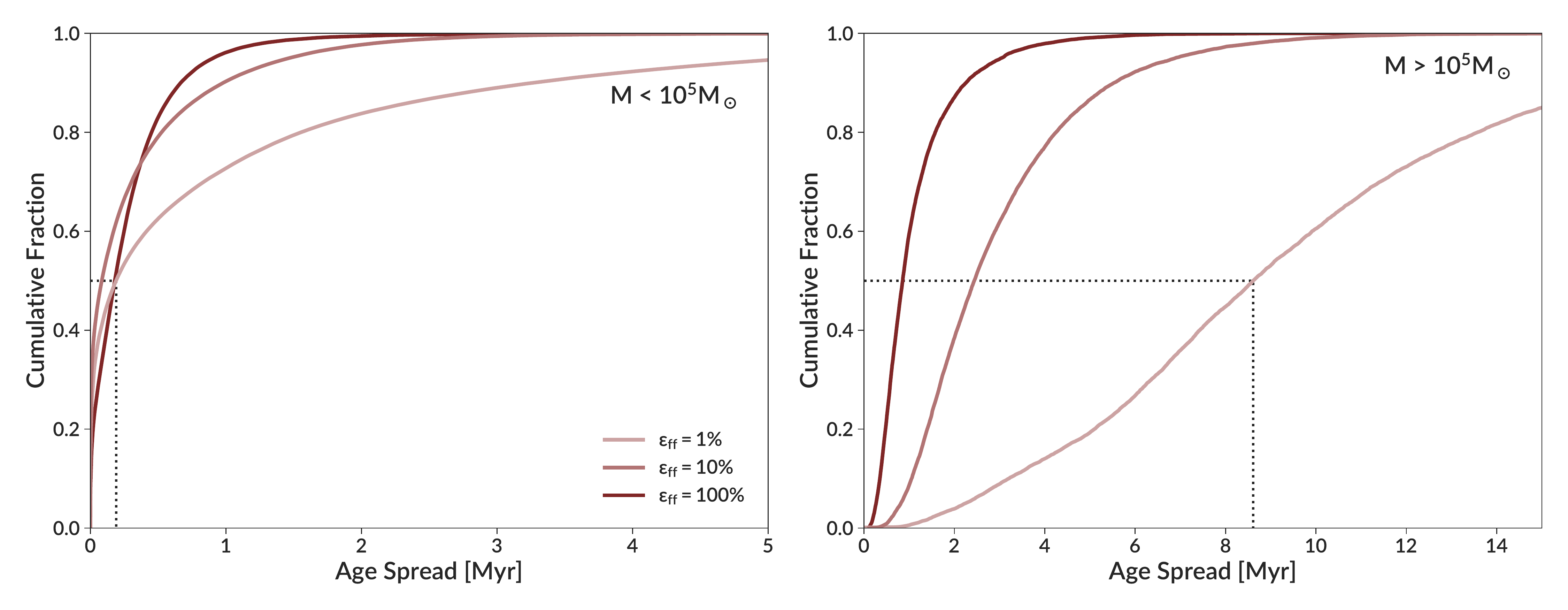}
    \vspace{-7mm}
    \caption{The cumulative distribution of the star particle internal age spread $\tau_{\rm spread}$ for the \huiic\ IC runs with varied $\epsff$. Note that this is not the duration of star formation as plotted in Figure~\ref{fig:age_duration_sn_timing}, it is the age spread as defined in Equation~\ref{eq:age_spread} evaluated at the end of cluster formation. The left panel shows clusters less massive than $10^5\Msun$, while the right panel shows clusters more massive than $10^5 \Msun$. Note the different range spanned by the two panels. The dotted line shows the longest median age spread. All runs use $\fboost=1$, $\fhno=0$, and show all clusters formed before $z=1.5$.}
    \label{fig:age_spread_oldic_sfe}
\end{figure*}

\begin{figure}
    \includegraphics[width=\columnwidth]{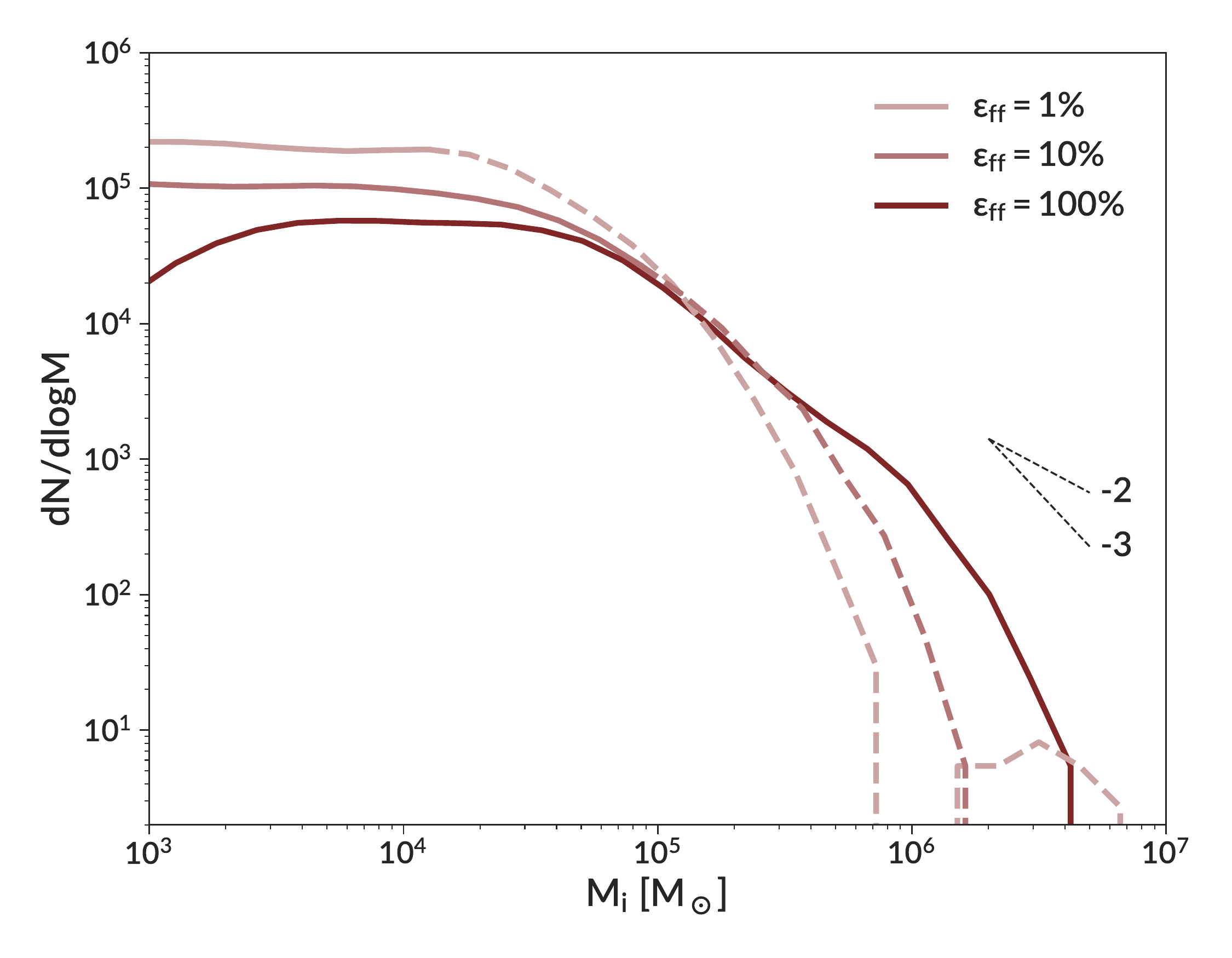}
    \vspace{-7mm}
    \caption{A comparison of the star particle initial mass function for the \huiic\ IC runs with varied $\epsff$. \revision{For runs with failed cluster formation, dashed lines indicate the range where more than 50\% of clusters have formation durations longer than 14~Myr.}
    Black dashed lines indicate power-low slopes of $-2$ and $-3$. The lower limit of the plot corresponds to one cluster. All runs use $\fboost=1$, $\fhno=0$, and show all clusters formed before $z=1.5$.}
    \label{fig:cimf_sfe_oldic}
\end{figure}

\begin{figure}
    \includegraphics[width=\columnwidth]{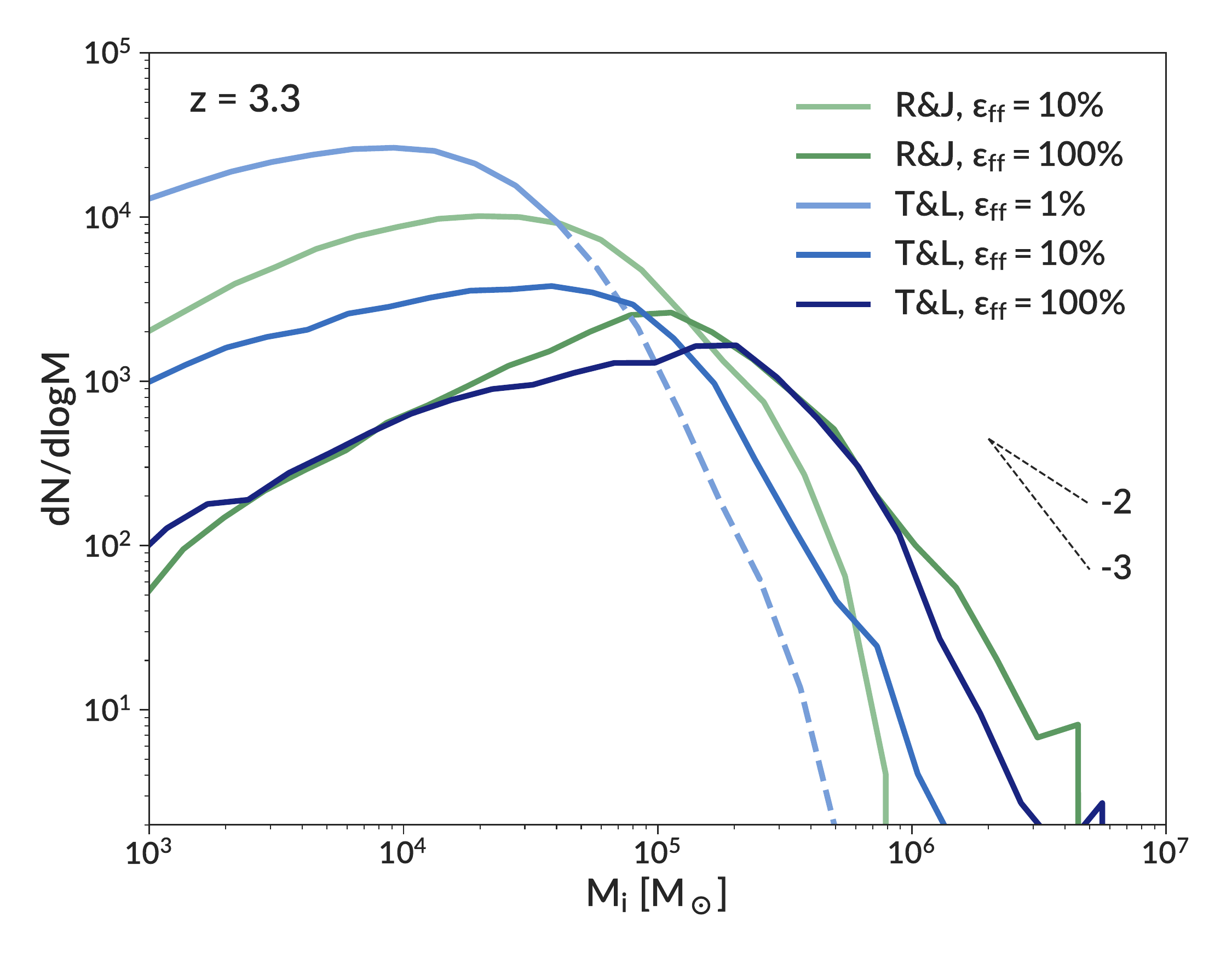}
    \vspace{-7mm}
    \caption{A comparison of the star particle initial mass function for the Local Group ICs with varied $\epsff$. 
    \revision{For the $\epsff=1$\% run with failed cluster formation, dashed lines indicate the range where more than 50\% of clusters have formation durations longer than 14~Myr.}
    Black dashed lines indicate power-low slopes of $-2$ and $-3$. The lower limit of the plot corresponds to one cluster. All runs use $\fboost=5$, $\fhno=20$\%, and show all clusters formed before \revision{$z=3.3$, the lowest redshift that all simulations have reached.}}
    \label{fig:cimf_sfe_lg}
\end{figure}

To illustrate the difference in the timescale of cluster formation, Figure~\ref{fig:age_spread_oldic_sfe} shows the cumulative distribution of age spread $\tau_{\rm spread}$ for runs using the \huiic\ IC. The dependence on $\epsff$ is clear. For massive clusters, the median age spread is 8.6~Myr for $\epsff=1$\%, while it is 2.4~Myr for $\epsff=10$\% and 0.9~Myr for $\epsff=100$\%. For $\epsff=1$\% many clusters have unphysically long age spreads, some longer than 15~Myr. We note that the age spread can be longer than the duration of star formation in some cases, as it is a measure of the variance in the star formation rate rather than simply its length. Atypical star formation histories, such as one with bursts of star formation at early and late times, can lead to large values of $\tau_{\rm spread}$. There is also a clear mass dependence. Clusters with masses below \revision{$10^5\Msun$} and $\epsff\ge10$\% have median age spreads less than 0.2~Myr, with all low-mass clusters from those runs having age spreads less than 2~Myr. However, for $\epsff=1$\% there is a clear tail to long age spreads even among low-mass clusters, with some clusters having age spreads as long as 10~Myr. 

We next investigate the effect of $\epsff$ on the cluster mass functions. Figure~\ref{fig:cimf_sfe_oldic} shows the initial cluster mass function for runs using the \huiic\ IC, and Figure~\ref{fig:cimf_sfe_lg} shows the same for the Local Group runs. Similar trends are seen in both plots. Higher values of $\epsff$ lead to more massive clusters and a higher maximum cluster mass, while lower values of $\epsff$ produce more low-mass clusters. The exception to this is a handful of very massive clusters that formed in the \huiic\ $\epsff=1$\% run, leading to a separate hump in the high-mass end of the mass function. The indicates that even with low values of $\epsff$, massive clusters are still possible, although typically rare. We note that we do not see such hump in the \tl\ run with $\epsff=1$\%.

The slope of the high-mass end of the mass function varies with $\epsff$, with the mass function being shallower for higher values of $\epsff$. As with all calculations of the mass function slope, we restrict our fit to clusters above $10^5\Msun$. For $\epsff=1$\%, the slope is between $-3.78$ and $-4.41$ for runs on the different ICs, while for $\epsff=10$\% it is between $-2.94$ and $-3.42$, and for $\epsff=100$\% it is between $-2.25$ and $-2.60$.

The exact shape of the mass function is somewhat different between the runs that use the \huiic\ IC and those that use the Local Group ICs, with the Local Group runs having fewer low-mass particles. These Local Group runs used $\fboost=5$, which decreases the number of low-mass clusters compared to lower values of $\fboost$ (see Figure~\ref{fig:cimf_fboost_oldic}). The different redshift of these runs also likely contributes. We find that the majority of massive particles form at very high redshift or in galactic mergers, when the star formation rate is high. This matches what was seen in \citetalias{li_etal_18_paper2}, and agrees with both observations and theoretical expectations \citep{portegies_zwart_etal_10,kruijssen_14}. In more quiescent epochs, high-mass particles do not form, giving proportionally more low-mass particles. As time progresses, more low-mass clusters are likely to form in the Local Group runs, possibly making their mass functions more similar to those seen in the \huiic\ IC.

\begin{figure}
    \includegraphics[width=\columnwidth]{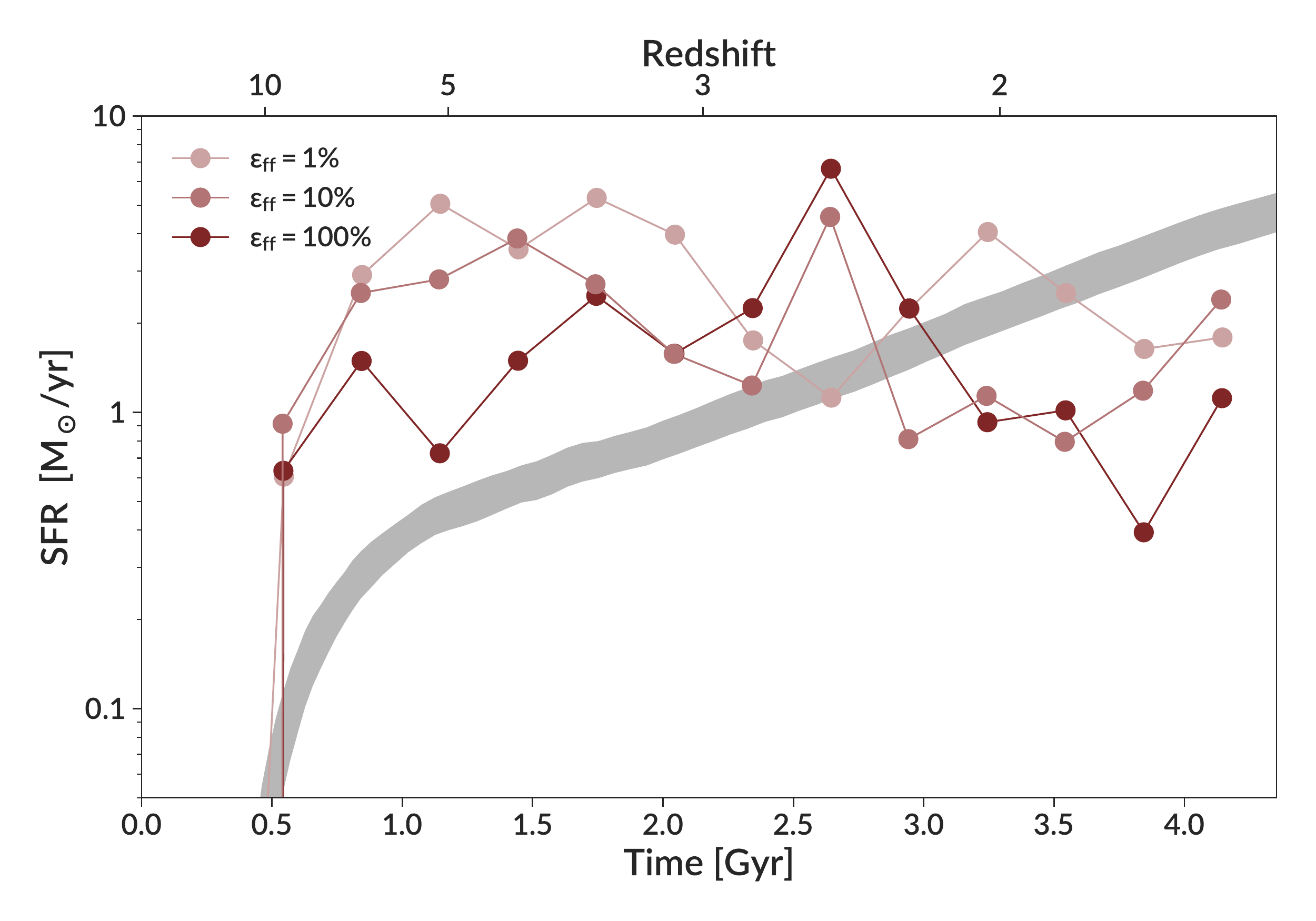}
    \vspace{-7mm}
    \caption{A comparison of the star formation rate of the central galaxy in the \huiic\ IC when varying $\epsff$. All runs use $\fboost=1$ and $\fhno=0$.}
    \label{fig:sfh_sfe}
\end{figure}

While $\epsff$ significantly affects cluster properties, it does not change the galactic star formation rate appreciably. Figure~\ref{fig:sfh_sfe} shows the star formation histories of runs when varying $\epsff$ while holding $\fboost=1$ and $\fhno=0$ constant. Here we find that lower values of $\epsff$ lead to somewhat higher star formation rates at early times. These star formation rates at $z\approx5$ are significantly higher than predicted by \um, and tend to decline with time rather than increase. However, we find opposite trends during the major merger at $z\approx2.6$, when the high $\epsff$ runs show a stronger burst. In the runs using the Local Group analogs, the star formation history does not change significantly with $\epsff$. 

In summary, we find that $\epsff$ does not have a significant impact on the galactic star formation rate, but does strongly influence star cluster properties. In particular, higher values of $\epsff$ lead to more massive star clusters and shorter timescales for cluster formation. These results confirm those found in \citetalias{li_etal_18_paper2}, indicating that they are robust predictions of our simulations.

\subsection{Virial criterion}
\label{sec:virial}

\begin{figure}
    \includegraphics[width=\columnwidth]{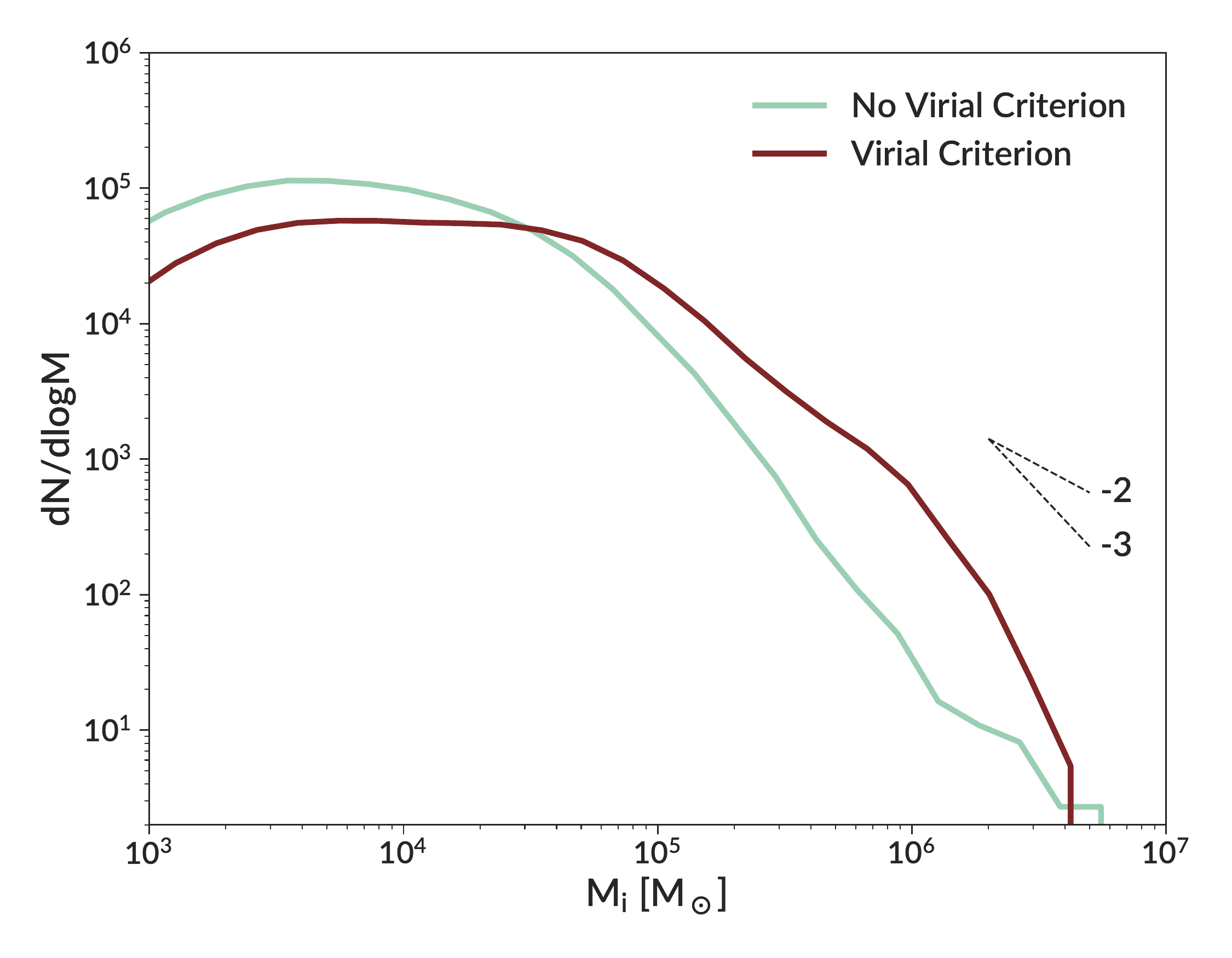}
    \vspace{-7mm}
    \caption{A comparison of the initial cluster mass function for runs with and without the virial criterion for seeding star formation. Black dashed lines indicate power-low slopes of $-2$ and $-3$. Both runs used the \huiic\ IC, $\epsff=100$\%, $\fboost=1$, $\fhno=0$, and show all clusters formed before $z=1.5$.}
    \label{fig:cimf_virial}
\end{figure}

One of the other changes to our star cluster formation prescription was the addition of a criterion restricting star-forming gas to be gravitationally bound (see the beginning of Section~\ref{sec:code_cluster_formation}). To investigate the difference this makes in cluster properties, we ran one simulation with the virial criterion turned off. While we find no significant differences in large scale galactic properties, we find differences in the star cluster populations. Figure~\ref{fig:cimf_virial} shows the mass function for runs with and without the virial criterion. The addition of the virial criterion leads to more high-mass clusters and fewer low-mass clusters. Quantitatively, the power-law slopes of the mass functions for clusters above $10^5\Msun$ are $-2.60$ for the run with the virial criterion and $-3.30$ for the run without it. While the maximum cluster mass is similar between the two runs, there are significantly more clusters with masses above $10^6\Msun$ when the virial criterion is enabled.

\begin{figure}
    \includegraphics[width=\columnwidth]{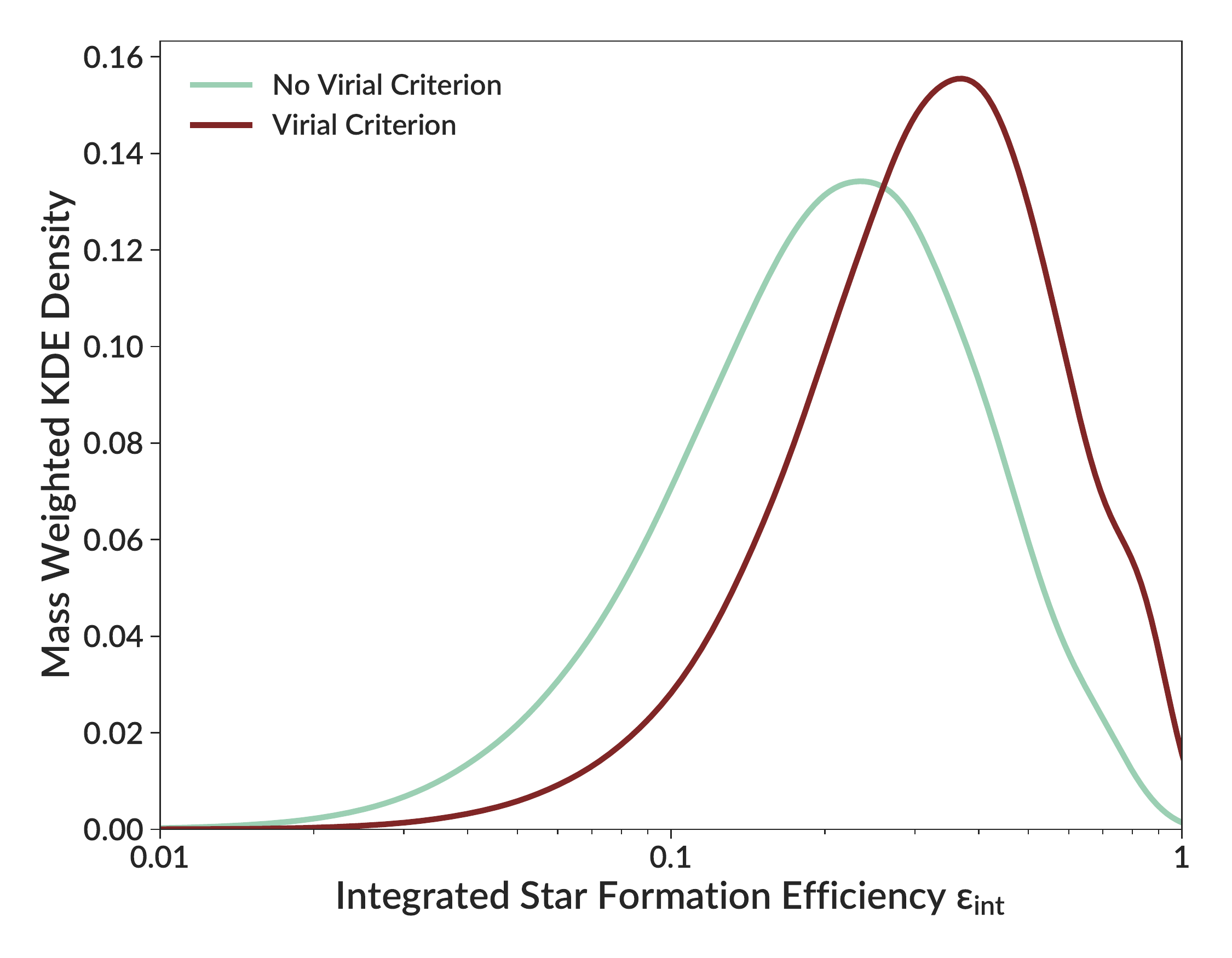}
    \vspace{-7mm}
    \caption{Kernel density estimation for the distribution of integrated star formation efficiency for clusters in the runs with and without the virial criterion. We use a Gaussian kernel with a width of 0.05 dex. Each curve is normalized to the same area for comparison purposes. 
    Both runs used the \huiic\ IC, $\epsff=100$\%, $\fboost=1$, $\fhno=0$, and show all clusters formed before $z=1.5$.}
    \label{fig:eps_int_virial}
\end{figure}

The increase in the number of high-mass clusters is expected, as Equation~\ref{eq:virial_2} shows lower gas densities lead to higher virial parameters. The cut on the virial parameter prevents these lower density GMCs from forming stars until they accrete more gas and collapse to higher density, leading to more total mass available for star formation. The later onset of star formation also delays stellar feedback, allowing more gas to accrete onto the cluster. These processes shift many low-mass clusters to higher masses, explaining the decrease in the number of low-mass clusters. In addition, as the virial criterion allows more gas accretion onto the GMC, its larger mass becomes more difficult to disperse with feedback, leading to longer durations of star formation. As a consequence of these effects, clusters have higher values of $\epsint$ when the virial criterion is enabled. In  Figure~\ref{fig:eps_int_virial} we show the distribution of $\epsint$ with and without the virial criterion. Both distributions have widths $\approx0.25$~dex, but the mean value for the run with the virial criterion is significantly higher (35\% compared to 21\%). 



In the run where we did not impose the virial criterion, we output the virial parameter $\alpha_{\rm vir}$ of each cluster as it formed. Using this information, we can postprocess the results to see if there are any correlations between the virial parameter and the resulting cluster properties. We find that clusters with $\alpha_{\rm vir} < 10$ tend to have higher initial masses, higher $\epsint$, and higher initial bound fractions than those with $\alpha_{\rm vir} > 10$. The virial criterion acts in a biased fashion to allow star formation to happen in regions that preferentially lead to higher mass clusters. Additionally, regions with $\alpha_{\rm vir} > 10$ are able to accrete more material over time until they pass the $\alpha_{\rm vir} < 10$ threshold, increasing the cluster mass that formed out of a given GMC.

In summary, we find that adding the requirement that star-forming gas have a virial parameter $\alpha_{\rm vir} < 10$ increases the number of massive clusters, gives clusters a longer formation timescale, and leads to higher values of $\epsint$. 

\section{Evolution of the cluster mass function}
\label{sec:jwst}

In the previous section we exclusively used the masses of the star particles at the end of their formation process. As not all stars are gravitationally bound to the newly formed cluster, we must incorporate the initial bound fraction to obtain the observable cluster masses. In addition, the plots in the previous section showed the distributions of initial masses for all clusters formed over the full time spanned by the simulation. This is not observable. In this section we include the cluster bound fraction and present the instantaneous cluster mass function at a given redshift to allow for more direct comparison with observations. While these are not true mock observations, the results shown here accurately represent the existing cluster populations at a given redshift in our simulations. 

\begin{figure}
    \includegraphics[width=\columnwidth]{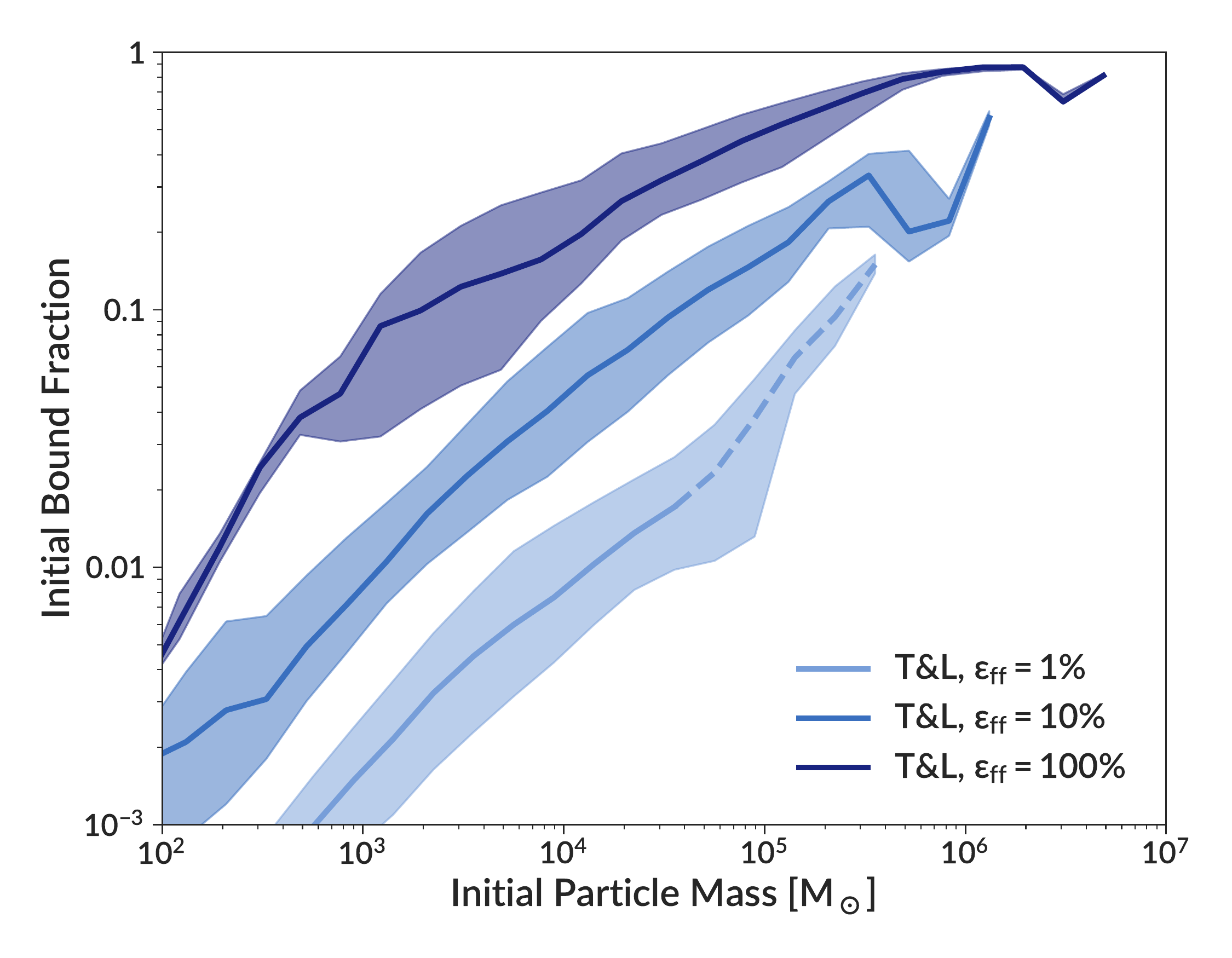}
    \vspace{-7mm}
    \caption{The initial bound fractions for runs using the \tl\ IC and varying $\epsff$. The solid line shows the median, with the shaded region showing the interquartile range of the distribution of the initial bound fraction at that mass. The mass plotted here is the particle mass at the end of cluster formation, not the bound cluster mass, so that the plotted variables are independent. 
    \revision{For the $\epsff=1$\% run with failed cluster formation, dashed lines indicate the range where more than 50\% of clusters have formation durations longer than 14~Myr.}
    We only show the \tl\ IC for clarity, but \rj\ and \huiic\ show the same behavior. All runs use $\fboost=5$, $\fhno=20$\%, and show all clusters formed before \revision{$z=3.3$, the lowest redshift that all simulations have reached.}}
    \label{fig:bound_fraction_tl_sfe}
\end{figure}

We start by examining the cluster initial bound fraction, which is needed to turn raw particle masses into bound cluster masses. Figure~\ref{fig:bound_fraction_tl_sfe} shows the initial bound fraction of clusters as a function of mass. As in \citetalias{li_etal_18_paper2}, we see the trend of higher mass clusters having higher bound fraction. Additionally, runs with higher $\epsff$ have higher bound fractions at a given particle mass. 


\begin{figure}
    \includegraphics[width=\columnwidth]{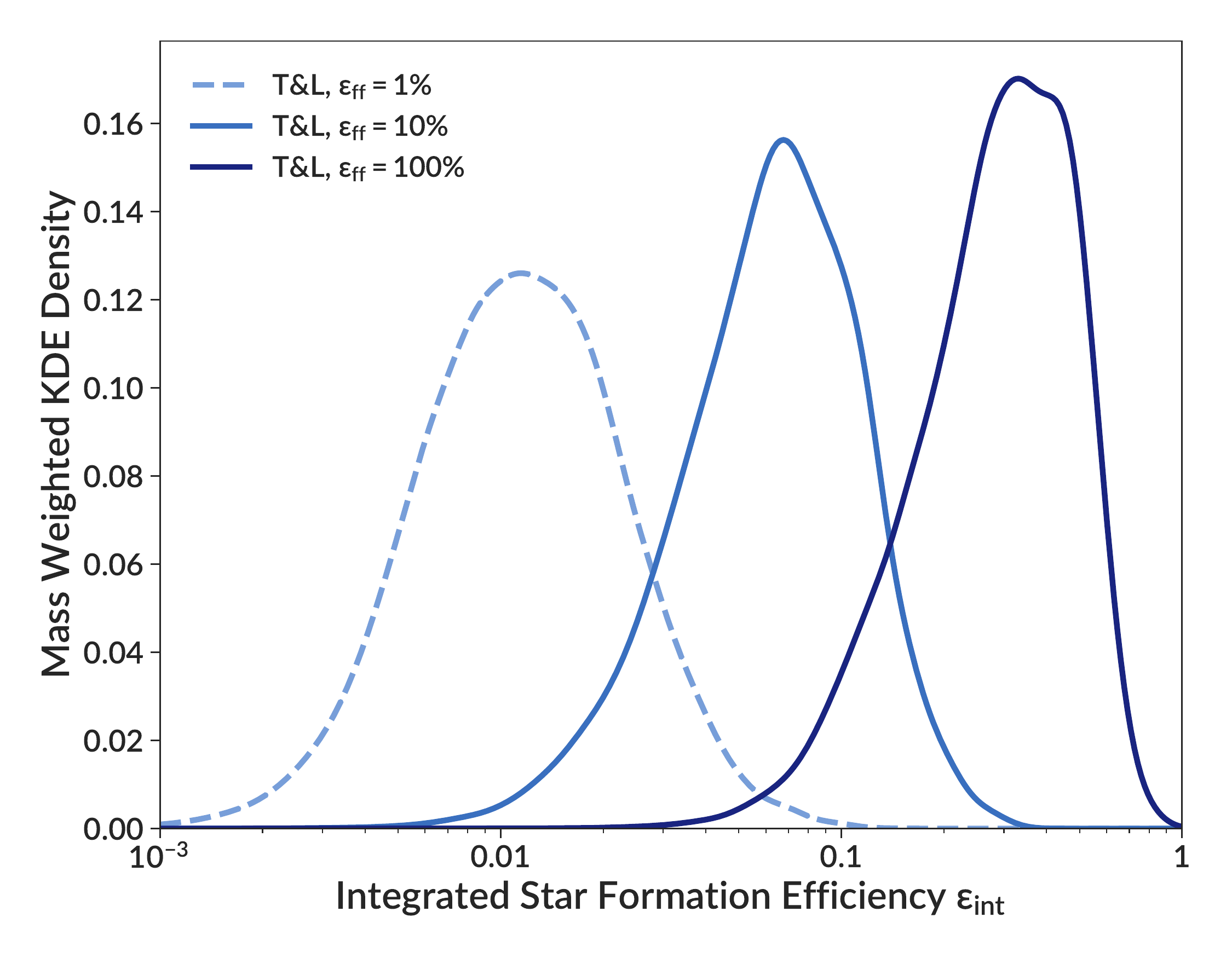}
    \vspace{-7mm}
    \caption{Kernel density estimation for the distribution of integrated star formation efficiency for clusters in the \tl\ IC with variations in $\epsff$. We use a Gaussian kernel with a width of 0.05 dex. Each curve is normalized to the same area for comparison purposes. All runs use $\fboost=5$, $\fhno=20$\%, and show all clusters formed before \revision{$z=3.3$, the lowest redshift that all simulations have reached.} We plot the $\epsff=1$\% run with a dashed line as that run had many clusters that failed to finish forming.}
    \label{fig:eps_int_tl_sfe}
\end{figure}

Our prescription for the initial bound fraction (Equation~\ref{eq:bound_fraction}) makes it solely dependent on the integrated star formation efficiency $\epsint$. In Figure~\ref{fig:eps_int_tl_sfe} we show the distributions of $\epsint$. Runs with lower $\epsff$ have lower $\epsint$. For a given run, the spread is due to trends with mass, where high-mass clusters have higher $\epsint$ than low-mass clusters. Quantitatively, the mean value of $\epsint$ takes values of 1.2\%, 7.2\%, and 30\% for $\epsff=1$\%, 10\%, and 100\%, respectively. As $\epsff$ increases, the widths of these distributions decrease, with values of 0.30, 0.24, and 0.17 dex, respectively.

The trend of higher $\epsint$ with higher $\epsff$ is a direct consequence of $\epsff$ controlling the cluster formation rate (Equation~\ref{eq:m_dot}). Higher $\epsff$ leads to higher star formation rates, allowing the cluster to accrete more of the gas from its surroundings. This is reflected in the duration of cluster formation in runs with different $\epsff$. A lower value $\epsff$ causes clusters to form more slowly. With a slow star formation rate, feedback also starts before the cluster has accreted a significant fraction of the surrounding gas, leading to lower $\epsint$. The different timescales also likely lead to the change in width of the distributions. As low values of $\epsff$ lead to longer timescales of cluster formation, there is more possibility for variation in the accretion history of the GMC. High values of $\epsff$ form quickly, so they are forming mostly out of the gas that was present at cluster birth. 

\begin{figure}
    \includegraphics[width=\columnwidth]{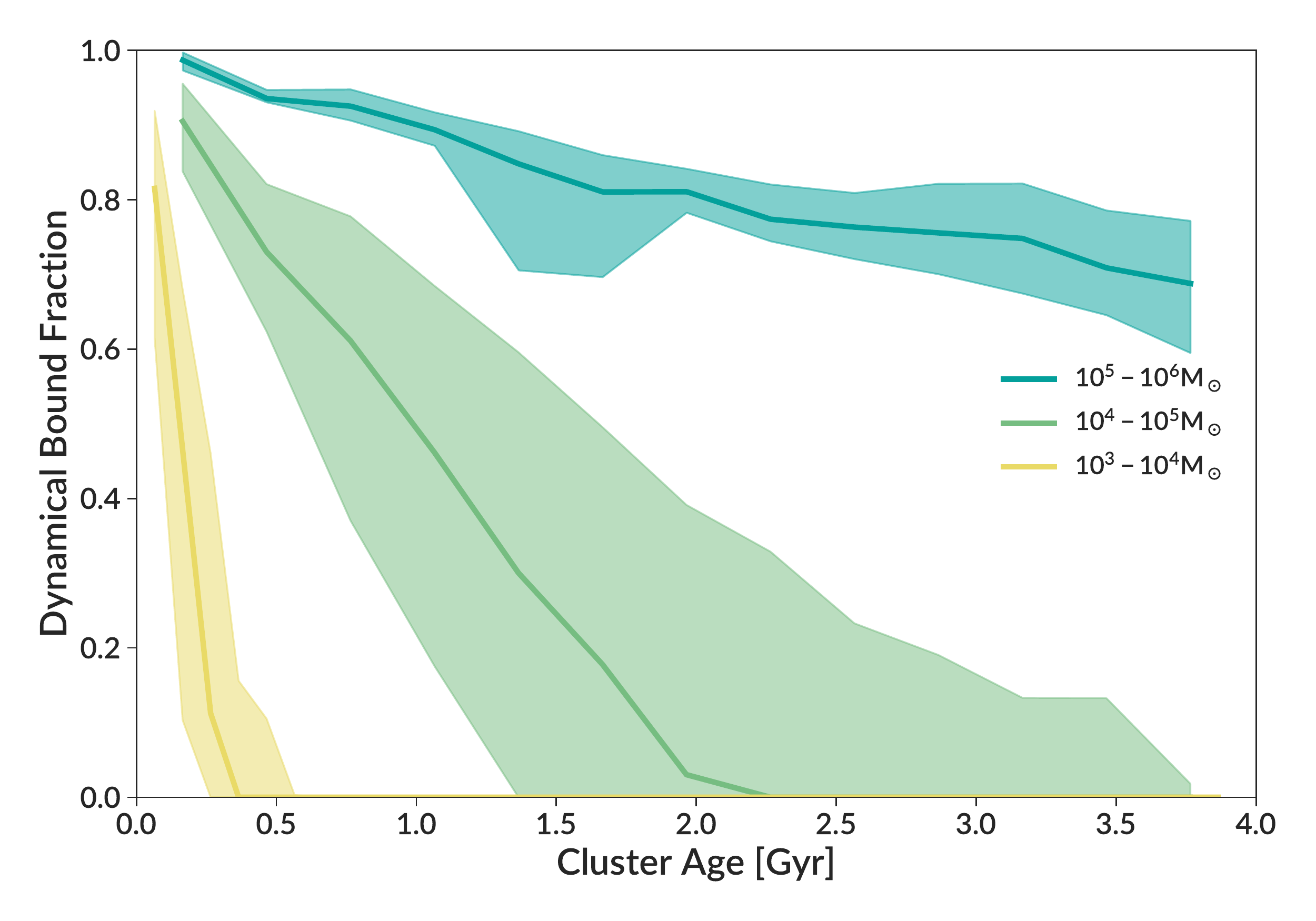}
    \vspace{-7mm}
    \caption{Evolution of the dynamical bound fraction $f_{\rm dyn}$ as a function of cluster age for clusters in different mass ranges. Lines show the median, with the shaded region showing the interquartile range. Clusters are grouped according to their initial bound mass at formation. The plot shows clusters in the central galaxy of the run using the \huiic\ IC, $\epsff=100$\%, $\fboost=1$, $\fhno=0$, and shows all clusters formed before $z=1.5$.}
    \label{fig:dynamical_disruption}
\end{figure}

In addition to the initial bound fraction, we also need to account for stellar evolution and dynamical disruption, which both cause clusters to lose mass with time. These processes are calculated in simulation runtime. In general, the mass $M_b$ bound to a cluster at time $t$ can be written as
\begin{equation}
    M_b(t) = M_i \, f_i \, f_{\rm se}(t) \, f_{\rm dyn}(t)
\end{equation}
where $M_i$ is the initial particle mass, $f_i$ is the initial bound fraction, $f_{\rm se}(t)$ accounts for mass loss due to stellar evolution, and $f_{\rm dyn}$ accounts for mass lost due to tidal stripping \citep{li_gnedin_19_paper3,meng_gnedin_22}. Our feedback scheme self-consistently decreases the stellar mass of the cluster whenever mass is ejected into the ISM, and dynamical disruption is calculated as described in Section~\ref{sec:cluster_disruption}. In Figure~\ref{fig:dynamical_disruption}, we show the impact of disruption on clusters of different mass, taking as an example the run using the \huiic\ IC, $\epsff=100$\%, $\fboost=1$, and $\fhno=0$. We choose a run using the \huiic\ IC as it reached a lower redshift, so clusters have more time to disrupt. Clusters with masses below $10^4\Msun$ are entirely disrupted within 500~Myr. Clusters of intermediate mass $10^4-10^5\Msun$ persist for a few Gyr, but do not survive until the present. However, clusters with masses above $10^5\Msun$ survive throughout the lifetime of the simulation. Tidal disruption only decreases the mass of these clusters by approximately 20\% over the 4~Gyr length of this simulation.

\begin{figure}
    \includegraphics[width=\columnwidth]{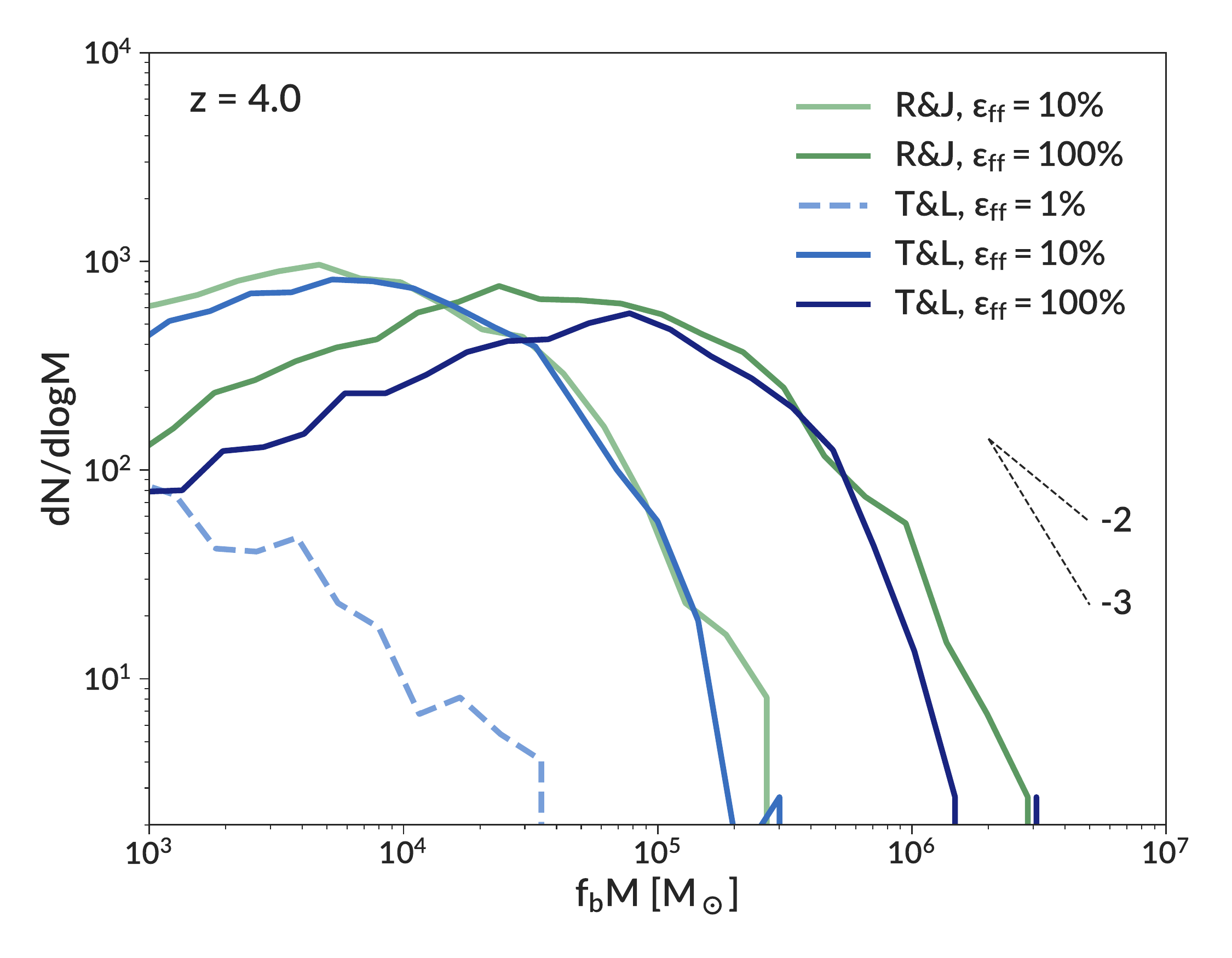}
    \vspace{-7mm}
    \caption{The bound mass function of all clusters present at $z=4$ using the Local Group ICs for different values of $\epsff$. 
    \revision{For the $\epsff=1$\% run with failed cluster formation, dashed lines indicate the range where more than 50\% of clusters have formation durations longer than 14~Myr.} Black dashed lines indicate power-low slopes of $-2$ and $-3$. The lower limit of the plot corresponds to one cluster. All runs use $\fboost=5$ and $\fhno=20$\%.}
    \label{fig:cimf_current_lg_sfe_z4}
\end{figure}

Using these disruption calculations, we now present the mass function of bound clusters at various redshifts. In Figure~\ref{fig:cimf_current_lg_sfe_z4}, we show the mass function of the surviving clusters at $z=4$ in the Local Group runs with varied $\epsff$. This figure shows trends similar to those seen in Figure~\ref{fig:cimf_sfe_lg}, with several trends more exaggerated now that bound cluster mass is included. First, we note similar shapes. Our mass functions have a sharp cutoff at high masses, a peak, and a shallower decrease to low masses. This shape is seen in all runs with $\epsff \geq 10$\%. The position of the peak depends strongly on $\epsff$. For $\epsff=100$\% it is at approximately $10^5\Msun$, while it is closer to $10^4\Msun$ for $\epsff=10$\%. This is due to a combination of three effects. First, as seen in Figure~\ref{fig:cimf_sfe_lg}, the initial particle masses are higher for higher values of $\epsff$. Second, higher values of $\epsff$ give higher bound fractions, as shown in Figure~\ref{fig:bound_fraction_tl_sfe}. The two effects magnify each other, such that higher values of $\epsff$ result in cluster mass functions that reach to significantly higher masses. The disparity is further increased by the effects of disruption, which preferentially removes low-mass clusters (Figure~\ref{fig:dynamical_disruption}). These three effects combine to produce dramatically different cluster mass functions when changing $\epsff$. Of note, the $\epsff=1$\% run has no existing clusters above $10^4\Msun$, while the $\epsff=10$\% runs have no clusters above $3\times10^5\Msun$.

We also note that, as described in Section~\ref{sec:feedback_strength}, the low-mass end of the mass function is sensitive to $\fboost$, with higher values of $\fboost$ decreasing the number of low-mass clusters. The runs shown in Figure~\ref{fig:cimf_current_lg_sfe_z4} were all run with $\fboost=5$. Lower values of $\fboost$ would increase the number of low-mass clusters and give it a shape more similar to that seen in the local universe. Similarly, massive clusters tend to form in epochs of intense star formation, while low-mass clusters dominate in more quiescent epochs. As only the \texttt{Thelma} IC has any significant mergers after the redshift shown in this plot, we expect there to be more low-mass clusters as time progresses. 

\begin{figure}
    \includegraphics[width=\columnwidth]{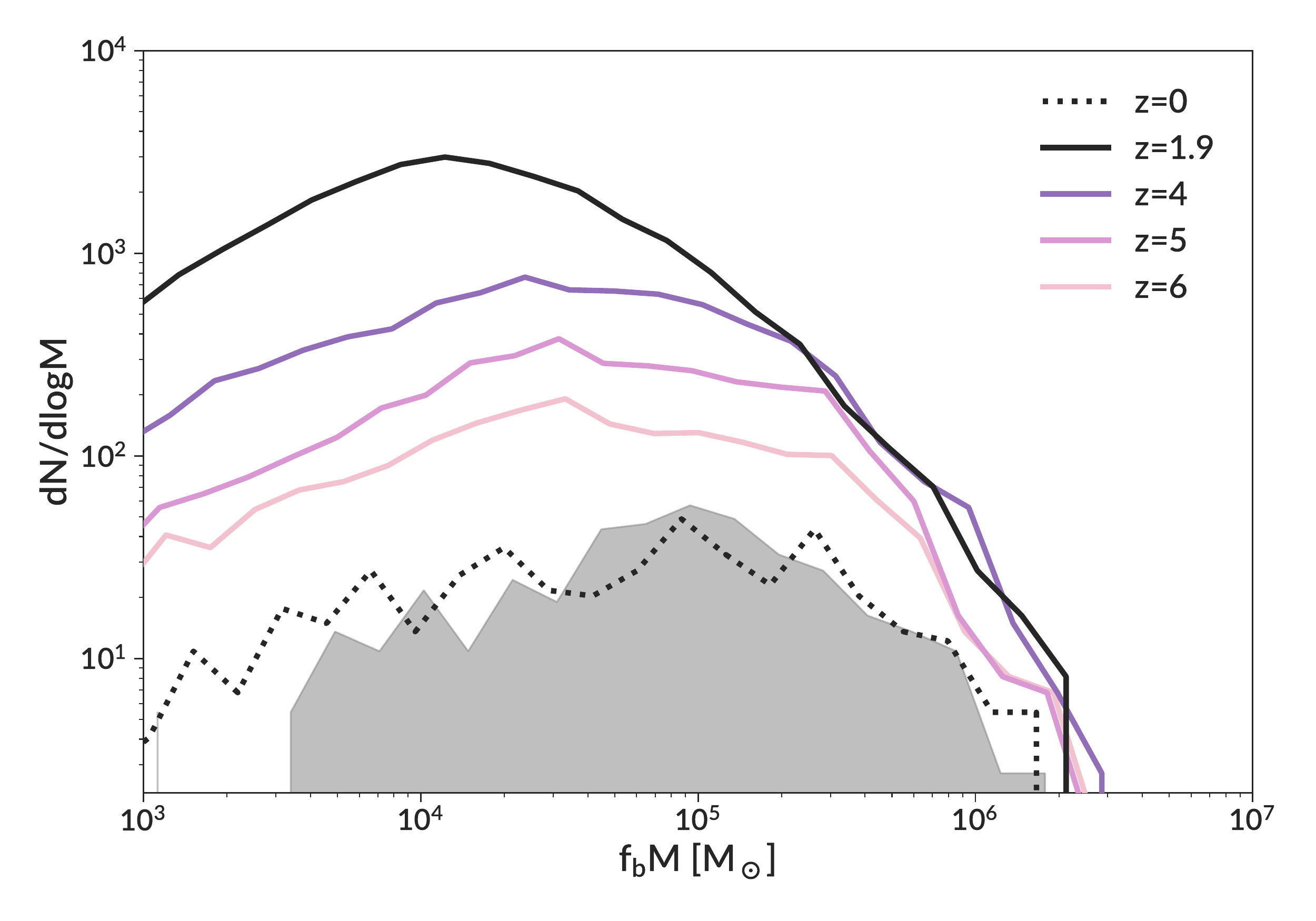}
    \vspace{-7mm}
    \caption{The bound mass function of all clusters present at a range of redshifts. All simulation lines are from the same run that uses the \rj\ IC, $\epsff=100$\%, $\fboost=5$, and $\fhno=20$\%. The \revision{dotted} line indicates the cluster population analytically evolved from $z=1.9$ to $z=0$. The shaded region shows the observed mass function of clusters in the MW.}
    \label{fig:cimf_rj_evolution}
\end{figure}

In Figure~\ref{fig:cimf_rj_evolution} we show the evolution of the bound cluster mass function from $z=6$ to $z=1.9$ for the run using the \rj\ IC, $\epsff=100$\%, $\fboost=5$, and $\fhno=20$\%. A significant fraction of clusters with masses above $2\times10^5\Msun$ are in place already at $z=6$. More massive clusters form by $z=4$, but we see little change in the massive end of the mass function beyond that redshift. At later epochs low-mass clusters dominate the mass function, particularly increasing the number of clusters around $10^4\Msun$. Clusters of low mass that appear in this plot are mainly from recent star formation. As Figure~\ref{fig:dynamical_disruption} shows, clusters with masses below $10^5\Msun$ disrupt within a few Gyr, and clusters below $10^4\Msun$ disrupt within several hundred Myr. 

We also show an analytical evolution of star clusters from the last output of this run at $z=1.9$ to the present, following \citet{li_gnedin_19_paper3}. The prescription for tidal disruption (Equations~\ref{eq:omega_tid}--\ref{eq:disruption_timescale}) depends on the galactic tidal field. In simulation runtime we calculate it self-consistently, but to extrapolate to $z=0$ we simply assume a constant value of $\Omega_{\rm tid}=175$~Gyr$^{-1}$. This value was chosen to produce the same final number of clusters as are observed in the MW. It results in the disruption of most clusters with masses below $10^5\Msun$, and decreases the masses of all surviving clusters, shifting the distribution to lower masses and decreasing the normalization. This calculation also assumes that no new clusters form after $z=1.9$. 

Our chosen value of $\Omega_{\rm tid}=175$~Gyr$^{-1}$ is quite high. It is equivalent to a maximum eigenvalue of the tidal tensor $\lambda_m \approx 10^5$~Gyr$^{-2}$. \citet{meng_gnedin_22} examined the tidal field for the \citetalias{li_etal_18_paper2} simulations, finding that clusters experience such strong tidal fields only shortly after their birth. As they migrate away from the high-density star-forming regions, the tidal field decreases significantly to typical values $\lambda_m \approx 3 \times 10^3$~Gyr$^{-2}$, or $\Omega_{\rm tid} \approx 30$~Gyr$^{-1}$. Choosing this low value of $\Omega_{\rm tid}$ would significantly increase the number of low-mass clusters surviving to $z=0$ in our simulations. However, this analysis was done at $z>1.5$. The value of the tidal field may increase over time as the galaxy grows. Our adopted value is also similar to that used by \citet{choksi_gnedin19} in an analytic model for cluster formation and destruction. These authors find that $\Omega_{\rm tid}=200$~Gyr$^{-1}$ can reproduce several observational results, including the GC mass function at $z=0$ and the relation between galaxy halo mass and mass of its globular cluster system. 

We compare our results with the distribution of masses of the observed MW GCs. We use the V-band absolute magnitude from \citet{harris_96} along with the luminosity dependent mass-to-light ratio
\begin{equation}
    \frac{M}{L_V} = 1.3 + \frac{4.5}{1 + \exp \left(2 M_V + 21.4 \right)}
\end{equation}
from \citet{harris_etal_17} to obtain GC masses. We find good agreement between the two mass functions. While we match the normalization by construction through our choice of $\Omega_{\rm tid}$, the similarity of the MF shape to that in the MW system is a test of the model. The maximum cluster mass matches the MW GCs well. We note that the \citet{harris_96} catalog includes both in-situ and ex-situ clusters in the MW. As the simulation $z=0$ result comes from analytic evolution of all clusters in the central galaxies at $z=1.9$, any later clusters that come in from later mergers would be missed. However, Figure~\ref{fig:cimf_rj_evolution} uses the \rj\ IC, which has quick early growth with no significant mergers after $z=1.9$ (Figure~\ref{fig:halo_growth}). We therefore expect few clusters from later infalling satellites, making a comparison to the full MW population reasonable. We also note that the \rj\ IC has more massive clusters than the \tl\ IC. This is likely becuase of its quick early growth (Figure~\ref{fig:halo_growth}), increasing the star formation density at early times and leading to the formation of more massive clusters. For the \tl\ runs, a lower value of $\Omega_{\rm tid}$ is required to reproduce the high-mass end of the Galactic GC mass function, leading to too many simulated low-mass clusters. 

In the runs with $\epsff=100$\%, our present-day mass functions have more clusters with masses above $3\times10^5\Msun$ than seen in \citetalias{li_etal_18_paper2}. This is a consequence of our initial mass functions extending to higher masses than in \citetalias{li_etal_18_paper2}. These changes are primarily driven by the addition of the virial criterion. As Figure~\ref{fig:cimf_virial} shows, the addition of this criterion significantly increases the number of massive clusters. The increase in the number of massive clusters allows us to increase the value of $\Omega_{\rm tid}$ from 50~Gyr$^{-1}$ (used by \citetalias{li_etal_18_paper2}) to 175~Gyr$^{-1}$. In that work higher values of $\Omega_{\rm tid}$ would have disrupted too many clusters. In the runs presented in this work, stronger disruption is required to produce an agreement for the massive end of the mass function while reducing the number of low-mass clusters.

Similarly to the mass function at $z=4$, the mass functions of surviving clusters at $z=0$ depend strongly on $\epsff$. For all runs with $\epsff\le 10\%$ (not shown), we find no clusters above $4\times10^5\Msun$, and the overall distributions shift to lower masses. That is clearly inconsistent with the observed mass function of MW GCs. 

\begin{figure}
    \includegraphics[width=\columnwidth]{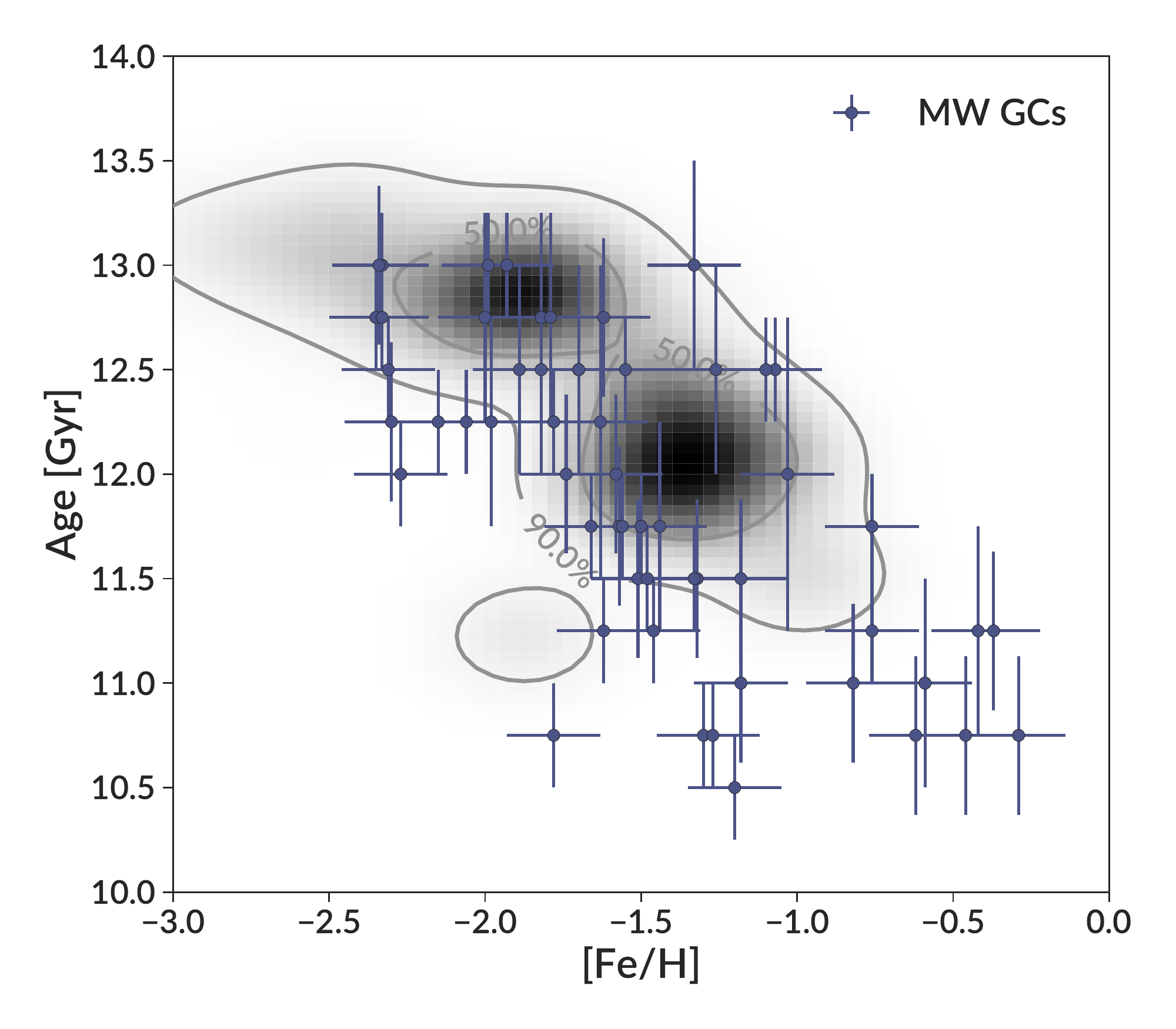}
    \vspace{-7mm}
    \caption{\revision{The age-metallicity relation for surviving simulated clusters and MW GCs. Data points show MW GCs from \citet{vandenberg_etal_13} and \citet{leaman_etal_13}. Grey shaded regions and contours indicate simulated clusters with masses above $3\times10^3\Msun$ at $z=0$ in the run using the \tl\ IC, $\epsff=100\%$, $\fboost=5$, and $\fhno=0$, with contours enclosing 50 and 90\% of the sample. Clusters from both central galaxies are included as there are no systematic differences between the two. The final output of this run corresponds to an age of 10~Gyr, meaning that all regions on the plot are accessible to the simulation.}}
    \label{fig:age_metallicity_z0}
\end{figure}

\revision{Another important relation found in observations is the age-metallicity relation of MW GCs \citep[e.g.][]{vandenberg_etal_2000,marinfranch_etal_09,dotter_etal_10}. Metal-rich clusters form systematically later than metal-poor clusters, as the galaxy enriches its interstellar medium with time.
In Figure~\ref{fig:age_metallicity_z0} we show the age-metallicity relation for simulated clusters that survive to $z=0$ in the run using the \tl\ IC, $\epsff=100$\%, $\fboost=5$, and $\fhno=0$ and compare to observations of MW GCs presented in \citet{vandenberg_etal_13} and \citet{leaman_etal_13}. We find broad agreement between the simulated cluster population and the MW GCs.}
While the plot shows only one run using the \tl\ initial condition, we see similar trends in all ICs. We note that the value of $\Omega_{\rm tid}$ used in the analytical disruption calculation slightly affects this result. A change in the disruption rate would affect which clusters that survive to the present. In particular, as most of the highest mass clusters form early, increased disruption tends to remove younger, higher metallicity clusters. While the shape of the age-metallicity relation changes little, the distribution of clusters within it does. 

\section{Discussion}
\label{sec:discussion}

\subsection{Timing of supernova feedback}
\label{sec:discussion_of_timing}

In Section~\ref{sec:timing} we discussed several prescriptions for the timing of SN, then in Section~\ref{sec:effect_of_timing} we examined how these prescriptions affect the properties of star clusters. We  find that later SN feedback leads to longer timescales for cluster formation and higher values of the integrated star formation efficiency. In this formalism, we assume that there is no difference in the formation time of low and high mass stars within a cluster. Individual stars of all masses have the same age. However, this assumption may be incorrect. For example, using a simulation of a star cluster forming out of a $2\times10^4\Msun$ GMC, \citet{grudic_etal_22} find that massive stars ($m > 10\Msun$) finish accreting 1~Myr later than the average star. \citet{padoan_etal_20} find a similar result using a simulation of $2\times10^6\Msun$ of gas in a (250~pc)$^3$ box with several star-forming regions. The delay in massive star formation in turn delays the onset of feedback. While our simulations account for the stellar age spread within the cluster when determining the timing of SN, they do not account for this systematic delay in the formation of individual stars.

As shown in Figure~\ref{fig:sn_timing}, our hybrid approach to the timing of SN feedback approximates well the delay in SN feedback due to the age spread of the stars, \revision{so it is our preferred model for future simulations.} However, it may need to be further refined to account for the delay in massive star formation. In particular, one possible approach would be to calibrate a subgrid model for the timing of cluster feedback to the results of GMC-scale simulations such as in \citet{grudic_etal_22}. Further delays in the onset of massive star feedback may increase the timescales of cluster formation and the integrated star formation efficiency, but these effects are likely to be small compared to the effects of other parameters, namely $\epsff$.

\subsection{Strength of stellar feedback}

In Figures~\ref{fig:sfh_feedback_oldic} and~\ref{fig:sfh_feedback_lg} we showed how the star formation rate of the central galaxies in our simulations changed when varying $\fboost$. We found that $\fboost=5$ produces too little star formation in the current simulation setup. In the \huiic\ runs we find that $\fboost=1-2$ matches the \um\ predictions fairly well, as do $\fboost=1-3$ in the \tl\ runs. As we discuss more in Appendix~\ref{appendix:hydro}, updates to the hydrodynamics are primarily responsible for the change in preferred values of $\fboost$. Such low values of $\fboost$ are unexpected. \citetalias{li_etal_18_paper2} calibrated $\fboost$, finding $\fboost=5$ to be their preferred value. Numerical tests in \citet{semenov_etal_17} have also shown that values of $\fboost\approx5$ best account for numerical losses of momentum as a SN shell moves across the simulation grid. Theoretical grounds for $\fboost >1$ also exist, with \citet{gentry_etal_17} finding that clustered SN can enhance momentum feedback by up to an order of magnitude relative to an isolated SN.

We also note that all of these runs, even with $\fboost=1$, show a large decrease in the star formation rate at $z<2$, in conflict with the abundance matching expectation. Both the hydrodynamics and feedback models have been updated to be more physically realistic than those used in \citetalias{li_etal_18_paper2}, but produce worse agreement in the star formation histories. This may indicate that there is additional relevant physics that needs to be included in our simulation.

Our model assumes that all stars above $8\Msun$ explode as SN. However, this assumption may not hold. Simulations of SN find that some progenitors collapse directly to a black hole without a SN explosion \citep{heger_etal_03,horiuchi_etal_14,pejcha_thompson_15}. If we were to include such scenarios in the feedback scheme, it would decrease the total energy and momentum from SN. We would therefore require a higher value of $\fboost$ to obtain reasonable star formation rates. Additionally, changing the minimum progenitor mass for SN makes a large difference in the energy injected by SN \citep{keller_kruijssen_22}. We assume $M_{\rm min} = 8 \Msun$, but this value is uncertain. Increasing it would decrease the number of SN, again requiring a higher $\fboost$ to compensate.

\subsection{Constraints on star formation efficiency}
\label{sec:discussion_epsff}

\begin{figure}
    \includegraphics[width=\columnwidth]{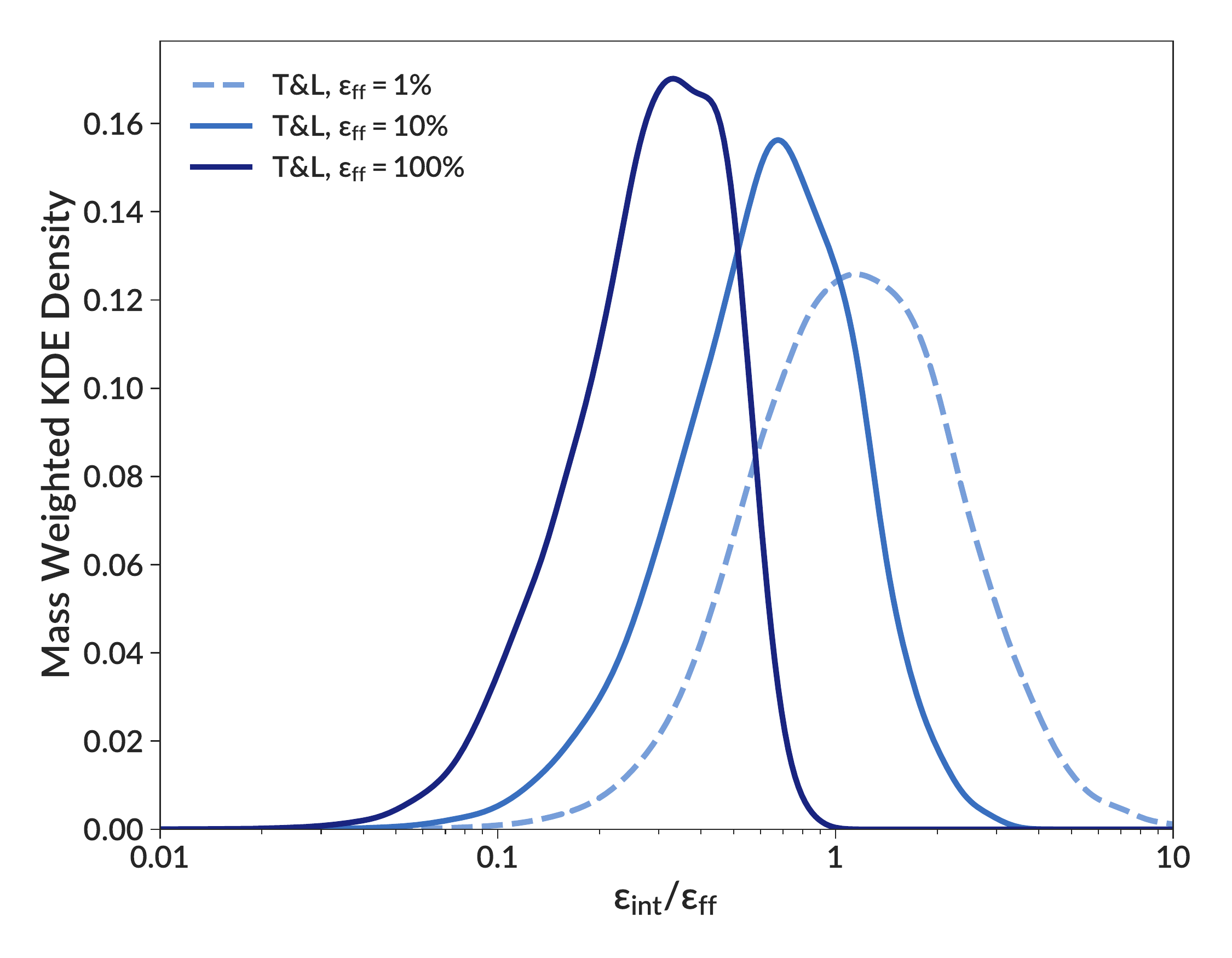}
    \vspace{-7mm}
    \caption{The distribution of $\epsint/\epsff$ for different values of $\epsff$ in the \tl\ IC. All runs used $\fboost=5$, $\fhno=20$\%, and show all clusters formed before \revision{$z=3.3$, the lowest redshift that all simulations have reached.} We plot the $\epsff=1$\% run with a dashed line as that run had many clusters that failed to finish forming. }
    \label{fig:eps_int_eps_ff_ratio}
\end{figure}

Figure~\ref{fig:eps_int_tl_sfe} shows the distribution of $\epsint$ for the runs varying $\epsff$. While we find a clear trend that decreasing $\epsff$ decreases $\epsint$, we can also examine the ratio $\epsint/\epsff$. Figure~\ref{fig:eps_int_eps_ff_ratio} shows this ratio for the \tl\ runs, which can be directly compared with Figure~8 of \citetalias{li_etal_18_paper2}. For all values of $\epsff$ we consider, we find higher values of $\epsint/\epsff$ than did \citetalias{li_etal_18_paper2}. Two of the changes discussed in Section~\ref{sec:results} are responsible. First, the modified SN feedback prescription delays SN feedback compared to \citetalias{li_etal_18_paper2}, which results in higher $\epsint$ (Figure~\ref{fig:eps_int_sn_timing}). Second, the introduction of the virial criterion leads to higher $\epsint$ at a given $\epsff$ (Figure~\ref{fig:eps_int_virial}). Combined, these two effects shift our distributions of $\epsff / \epsint$ to higher values.

We still see the same trend with $\epsff$ as did \citetalias{li_etal_18_paper2}, where higher values of $\epsff$ lead to smaller $\epsint/\epsff$ ratios. Quantitatively, the mean value of this ratio drops from 1.15 to 0.57 to 0.30 for $\epsff=1$\%, 10\%, and 100\%, respectively. Conceptually, this ratio is proportional to the number of freefall times over which the cluster accreted material. As discussed in Section~\ref{sec:results_epsff}, lower values of $\epsff$ lead to longer formation timescales, in accordance with this result. 

As we have discussed throughout Section~\ref{sec:results}, the duration of a star formation episode is sensitive to cluster feedback and formation prescriptions. While difficult to constrain precisely, current observations indicate that the age spread within clusters is less than $\approx 6$~Myr (see the compilation of age data in \citetalias{li_etal_18_paper2}). These age spreads can still be significantly larger than $t_{\rm ff}$. One example is the Orion Nebula Cluster (ONC), where star formation appears to have occurred over several freefall times \citep{da_rio_14,caldwell_chang_18,kounkel_etal_18}. In contrast, simulations of individual molecular clouds generally show star formation ending after one $t_{\rm ff}$ \citep[e.g.][]{grudic_etal_22}. Our simulated age spreads, shown in Figure~\ref{fig:age_spread_oldic_sfe}, are consistent with the observations for $\epsff \ge 10\%$. We see a strong \revision{mass} trend, but even for massive clusters the vast majority have age spreads smaller than 6~Myr. However, our results rule out $\epsff=1$\%, which has unphysically long age spreads for clusters of all masses.

The shape of the initial cluster mass function is another key observable. YMCs in the MW and nearby galaxies are found to follow the functional form of \citet{schechter_76}, with a power law slope of $-2$ at the low-mass end \citep{portegies_zwart_etal_10}. Our mass functions have a positive power law slope at low mass, peak at a mass that depends on $\epsff$ ($10^5\Msun$ for $\epsff=100$\%), then decline in a manner consistent with a power-law. In essence, our simulations are missing low-mass clusters. While our cluster formation algorithm only seeds clusters if they have an expected mass of $6\times10^3\Msun$, runs with $\epsff=100$\% show the increasing mass function above this mass. This may indicate that $\epsff=100$\% forms stars too efficiently, leading to too few low-mass clusters. However, $\epsff \leq 10$\% results in too few massive clusters, with no clusters projected to reach $z=0$. 

Lastly, $\epsff$ has been measured in observations with several methods \citep{evans_etal_14,usero_etal_15,lee_etal_16,ochsendorf_etal_17,utomo_etal_18}. While the observations have somewhat different medians, uncertainties, and intrinsic scatter in $\epsff$, a value of $\epsff\approx1$\%  is typical. However, we find that this value does not produce reasonable star cluster properties in our simulations. The timescales of cluster formation reach our algorithmically imposed limit of 15~Myr. Such timescales are in conflict with observations. Low values of $\epsff$ also produce few massive clusters. Even a value of $\epsff=10$\% produces few clusters with high enough mass to reach $z=0$ as GCs. Our simulations prefer higher values of $\epsff$. Among the runs presented here, $\epsff=100$\% produced the most realistic cluster properties, as it did in \citetalias{li_etal_18_paper2}.


\revision{To compare with observations more directly, we postprocess the simulations to calculate an effective value of $\epsff$ in a way analogous to how it is derived in observations. First we identify clusters that are actively forming in several simulation snapshots. Within a sphere of radius $r$ centered on the cluster, we calculate the inferred value of $\epsff$ as
\begin{equation}
    \epsffinf(r) = \frac{\bar{t}_{\rm ff}(r) \, \dot{M}(<r)}{M_{\rm gas}(<r)}
\end{equation}
where $\bar{t}_{\rm ff} \equiv \sqrt{3 \pi / 32 G \bar{\rho}}$ is calculated using the mean density $\bar{\rho}$ within the sphere. In the rest of this section we will use $\epsffinf$ to refer to the inferred value from this procedure, while $\epsff$ will refer to the value used in runtime of the simulation.
To calculate $\dot{M}$, we use a procedure analogous to that used in studies that determine $\epsff$ by counting young stellar objects (YSOs) to determine the star formation rate within a cloud \citep{evans_etal_14,heyer_etal_16,ochsendorf_etal_17}. 
These studies use YSOs to estimate the mass of recently formed stars, then divide it by the lifetime of the YSO phase typically set to a fixed time of 0.5~Myr. As we do not store the full accretion histories of simulated clusters, we cannot directly obtain the star formation rate over the last 0.5~Myr. Instead, we approximate it with the average star formation rate over the relevant timescale:
\begin{equation}
    \dot{M} = \frac{M}{\max\left(\tau_{\rm spread}, 0.5\rm{\ Myr}\right)}
    \label{eq:mdotav}
\end{equation}
where $M$ is the current mass of the actively forming cluster. For clusters with large age spreads this prescription gives the average star formation rate, while for clusters with short age spreads this matches the rate inferred observations assuming an 0.5~Myr timescale. We choose to use the cluster age spread rather than the full duration as it more accurately reflects the timescale over which the bulk of cluster formation happens. The total $\dot{M}$ within a given sphere is the sum of $\dot{M}$ from all actively forming clusters in the sphere.}

\revision{This calculation of $\epsffinf$ involves significant averaging both in time and space, compared to the application in simulation runtime. A typical local timestep at the highest refinement levels is 100--1000~yr, orders of magnitude shorter than even 0.5~Myr. Therefore, the finite difference calculation of the star formation rate $\dot{M}$ from Equation~\ref{eq:m_dot} is a much closer approximation to the true derivative than Equation~\ref{eq:mdotav}. Considering spheres of radius $r > 5$~pc also introduces averaging of the stellar and gas mass on a larger scale than our adopted GMC radius. Both of these effects tend to shift $\epsffinf$ to smaller values than the input $\epsff$.}

\begin{figure}
    \includegraphics[width=\columnwidth]{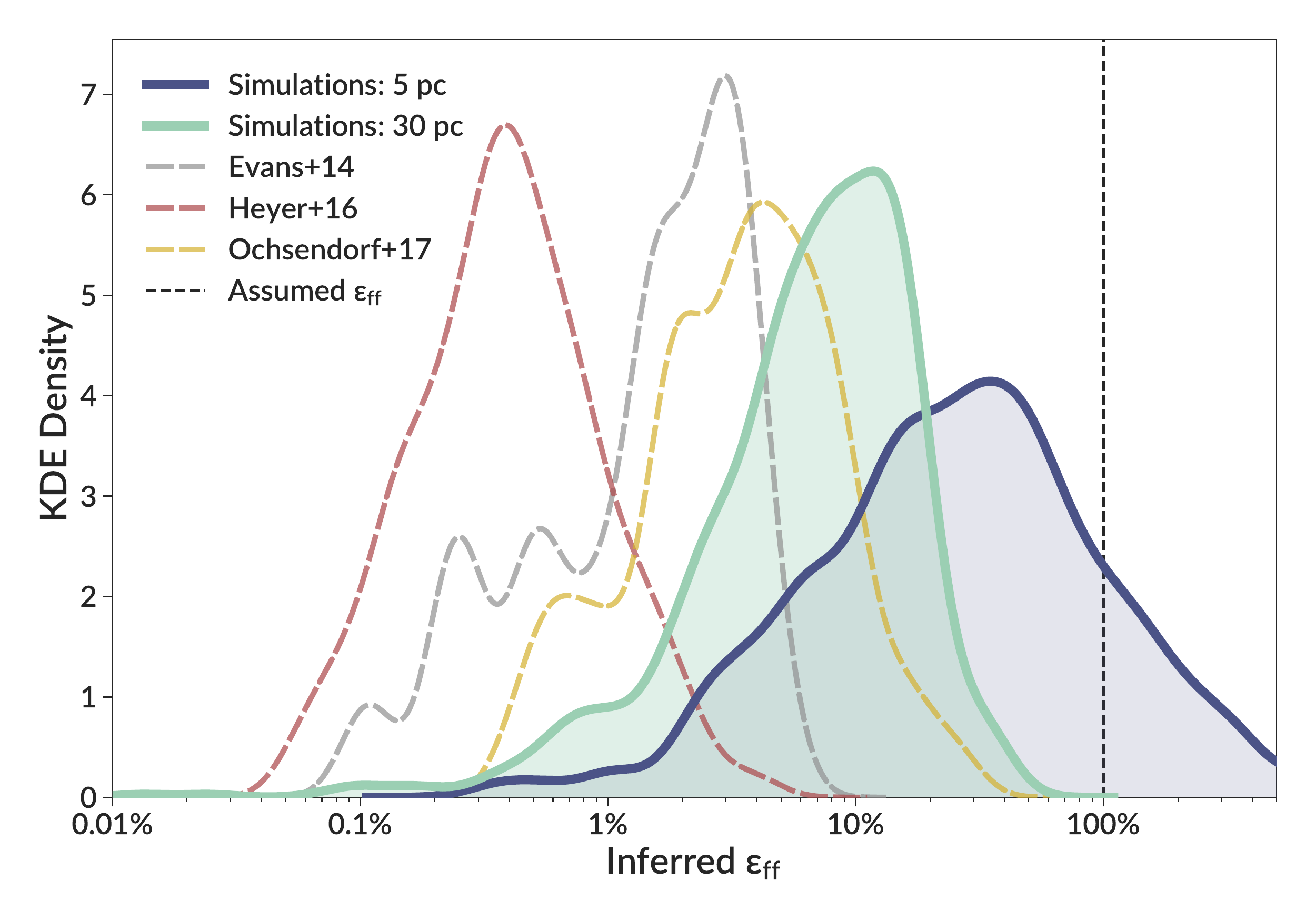}
    \vspace{-7mm}
    \caption{\revision{Kernel density estimation of the distribution of $\epsff$ inferred from postprocessing simulated star clusters and from observations. We use a normalized Gaussian kernel with a width of 0.15 dex. The simulated clusters are from the run with the \huiic\ IC, $\epsff=100$\%, $\fboost=1$, and $\fhno=0$. The 5~pc line shows $\epsffinf$ as inferred from the region actively participating in cluster formation, while the 30~pc line shows the value inferred for larger star-forming complexes.}}
    \label{fig:inferred_epsff}
\end{figure}

\revision{In Figure~\ref{fig:inferred_epsff} we show the distribution of values of $\epsffinf$ calculated for two choices of the averaging radius: 5~pc and 30~pc. As there are few clusters actively forming in any given snapshot, we use all snapshots from $z=9-1.5$ in the run using the \huiic\ IC, $\epsff=100$\%, $\fboost=1$, and $\fhno=0$, giving a sample of 748 actively forming clusters.
The radius of 5~pc matches the GMC sphere actively participating in star formation. The inferred values peak at around 30\% with large scatter but are significantly below the simulation input $\epsff=100$\%. 
The procedure to infer $\epsffinf$ uses the cluster formation timescale to average the star formation rate, which creates the wide spread and systematic shift.
In addition, this procedure calculates $t_{\rm ff}$ and $M_{\rm gas}$ at one instant, which may not reflect typical conditions over the course of the cluster's growth.
}

\revision{
Considering a larger sphere radius of 30~pc adds also spatial averaging. For an isolated cluster, increasing the size of the sphere would simply include more surrounding gas without increasing $\dot{M}$, leading to smaller inferred values of $\epsffinf$. 
However, we find that clusters often form in larger star-forming complexes with many clusters in close proximity of each other. Our choice of 30~pc corresponds to the typical radius of these star-forming complexes. 
Figure~\ref{fig:inferred_epsff} shows that these complexes have a peak value of $\epsffinf \approx 10\%$, with less scatter than the values inferred on 5~pc scales. The lower mean value is due to the inclusion of more gas not participating in star formation, while the decreased scatter comes from averaging together multiple clusters within each region.}

\revision{This exercise shows that the inferred values of $\epsffinf$ are a factor of 10 lower than the simulation input. Still, for this run typical $\epsffinf$ are higher than those seen in observations. In Figure~\ref{fig:inferred_epsff} we include observations from \citet{evans_etal_14}, \citet{heyer_etal_16}, and \citet{ochsendorf_etal_17}, which all use the YSO method but do so on different scales. \citet{evans_etal_14} and \citet{heyer_etal_16} use clumps with typical radii of a few pc, while \citet{ochsendorf_etal_17} uses star-forming complexes with radii around 40~pc. Even with these differences of scale, all studies measure mean values of $\epsff$ consistent with $\sim$1\%. However, we note that we cannot make a direct comparison between these observations and our inferred values of $\epsffinf$. Each ingredient of the calculation of $\epsff$ has systematic differences. The mass of recently formed stars is calculated differently, as we do not directly model the number of observable YSOs in each cluster. The timescales for calculating the star formation rate are also different, as many of our clusters have $\tau_{\rm spread}$ longer than the 0.5~Myr used in observations. Lastly, detailed modeling of CO and HCN abundances and ionization states is needed to calculate $M_{\rm gas}$ exactly as is done in observations. To resolve these differences would require a further analysis in the simulation runtime. Nevertheless, Figure~\ref{fig:inferred_epsff} demonstrates that the discrepancy with observations is substantially smaller than appears from a straightforward comparison with the simulation input.}

\subsection{Failed cluster formation}
\label{sec:failed}

In Section~\ref{sec:results_epsff} we showed that in some runs with low $\epsff$, clusters fail to finish formation before it is automatically ended at 15~Myr. In this section, we investigate the reasons for these failed clusters.

We find that no runs with the high value of $\epsff=100$\% have failed cluster formation, all runs with the low value of $\epsff=1$\% fail, and among the runs with the intermediate value $\epsff=10$\%, only the run using the \huiic\ IC and $\fboost=1$ failed. All other runs with $\epsff=10$\% used higher $\fboost$ and did not fail. \revision{In total, 4 of our 29 runs experience failed cluster formation.}

These trends are due to an interplay between $\epsff$ and $\fboost$. When $\epsff$ is low, cluster formation progresses slowly, leaving significant amounts of gas. We find low values of $\epsint$ for low $\epsff$ (Figure~\ref{fig:eps_int_tl_sfe}), meaning that at the end of cluster formation, only a small fraction of gas has been turned into stars. This applies in the midst of cluster formation, too. We examine the gas densities of the host cells of clusters as they form and find that for lower values of $\epsff$ there is more gas near the cluster at a given time after the beginning of cluster formation, meaning that GMCs are more massive with low values of $\epsff$. In addition, the slower star formation with low $\epsff$ leads to fewer stars to provide feedback. When SNe begin, they must first disperse the gas within the cluster. Higher values of $\fboost$ make this process more efficient. Therefore, higher values of $\fboost$ lead to shorter timescales for cluster formation when $\epsff$ is low. In contrast, when $\epsff=100$\%, clusters consume a high fraction of the gas within their GMC. SN feedback of any $\fboost$ is able to clear the smaller amounts of gas that remain. 

\begin{figure}
    \includegraphics[width=\columnwidth]{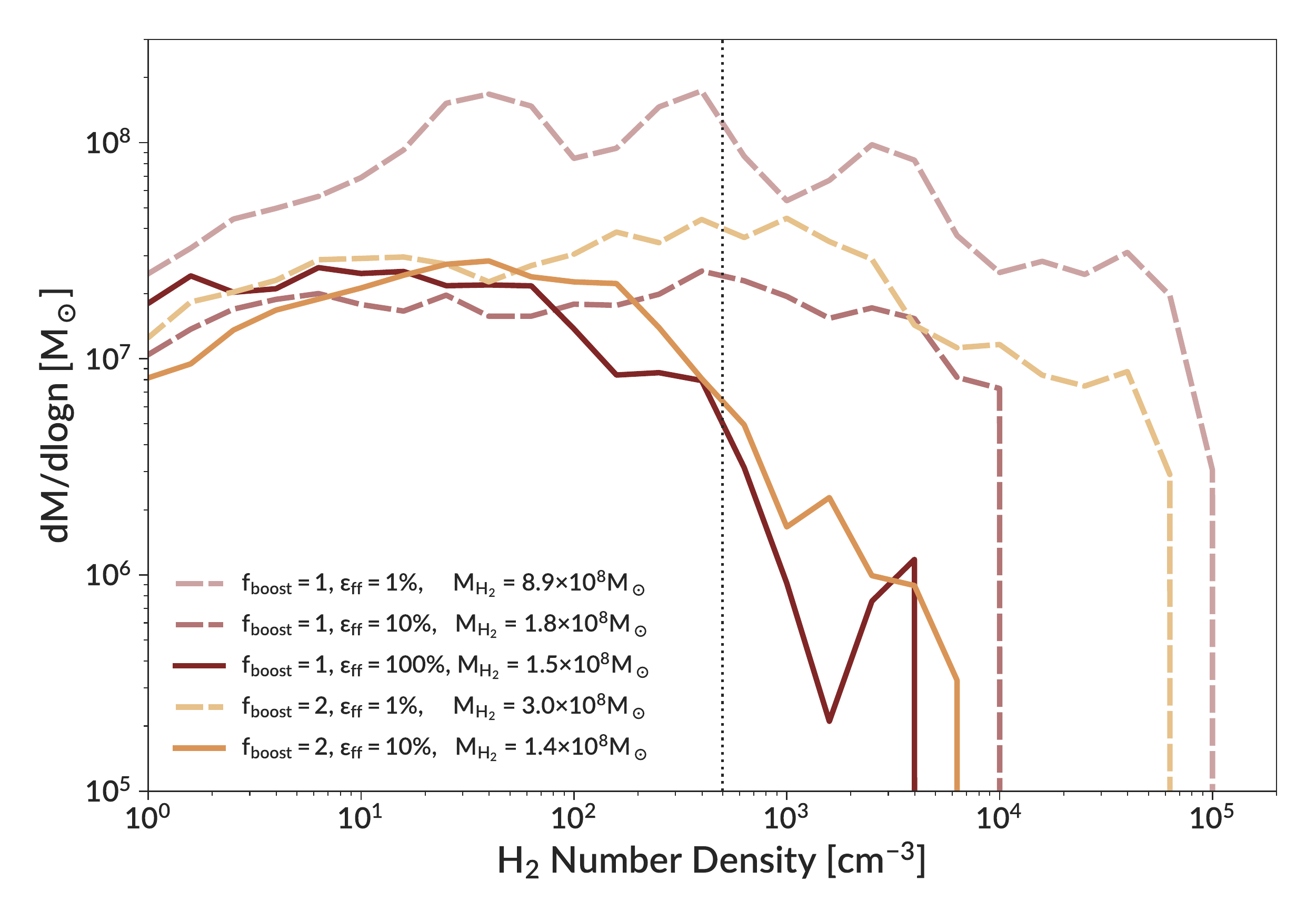}
    \vspace{-7mm}
    \caption{The distribution of cell mass-weighted molecular gas densities for different combinations of $\epsff$ and $\fboost$ in the \huiic\ IC. All runs used $\fhno=0$ and show the gas within 10~kpc of the galactic center. Dashed lines indicate runs with failed cluster formation. Runs with $\fboost=1$ are shown at $z=1.56$, while those with $\fboost=2$ are shown at $z=2.57$. The dotted line at 500~cm$^{-3}$ indicates the minimum density for star formation.}
    \label{fig:h2_pdf}
\end{figure}

In Figure~\ref{fig:h2_pdf} we illustrate this effect by presenting instantaneous distribution of the molecular gas density within galaxies with different combinations of $\epsff$ and $\fboost$. Runs with failed cluster formation have distributions that extend to higher densities than runs without failed cluster formation. As feedback cannot terminate star formation, gas continues to accrete onto the GMC, increasing the density. Of particular note is the gas at densities higher than the star formation threshold. The total gas number density must be greater than $10^3$~cm$^{-3}$ with a molecular fraction of 0.5, giving a minimum molecular number density of 500~cm$^{-3}$. Above this threshold, the failed runs have significantly more gas than runs that successfully terminate star formation. 

The prescription for SN feedback also contributes to why these runs had failed cluster formation while the runs of \citetalias{li_etal_18_paper2} did not. In the 15~Myr timescale for cluster formation, the prescription of \citetalias{li_etal_18_paper2} injects significantly more energy than the updated model (see Figure~\ref{fig:sn_energy}). For low hypernova fractions typical of most clusters, the new prescription injects only 26\% of the energy of \citetalias{li_etal_18_paper2} within the first 15~Myr, increasing to about 50\% after 40~Myr. This is exacerbated by the lower $\fboost$ used in the updated runs. The total momentum injected by SN feedback during the 15~Myr of cluster formation can be more than an order of magnitude lower than in \citetalias{li_etal_18_paper2}. For low values of $\epsff$, this results in SN feedback being unable to disperse the GMC.

The timing of SN also contributes to failed cluster formation. Compared to \citetalias{li_etal_18_paper2}, SN start later in the new prescription due to the stellar lifetimes chosen (Figure~\ref{fig:sn_energy}). These runs also use the average approach for determining cluster feedback timing, as they were run before the hybrid approach was finalized. This average approach further delays the onset of SN (Figure~\ref{fig:sn_timing}). In addition, we find that runs with lower $\epsff$ have later average times of cluster formation, $t_{\rm ave}$, meaning that SN is delayed even further in these runs. These delays in the onset of SN gives the gas outside the GMC more time to accrete onto the GMC, leading to higher gas masses that SN feedback then needs to disperse. This combines with the effect described in the previous paragraphs to make GMCs more difficult to disperse for lower $\epsff$, further explaining why we find that $\fboost$ affects the timescales of cluster formation for low $\epsff$ but not for $\epsff=100$\%.

To summarize our understanding of why massive star clusters fail to finish forming when $\epsff \le 10$\%, lower values of $\epsff$ turn gas into stars at a slower pace. At a given time after the beginning of cluster formation this results in fewer stars, and therefore less stellar feedback, embedded in a more massive GMC. The onset of SN is delayed compared to \citetalias{li_etal_18_paper2} due to our choice of stellar lifetimes (see Figure~\ref{fig:sn_energy}), and then is delayed further after accounting for the age spread within the cluster. This allows more material to accrete onto the GMC, making it even more difficult for feedback to disperse. Once SN feedback starts, the updated feedback prescription injects less momentum than \citetalias{li_etal_18_paper2}. The new prescription has fewer SN and is further exacerbated if low values of $\fboost$ are chosen. Although lower values of $\fboost$ produce more reasonable star formation rates for $z > 2$, these low values fail to provide enough feedback to disperse GMCs when $\epsff$ is low. This may indicate that another source of feedback is needed at early times to help disperse GMCs \revision{or that the combination of $\epsff\leq10$\% and $\fboost=1$ is ruled out by our simulations}.

\section{Conclusions}
\label{sec:conclusions}

We have described improvements to the implementation of star cluster formation and feedback in the ART code. We introduced a new criterion for the seeding of cluster particles, requiring the star-forming gas to be gravitationally bound. We also implemented a new prescription for the initial bound fraction of clusters based on simulations of individual GMCs. We added runtime tracking of C, N, O, Mg, S, Ca, and Fe, with enrichment coming from SNIa, SNII, stellar winds, and AGB stars. We updated the SN feedback prescriptions significantly. We now implement SN as discrete events, with rates based explicitly on the stellar lifetimes and IMF. We also explored effects of hypernovae, which inject more energy and have different elemental yields. Lastly, we improved our prescription for the timing of SN to account for the age spread of stars within a cluster. 

With these code updates, we ran 20 simulations using the initial condition from \citetalias{li_etal_18_paper2} and 9 simulations using two Local Group-like ICs from the ELVIS project. These runs have a range of parameters, including variations in $\epsff$, $\fhno$, $\fboost$, and the timing of SN feedback. We explored how these parameters affect the properties of galaxies as well as the populations of star clusters within them. Our results are summarized as follows.

$\bullet$ Delaying the onset of SN (without changing the total energy injection) results in longer formation timescales for massive clusters and higher $\epsint$ (Figures~\ref{fig:age_duration_sn_timing}, \ref{fig:eps_int_sn_timing}), but does not significantly change the galaxy star formation rate. 

$\bullet$ Higher values of the momentum boosting factor for SN greatly decrease the galactic star formation rate (Figures~\ref{fig:sfh_feedback_oldic}, \ref{fig:sfh_feedback_lg}). While no value of $\fboost$ can reproduce the abundance matching expectation for the full redshift range explored here ($z>1.5$),  we find that the range $1 < \fboost < 3$ produces reasonable agreement for $z > 2$. Higher values of $\fboost$ decrease the total stellar mass by decreasing the number of low-mass clusters that form, without changing the number of massive clusters (Figure~\ref{fig:cimf_fboost_oldic}).

$\bullet$ The hypernova fraction $\fhno$ makes little difference to galaxy or cluster properties (Figures~\ref{fig:sfh_feedback_oldic}, \ref{fig:cimf_fboost_oldic}). The strong decrease in $\fhn$ with metallicity (Equation~\ref{eq:hn_fraction}) results in limited change in the total energy injected by SN (Figures~\ref{fig:sn_energy}, \ref{fig:age_metallicity_hn}). 

$\bullet$ The local star formation efficiency per freefall time does not have a strong impact on the galactic star formation rate (Figure~\ref{fig:sfh_sfe}). However, it strongly changes cluster properties. Higher values of $\epsff$ lead to more massive clusters (Figures~\ref{fig:cimf_sfe_oldic}, \ref{fig:cimf_sfe_lg}), shorter timescales for cluster formation (Figure~\ref{fig:age_spread_oldic_sfe}), higher initial bound fractions (Figure~\ref{fig:bound_fraction_tl_sfe}), and higher $\epsint$ (Figure~\ref{fig:eps_int_tl_sfe}). 

$\bullet$ Adding the virial parameter criterion to require star-forming gas be gravitationally bound produces more high-mass clusters (Figure~\ref{fig:cimf_virial}), longer timescales for cluster formation, and higher $\epsint$ (Figure~\ref{fig:eps_int_virial}). 

$\bullet$ In runs with low values of $\epsff$, we find a population of clusters that fail to finish forming after 15~Myr. Low values of $\epsff$ form stars slowly, leaving massive GMCs that are difficult for feedback to disperse, especially with low values of $\fboost$.

$\bullet$ We present the evolution of the observable mass function of clusters at various redshifts (Figure~\ref{fig:cimf_rj_evolution}). Most massive clusters form at high redshifts when the star formation density is high, with low-mass clusters dominating in quiescent epochs. 

$\bullet$ We analytically extrapolate the dynamical disruption of clusters from the last available output to $z=0$ (Figure~\ref{fig:cimf_rj_evolution}). We can match the observed mass function of MW GCs by assuming a high value for the cluster disruption rate. The surviving clusters also match the age-metallicity relation of MW GCs (Figure~\ref{fig:age_metallicity_z0}).

$\bullet$ Among the values of $\epsff$ we explored, only $\epsff=100$\% can match the MW GC mass function. Runs with  $\epsff=1$\% produces clusters with unphysically long age spreads (Figure~\ref{fig:age_spread_oldic_sfe}), and runs with $\epsff=10$\% produce too few high-mass clusters (Figures~\ref{fig:cimf_sfe_oldic}, \ref{fig:cimf_sfe_lg}, \ref{fig:cimf_current_lg_sfe_z4}).

This exploration emphasizes the importance of well-calibrated subgrid models for modeling star clusters in simulations of galaxy formation. Some modeling choices, such as the optimal value of $\epsff$, whether to enforce a virial criterion when seeding star clusters, and different prescriptions for the timing of SN feedback all affect the resulting cluster populations without significantly impacting global galaxy properties. A successful model of star formation and feedback in simulations must be able to reproduce not only galaxy-scale properties, but also the small-scale properties of individual star clusters.

\section*{Acknowledgements}

We thank Vadim Semenov, Hui Li, Eric Bell, Mateusz Ruszkowski, and Gus Evrard for helpful discussions. We also thank the referee for suggestions that improved the paper. GB and OG were supported in part by the U.S. National Science Foundation through grant 1909063 and by NASA through grant HST-AR-16614.001-A.

\section*{Data Availability}

The data underlying this article will be shared on reasonable request to the corresponding author.


\bibliographystyle{mnras}
\bibliography{bibfile} 


\appendix
\section{Hydrodynamics}
\label{appendix:hydro}

\begin{figure}
    \includegraphics[width=\columnwidth]{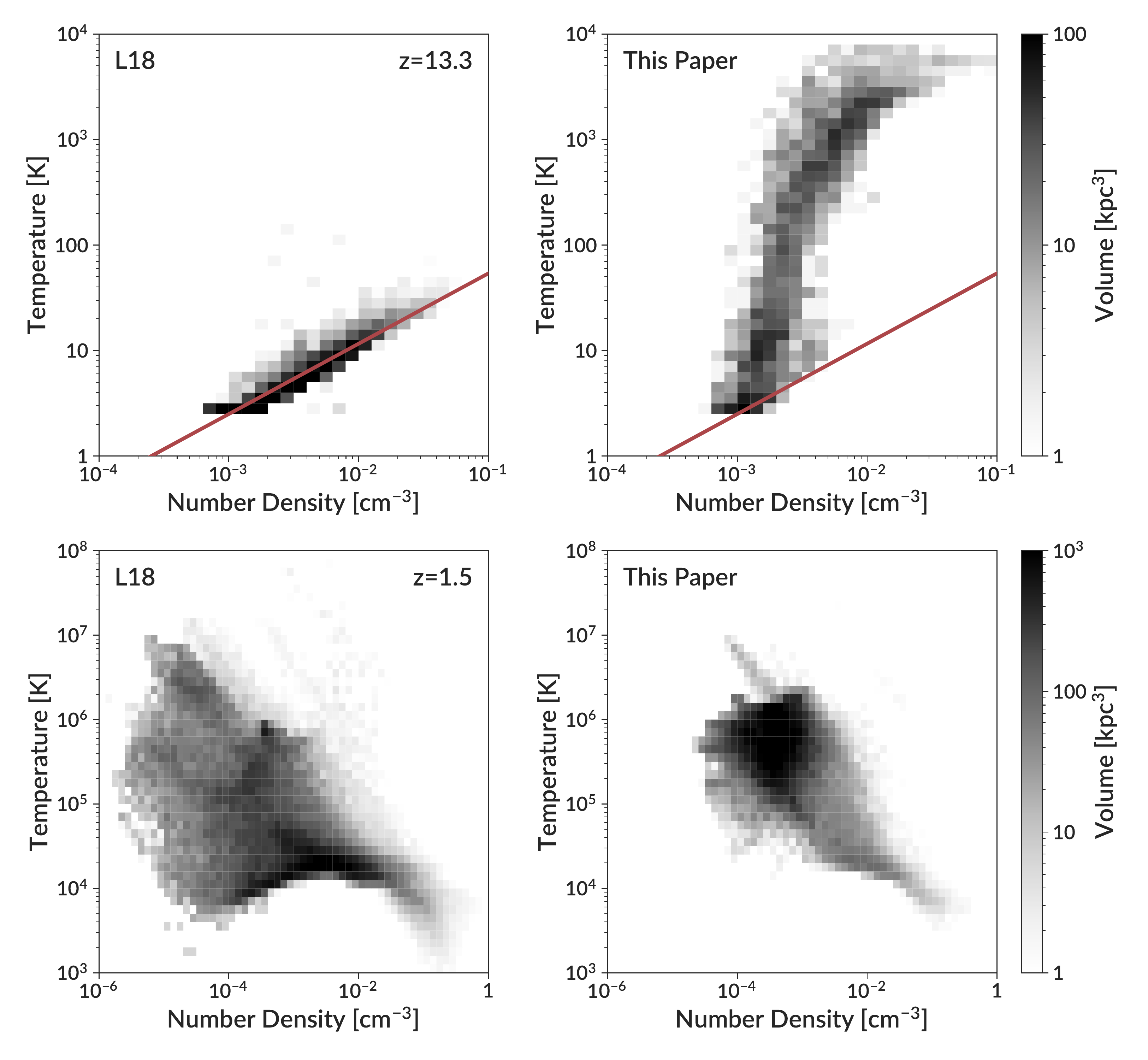}
    \vspace{-5mm}
    \caption{Heatmap showing the temperature and density of gas within the virial radius of the largest halo. In each panel, the shading shows the volume of gas at the given temperature and density. The left column shows a run using the hydrodynamic scheme of \citetalias{li_etal_18_paper2}, $\epsff=100$\%, $\fboost=5$, and $\fhno=0$, while the right column shows the run with \revision{the updated} energy-based hydrodynamics scheme,
    $\epsff=100$\%, $\fboost=1$, and $\fhno=0$. The top row show these runs at $z=13.3$ before any stars formed, while the bottom row shows the runs at $z=1.5$. In the top panels, the red line shows the expected behavior for pure adiabatic compression. The code version of L18 exactly follows this line, while the updated version has extra heating from proper treatment of shocks.}
    \label{fig:app_gas_phase}
\end{figure}

When updating from the version of the ART code used in \citetalias{li_etal_18_paper2}, we changed the model of how internal energy is calculated in the presence of subgrid turbulence. The hydro solver independently tracks total energy, thermal energy, and energy of unresolved subgrid turbulence. The thermal energy and subgrid turbulence are assumed to evolve adiabatically (other than energy injection from sources such as stellar feedback). As these are calculated independently, there is no initial restriction for the sum of thermal, kinetic, and turbulent energies to equal the total. As the adiabatic assumption is not always correct for thermal energy (particularly in shocks), the new version calculates the thermal energy as $E_{\rm th} = E_{\rm tot} - E_{\rm kinetic} - E_{\rm turb}$. This energy synchronization allows for shocks to transfer energy from kinetic to thermal, as should happen. The adiabatic assumption is only used in cases where the gas is highly supersonic, such that $E_{\rm tot} \approx E_{\rm kinetic}$. In this case, the subtraction would be susceptible to numerical errors, so we revert to the adiabatic assumption. In the old version of the code, which \textit{always} relied on the adiabatic assumption, shocks were not treated properly and energy that should have been transferred from kinetic to thermal was simply lost. This is visualized in the top row of Figure~\ref{fig:app_gas_phase}, where we show the phase diagram of gas within the virial radius at $z=13.3$ before stars have formed. The hydrodynamic scheme of \citetalias{li_etal_18_paper2} follows what is expected for pure adiabatic compression, while the new scheme shows gas being heated by virial shocks.

\begin{figure}
    \includegraphics[width=\columnwidth]{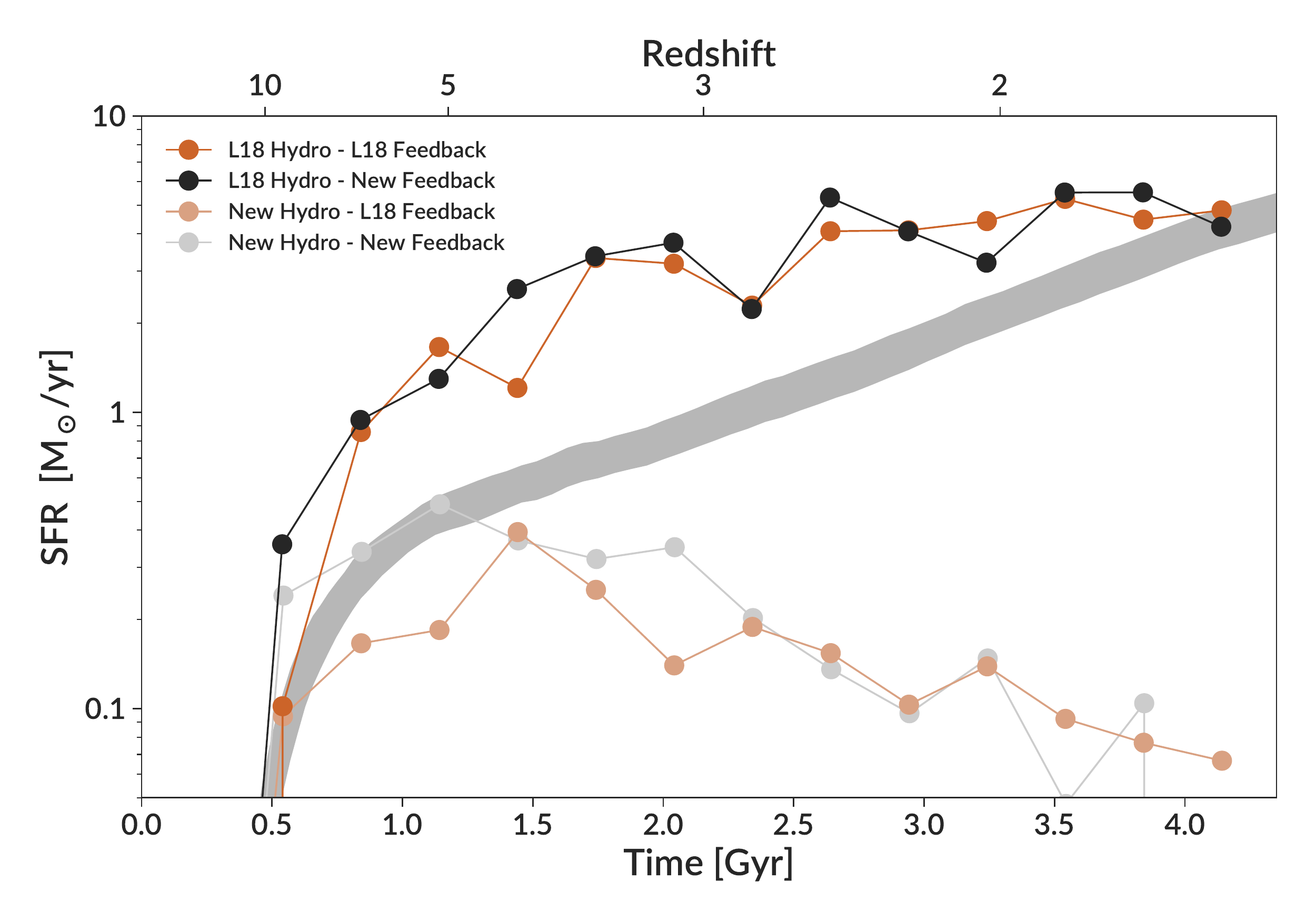}
    \vspace{-7mm}
    \caption{The star formation history for galaxies in our test runs with varying hydrodynamics and feedback. We compare the feedback model of \citetalias{li_etal_18_paper2} to the feedback model presented in this paper. All runs use $\epsff=100$\%, $\fboost=5$, and $\fhno=0$. We compare to the \um\ model \citep{behroozi_etal_19}. The change in hydrodynamics is solely responsible for the change in star formation rate, while our updates to feedback have little effect.}
    \label{fig:app_sfh_hydro}
\end{figure}

While the newer version of the code is better physically motivated, it significantly changed the structure of modeled galaxies. We find large differences in temperature distributions of the gas. The bottom panel of Figure~\ref{fig:app_gas_phase} shows the phase diagram of gas within the virial radius at $z=1.5$. Here the run with the updated hydrodynamics has significantly more hot, low-density gas in the halo. This hot gas prevented cold gas from accreting onto the disc, effectively reducing star formation. We show this star formation in Figure~\ref{fig:app_sfh_hydro} using test runs that vary both the hydro and feedback schemes. We test the stellar feedback model presented in this paper as well as the model used by \citetalias{li_etal_18_paper2}. All runs use $\fboost=5$, yet runs with the new hydro scheme produce dramatically lower star formation rates. 

\bsp	
\label{lastpage}
\end{document}